\documentclass{cernyrep} 
\usepackage{texnames}
\usepackage[T1]{fontenc}
\usepackage{amsmath,amssymb,hyperref,url,color}
\newcommand{\Half}{\frac{1}{2}}

\newcommand{\comment}[1]{}
\newcommand\dsl{{\partial\kern-6.5pt/}}
\newcommand\Dsl{{D\kern-6.5pt/}}
\def\GTS{\raise.3ex\hbox{$>$\kern-.75em\lower1ex\hbox{$\sim$}}}
\def\LTS{\raise.3ex\hbox{$<$\kern-.75em\lower1ex\hbox{$\sim$}}}
\begin{document}
\title{Field Theory and EW Standard Model.}\footnote{Notes of lectures given at the second Asia Europe Pacific School in High Energy Physics (AEPSHEP) school held in Puri, India, 4-17 November, 2017. To appear in a CERN yellow report.}

\author{Rohini M. Godbole}

\institute{Centre for High Energy Physics, Indian Institute of Science, Bangalore, 560012, India.}

\maketitle 

\begin{abstract}
In this set of four lectures I will discuss  some aspects of the Standard Model (SM) as a quantum field theory and related phenomenological observations which have played a crucial role in establishing the $SU(2)_{L} \times U(1)_{Y}$ gauge theory as the correct description of Electro-Weak (EW) interactions. I will first describe in brief the idea of EW unification as well as basic aspects of the Higgs mechanism of spontaneous symmetry breaking. After this  I will discuss anomaly cancellation, custodial symmetry and implications of the  high energy behavior of scattering amplitudes for the particle spectrum of the EW theory. This will be followed up by a discussion of the  'indirect' constraints  on the SM particle masses such as $M_{c}, M_{t}$ and $M_{h}$ from various precision EW measurements. I will end by discussing the theoretical limits on $M_{h}$ and implications of the observed Higgs mass for the SM and beyond.
\end{abstract}

\section{Introduction}
 I am asked to discuss 'Field Theory and the EW Standard Model' in these four lectures. The title  encompasses developments of the last 60-70 years. These lectures are happening on the backdrop of the  discovery of the Higgs at the LHC~\cite{bib:higgs-discovery}, the concluding finale of the establishment of the correctness of the Standard Model as the theoretical description of  EW interactions.  To cover this entire journey in four lectures, clearly I have had to pick and choose a few topics. I have done after sharing a questionnaire with all of you.

I would like to focus on the salient and non negotiable aspects of EW phenomenology which helped establish the $SU(2)_{L} \times U(1)_{Y}$ gauge field theory as the correct theory of the EW interactions. In this I will like to tell the story of how requirements of consistency of EW theory itself have guided us in the development of Standard Model (SM), as we know it today, by setting up the goal posts for theory and experiments. I will begin by discussing some aspects of the pre-gauge theory description  of weak interactions in terms of a current-current Lagrangian. As we understand today this is the effective theory which results from the $SU(2)_{L} \times U(1)_{Y}$ description, when the heavy gauge boson fields have been integrated out. It is interesting to understand the role that various features of this effective description  have played in helping us 'infer' the more fundamental theory which is the SM. I will try to point out some of these. I will then begin a discussion of SM as a gauge theory,  by first setting up the notation of the SM Lagrangian followed by a somewhat brief discussion of the Higgs mechanism. Then I give a very brief summary of the successes of the SM all the way from its formulation till date. I will then discuss relationship between the particle spectrum of the SM  and the twin issues of  anomaly cancellation and custodial symmetry. I will then sketch how one can understand the development of the SM as a theory in terms of taming bad high energy behavior of scattering amplitudes. Then will come a discussion of the GIM mechanism and  'prediction' of the mass of the charm quark $M_{c}$
from the measured mass difference between $K_0$ and $\bar K_{0}$. This  will be followed by a discussion of the experimental measurements which established the EW part of the SM as a quantum gauge  field theory based on the gauge group $SU(2)_{L} \times U(1)_{Y}$, albeit where the symmetry is broken spontaneously. I will assume essentially that people are aware of some of the details of the Spontaneous Symmetry Breaking (SSB)  and hence will only sketch it here. As we know establishing the $SU)2)_{L} \times U(1)_{Y}$ theory with SSB as the correct theory of EW interactions was done  by testing the precision measurements of various EW observables against the predictions for the same including radiative corrections.  Inclusion of these radiative corrections  is possible only in a renormalisable quantum field theory. In particular I will discuss the history of  determination of $M_{t}$ and $M_{h}$ from 'indirect' effects on observables through loop corrections. In the last lecture I will discuss various theoretical bounds on the Higgs mass  and also the theoretical implications of the observed mass of the Higgs at the LHC~\cite{bib:one,bib:PDG} for the SM. 

\section{Preliminaries}
\label{prelim}
\subsection{Periodic table of particle physics}
\label{periodic-table}
The SM stands on the joint pillars of relativistically invariant quantum field theories and gauge symmetries. The SM is a quantum gauge field theory based on the gauge group  $SU(3)_{C} \times SU(2)_{L} \times U(1)_{Y}$ which describes the strong and electro-weak(electromagnetic and weak) interactions. The subject matter of these lectures is going to cover only the EW part of the SM. Gauge theory of strong interactions, QCD, will be discussed in a different set of lectures at this school. 

As things stand today, the periodic table of the SM is complete. 
One part of this periodic table  are the spin-$\Half$ matter particles: the quarks and the leptons and their anti-particles. Table~\ref{tab:tab1} summarises the details of the currently available information on all the matter fermions. 

\begin{table}[htb]
\caption{Elementary fermions of the Standard Model, all of spin $\Half$. The three quark colours are indicated explicitly, while leptons are colourless. Electric charges in units of the positron charge, are displayed on the left side. The anti-particles form a similar table with opposite charges.}
\label{tab:tab1}
\begin{tabular}{ll}
\hline
&\\
Quarks & Leptons \\
&\\
\hline\hline
&\\
$\begin{array}{c} {\scriptstyle\phantom{-} 2/3}\\ \scriptstyle{-1/3}\\ \end{array}$
${\color{red}\left( \begin{array}{c} u\\ d\\ \end{array} \right)}$
\hspace{0.1cm}
${\color{red}\left( \begin{array}{c} c\\ s\\ \end{array} \right)}$
\hspace{0.1cm}
${\color{red}\left( \begin{array}{c} t\\ b\\ \end{array} \right)}$
\qquad\qquad&
$\begin{array}{c} \scriptstyle{\phantom{-}0}\\ \scriptstyle{-1}\\ \end{array}$
$\left( \begin{array}{c} \nu_e\\ e\\ \end{array} \right)$
\hspace{0.1cm}
$\left( \begin{array}{c} \nu_\mu\\ \ \mu \\ \end{array} \right)$
\hspace{0.1cm}
$\left( \begin{array}{c} \nu_\tau \\ \tau\\ \end{array} \right)$
\\[5mm]
$\begin{array}{c} {\scriptstyle\phantom{-} 2/3}\\ \scriptstyle{-1/3}\\ \end{array}$
${\color{blue}\left( \begin{array}{c} u\\ d\\ \end{array} \right)}$
\hspace{0.1cm}
${\color{blue}\left( \begin{array}{c} c\\ s\\ \end{array} \right)}$
\hspace{0.1cm}
${\color{blue}\left( \begin{array}{c} t\\ b\\ \end{array} \right)}$
\\[5mm]
$\begin{array}{c} {\scriptstyle\phantom{-} 2/3}\\ \scriptstyle{-1/3}\\ \end{array}$
${\color{green}\left( \begin{array}{c} u\\ d\\ \end{array} \right)}$
\hspace{0.1cm}
${\color{green}\left( \begin{array}{c} c\\ s\\ \end{array} \right)}$
\hspace{0.1cm}
${\color{green}\left( \begin{array}{c} t\\ b\\ \end{array} \right)}$
&\\[5mm]
\hline
$M_u = 2$ MeV & $M_{\nu_1} = 0-0.13 \times 10^{-6}$ MeV\\
$M_d = 5$ MeV & $M_e = 0.511$ MeV\\
$M_c = 1,300$ MeV & $M_{\nu_2} = 0.009 - 0.13 \times 10^{-6}$  MeV\\
$M_s = 100$ MeV & $M_{\mu} = 106$  MeV\\
$M_t = 173.000$ MeV & $M_{\nu_3} = 0.04 - 0.14 \times 10^{-6}$ MeV\\
$M_{b} = 4.200$ MeV & $M_{\tau}=1.777$ MeV\\
\hline\hline
\end{tabular}
\end{table}
Of course, a gauge field theoretic description of the interactions among these elementary particles needs in the SM particle spectrum, also the gauge bosons which would be the carrier of the various interactions. This leads to the second set of members of the 'periodic table' of particle physics, viz. the spin-1  gauge bosons: the photon, $W$ and $Z$ bosons and gluons.  
Their details are indicated in Table~\ref{tab:tab2}.

\begin{table}[h]
\begin{center}
\caption{Elementary bosons of the Standard Model. There are no separate anti-particles: $W^-$ is the anti-particle of $W^+$ and the rest are neutral. $Q$ indicates the electromagnetic charge of the boson in units of positron charge.}
\label{tab:tab2}
\begin{tabular}{lll}
\hline
&&\\
Electromagnetic and weak & Strong & Higgs\\
(Spin 1) & (Spin 1) & (Spin 0)\\
&&\\
\hline\hline
&&\\
$\gamma$ (photon) & 
$g$ (gluons) & 
$h$ (Higgs)\\[2mm]
$W^\pm$, $Z$ (weak bosons) 
&&\\[2mm]
\hline
$M_\gamma=0,~ Q_\gamma=0$ & $M_g=0,~ Q_g=0$ &
$M_h=125.4\pm$ GeV, $Q_h=0$\\
$M_{W}= 80.404$ GeV, $Q_W=\pm 1$ &&\\
$M_{Z}= 90.1876$ GeV, $Q_Z=0$ 
&&\\
\hline\hline
\end{tabular}
\end{center}
\end{table}

As we will discuss in detail later, gauge invariance, which guarantees the renormalisability  of this theory, would require that all of the gauge bosons  should be massless. Not only that, the same invariance would require the matter fermions also to be massless. However, other than the gluon and the photon all the other members of this periodic table (cf. tables ~\ref{tab:tab1} and ~\ref{tab:tab2}) are patently massive. In fact, it is the mechanism of Spontaneous Symmetry Breaking (SSB), which allows these particles to have non zero masses and helps keep the theory still consistent with gauge invariance. SSB of the EW gauge symmetry via the Higgs mechanism (or Brout-Englert-Higgs mechanism for the purists)\cite{bib:two}, is the key ingredient of renormalisable gauge theories of the EW interaction.  This requires existence of yet another member of the periodic table, which is the Higgs boson. This too has been included in the list of the SM bosons in Table~\ref{tab:tab2}, now that its existence has been established firmly and the discovery awarded a Nobel prize!    

\subsection{Weak interactions: pre-gauge theory}
\label{pregauge}
Fermi's theory of $\beta$ decay~\cite{fermi-betadecay}, was the blueprint of the early theoretical  description of the weak interactions which are  responsible not just for the radioactive $\beta$ decays of nuclei but also for the  strangeness conserving and strangeness changing weak decays of the mesons and baryons. This culminated in the famous V-A theory of weak interactions~\cite{VA1,VA2}. According to this theory, the $\mu$ decay $\mu^{-}  \rightarrow \nu_{\mu} e^{-} \bar \nu_{e}$ for example, could be described by an effective Hamiltonian 
\begin{equation}
{\cal H}^{\mu~decay}_{eff} = - \frac{G_{\mu}}{\sqrt{2}} \left[ J_{\nu e}^{\rho + \dag}  J_{\nu \mu, \rho}^{+} + h.c. \right],
\label{eq:mudecay}
\end{equation}
where  
\begin{equation}
J^{\rho +}_{12} =\bar \psi_{1} \gamma^{\rho} (1 - \gamma_{5}) \psi_{2} \equiv J^{\rho CC}_{12}.
\label{eq:cccurrent}
\end{equation}
In the same way, the $\beta$ decay of the neutron could be described by an effective interaction given by
\begin{equation}
{\cal H}^{\beta decay}_{eff} = - \frac{G_{F}}{\sqrt{2}} \left[ J_{e \nu}^{\mu + \dag}  J_{\mu, pn}^{+} + h.c. \right],
\label{eq:ndecay}
\end{equation}
with 
\begin{equation}
J_{pn}^{\mu +} =  \bar \psi_{p} (1 - 1.26 \gamma_{5}) \gamma^{\mu} \psi_{n}
\label{eq:jnp}
\end{equation} 

\begin{figure}[htb]
\begin{center}
\includegraphics*[width=6cm,height=4.2cm]{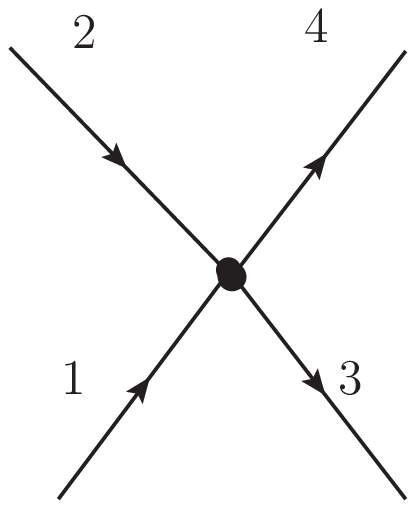}
\hspace{2cm}
\includegraphics*[width=6cm,height=4cm]{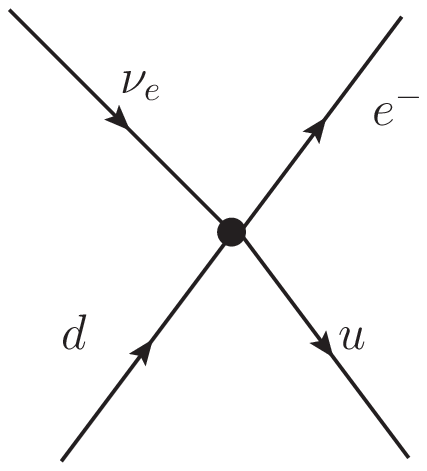}
\caption{Generic four fermion interaction responsible for the weak processes (left panel) and the basic process describing the $\beta$ decay (right panel)}
\label{fig:fig1}
\end{center}
\end{figure}
In fact, it was established that when written in terms of the quarks which make up the mesons and baryons, all the weak processes could be described in terms of a four fermion, current-current interaction depicted in left panel of Figure \ref{fig:fig1} which shows a transition $1 \rightarrow 
\bar 2 + 3  + 4$. For example, the basic transition describing the $n$ decay $n(udd) \rightarrow p (uud) + e^{-} + \bar\nu_{e}$,  is given by the current-current interaction depicted in the right panel. The crux of $V-A$ theory is that only the left chiral fermions are involved in this weak interaction Hamiltonian. The effective Hamiltonian is then written as
\begin{equation}
{\cal H}^{4 fermion}_{eff} = - \frac{G_{\mu}}{\sqrt{2}} \left[ J_{24}^{\mu + \dag}  J_{31, \mu}^{+} + h.c. \right]
=  - 4 \frac{G_{\mu}}{\sqrt{2}} \left[ \left(\bar \psi_{3L} \gamma^{\mu} \psi_{1L} \right)  \left(\bar \psi_{4L} \gamma_{\mu} \psi_{2L} \right) + h.c. \right] 
\label{eq:4fermion}
\end{equation}
The appearance  $\psi_{L} = 1/2 (1- \gamma_{5}) \psi$ in the 
Eq.~\ref{eq:4fermion}, indicates that only left chiral fermions are involved in this  charged weak current. As we will see later, it is this fact that decides the representation of the $SU(2)_{L}$ gauge group to which  the various fermion fields belong.

\begin{figure}[htb]
\begin{center}
\includegraphics*[width=4cm,height=4cm]{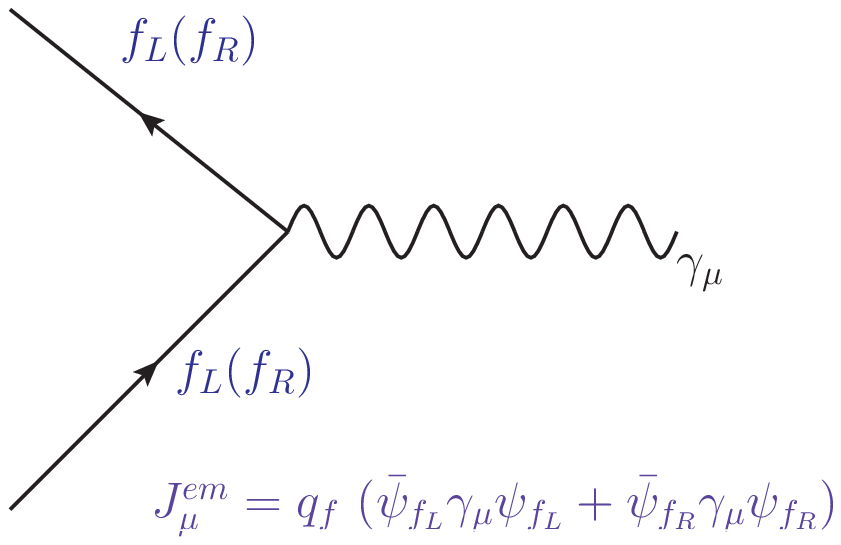}
\hspace{1.0cm}
\includegraphics*[width=4cm,height=4cm]{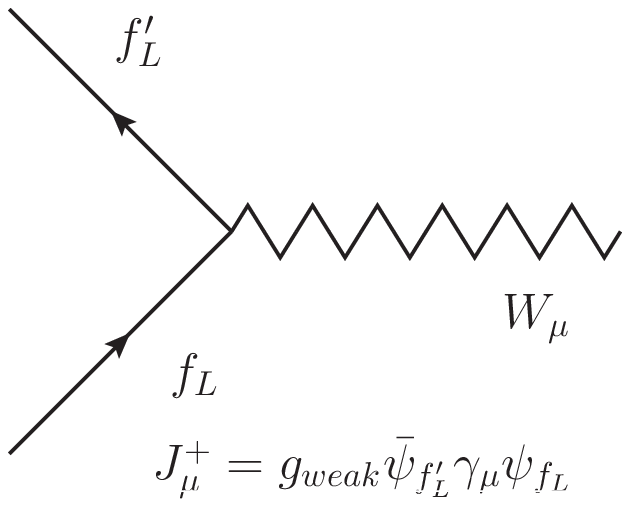}
\caption{The left panel shows the usual QED vertex depicting the interaction with the QED gauge boson $\gamma_{\mu}$ and the right panel shows the generic vertex describing the universal weak interaction among quarks and leptons.}
\label{fig:weakvertex}
\end{center}
\end{figure}

We understand the electromagnetic interaction in terms of the electromagnetic  current $J_{\mu}^{em}  = \bar \psi_{L} \gamma_{\mu} \psi_{L} + \bar \psi_{R} \gamma_{\mu} \psi_{R}$ and the electromagnetic field $A_{\mu}$.  The corresponding vertex is depicted in the left panel of Fig.~\ref{fig:weakvertex}.  Eq.~\ref{eq:4fermion} means that one can similarly think of the weak current $J_{\mu}^{+}$ (for example) coupled to a charged gauge boson (a weak boson $W$) $W_{\mu}^{+}$. The basic transition brought about by the charged current could then be depicted as shown in the right panel of Fig.~\ref{fig:weakvertex}.   
The electromagnetic charge of $f'$  differs from that of $f$ by one unit and in case $f$ is strange quark, the strangeness changes by one unit as well. In that case this current indicates a transition which brings about $\Delta S = \Delta Q = 1$, where $S$ and $Q$ stand for the strangeness and the electromagnetic charge respectively. While the decay of a  neutron $n$ involves the current $J_{ud}^{\mu +}$, the decay of $\Lambda$ for example, involves the current $J_{us}^{\mu +}$. The strength of the four-fermion interaction is then decided by $g_{weak}$ of Fig.~\ref{fig:weakvertex}. Experimentally  measured values of $G_{\mu}$ and $G_{F}$ of Eqs.~\ref{eq:mudecay}, \ref{eq:ndecay} were somewhat different from each other, though very close, $G_{F} \sim 0.98 G_{\mu}$. For the effective Hamiltonian for $\Lambda$  decay, for example, the corresponding coefficient was yet again different from both $G_{\mu}, G_{F}$,  $G_{\Lambda}$ being $0.20 G_{\mu}$ \footnote{The very near equality between $G_{\mu}$ and $G_{F}$ was an indication that the vector current was not affected by the strong interactions of the $n$ and $p$ and is the same for $e^{-}$ to $\nu$ transition as for $n$ to $p$. This was called the 'Conserved Vector Current hypothesis' (CVC). In all the discussions regarding the mixing angle, we are referring to the coefficient of this conserved vector part of the current at the hadron level.}.
It was Cabibbo's observation~\cite{bib:CCKM} that all this could be consistent with a completely universal charged weak current \ie a current which has the same strength for the leptons as well as the quarks and also for $\Delta S =0$ and $\Delta S = 1$ alike, if in case of quarks, the basic charged current in Fig.~\ref{fig:weakvertex}    describes a transition with $f' = u$, $f = d' =  d \cos \theta_{c} + s \sin \theta_{c}$, with $\sin \theta_{c} \sim  12^{\circ}$. This means that the interaction  eigenstate $d'$ is a linear combination of the mass eigenstates $d$ and $u$. Clearly, the orthogonal combination $s' = -d \sin \theta_{c} + d \cos \theta_{c}$, is an interaction eigenstate coupling with a $W^{\pm}$ and a {\it new } quark with charge $+ \frac{2}{3}$. This thus indicates existence of the fourth quark : the charm quark $c$. As we will see later its existence ensures flavour conservation of the weak neutral currents at tree level automatically. This then helps one understand the experimentally observed suppression of the Flavour Changing Neutral Currents (FCNC) which will be discussed in detail later. Thus the states to be identified with the interaction eigenstates would be:
\[
\left(\begin{array}{c}  u'\\  d' \\ \end{array}\right) = \left(\begin{array}{c}  u\\  d \cos \theta_c + s \sin \theta_c\\ \end{array}\right);~~ \left(\begin{array}{c}  c'\\  s' \\ \end{array}\right) = \left(\begin{array}{c}  c\\  - d \sin \theta_c + s \cos \theta_c\\ \end{array}\right)
\]

At this point let us also mention one  more feature of the phenomenology of quark mixing which will be relevant later.
In fact, the physics of the $K_{0}, \bar K_{0}$ mesons not only revealed the existence of suppressed nature of the FCNC but also CP violation in $K_{0}$--$\bar K_{0}$ system. This CP violation can also be understood as coming from the above quark-mixing but ONLY if the mixing matrix involves a phase. For this to be possible we have to have at least three generations of quarks. This was noted by Kobayashi-Maskawa~\cite{bib:KMCKM}. This makes it possible to understand the CP violation observed in the neutral meson system, in the context of a gauge theory of EW interactions, in terms of the mixing in the quark sector. However, this requires existence of at least three generations. Thus one sees that in some sense, the need to understand the observed phenomenology of FCNC and CP violation, in the framework of a gauge theory, predicted the existence of the $c$ and the $t$ quark respectively. 

For future reference note that the connection between the mass eigenstates $u,d,c,s,t$ and $b$ and  the interaction eigenstates $u',d',c',s',t'$ and $b'$ is given by $u'=u,c'=c,t'=t$ and
\begin{equation}
\left( \begin{array}{c}
d' \\ s' \\ b'
\end{array}\right)
=
\left(
\begin{array}{c c c}
V_{ud} & V_{us} & V_{ub}\\
V_{cd} & V_{cs} & V_{cb}\\
V_{td} & V_{ts} & V_{tb}
\end{array}
\right)
\left(
\begin{array}{c}
d \\ s \\ b
\end{array}
\right),
\label{eq:CKM}
\end{equation}
where $V_{ud}$ etc. are elements of the CKM matrix $\bf V$ \Refs~(cf. \cite{bib:CCKM}--~\cite{bib:KMCKM}). This describes the  interaction eigenstates in terms of the mass eigenstates.

\begin{figure}[htb]
\begin{center}
\includegraphics*[width=14cm,height=5cm]{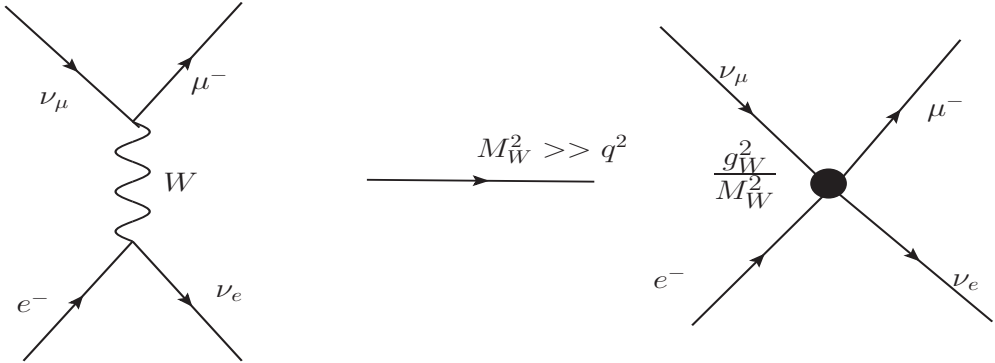}
\caption{Contact interaction resulting from $M_{W} \rightarrow \infty$ limit} 
\label{fig:fig1p}
\end{center}
\end{figure}

At this point let us also note that the same four fermion interaction that describes the decay $\mu^{-} \rightarrow e^{-} + \bar \nu_{e} + \nu_{\mu}$ can also describe, for example,  the scattering  processes such as $\nu_\mu + e^- \to \nu_e + \mu^-$,  
corresponding to  $1 = e^{-}, 2 =\nu_{\mu}, 3 = \nu_{e}$ and $4 = \mu^{-}$ in the left panel of Fig.~\ref{fig:fig1}. The same effective Hamiltonian as in Eq.~\ref{eq:4fermion} then also describes this scattering process as well. If one calculates the 
total cross-section one gets,
\begin{equation}
\sigma_{\rm tot} = \frac{G_\mu^2 s}{\pi} = 
\frac{2 G_{\mu}^{2} m_{e} E_{\nu_{\mu}}}{\pi}.
\label{eq:fermics}
\end{equation}
This linear rise of scattering cross-section with $s$, the square of the centre of mass energy or alternatively $E_{\nu_{\mu}}$,  is a reflection of the 'pointlike' nature of the Fermi interaction of Eq.~\ref{eq:4fermion}. It can be seen, by doing a partial wave analysis of the scattering amplitude, that this behaviour implies violation of unitarity when $\sqrt{s} \geq 300 $ GeV. Of course, in practical terms it corresponds to a $E_{\nu_{\mu}} \geq 10^{8}$ GeV and hence perhaps not very relevant. However, it is the principle that matters. A cure to this problem of the current -current interaction was indeed offered by postulating the existence of a massive, charged boson (called the weak-boson $W^{\pm}$) by Schwinger. This is the same $W^{\pm}$ we have already introduced while writing the weak vertex in Fig.~\ref{fig:weakvertex}.  Thus the point interaction of Eq.~\ref{eq:4fermion} can be understood as an interaction resulting from the exchange of a $W^{\pm}$ boson, in the limit of the said mass $M_{W}$ being  much bigger than all the energies in the system. This  is depicted in Fig.~\ref{fig:fig1p}.
The observed short range of the weak force causing the $\beta$ decay, indicated that the $W^{\pm}$ boson is massive, unlike the
photon mediating the electromagnetic interaction which is massless. The success of the effective Hamiltonian of Eq.~\ref{eq:4fermion} implies a lower bound much bigger than $\UMeV$ and hence $\sim {\cal O}$ (GeV). To summarize, we see that the requirement that unitarity bound be respected, indicates the existence of a massive charged vector boson $W^{\pm}$ and the  four-fermion weak interactions can be understood as caused by an exchange of this massive boson. The 'massive' nature of the exchanged boson was also consistent with the observed 'short' range of the weak interactions. However, if it is a gauge boson, then the massive nature will also break gauge invariance! Further, the massive nature of the gauge boson causes problems such as bad high energy behavior of scattering amplitudes as well as non renormalisability of the theory.  How a massive gauge boson is to be accommodated in the framework of a gauge theory is going to be the topic of discussion in the next section.

\subsection{Observations meet predictions of the SM}
Before beginning with a discussion of details of a gauge theory, let us just briefly take a look how the establishment of the SM has been a synergistic activity between theoretical and experimental developments. We saw already how the form of the 
pre-gauge theory, effective Hamiltonian  description of weak interactions, obtained phenomenologically from the data  hinted at a possible gauge theoretic description of the same. Equally interesting are the hints at existence of new particles given by the  theory.  While some of the members of this periodic table, like the $\mu$, were unlooked for and some like the $\nu$ were met with quite a bit of disbelief when postulated  theoretically, for most of the recent additions their existence and in some cases even their masses were predicted if the EW interactions were to be described by a  renormalisable gauge theory.

In fact, the existence of strange particles which contain the strange quarks, coupled with experimental features such as the suppression of the FCNC in EW processes alluded to before, indicated the existence of the charm quark, as already indicated above. Further, the small mass difference between $K_{L}$ and $K_{S}$ (or alternatively the $K_{0}$--$\bar K_{0}$ mixing) could be used to obtain an estimate of its mass. 
Accidental discovery of some members of the third  lepton and quark family, combined with the requirement of anomaly cancellation, an essential feature for a renormalisable theory, meant that the remaining members of the same family had to exist.  Hence  $t$ and the $\nu_{\tau}$ were hunted for very actively once the $b$ and the  $\tau$ made their appearance! The properties of a renormalisable quantum field theory were the essential reasons behind the belief in these predictions. The mass of the $t$ quark could also be predicted in the SM, using experimental information on neutral B meson mixing and properties of the $Z$ boson, as we will see below.

The story is not very different for the EW gauge bosons. As was already mentioned,  requiring consistency of the pre gauge theory description of the weak interactions with unitarity,  had indicated a nonzero mass for the charged $W^{\pm}$ but had not indicated what the mass would be, except that it should be much larger than the typical energy scales involved in the weak decays $\sim \UMeV$.  It is the unified description of the EW interactions of the Glashow-Weinberg-Salam (GSW)
model~\cite{bib:three}  that actually gave a lower limit on its mass. Note that  the  correctness of the $V-A$ nature of weak interactions and pure vector nature of the electromagnetic interactions predicted existence of a neutral boson other than the photon $\gamma$. In the GSW model, the masses of the $W$ and the new $Z$ boson required in the unified EW theory, were all predicted in terms of the life time of the $\mu$ and the weak mixing angle $\theta_{W}$ which was a free parameter in the model.  This could  be determined from measurements of rates of various weak processes. 

Not just this, the SM also predicted existence of yet another boson, this time spin $0$; viz. the Higgs boson. The mass of the said Higgs boson, however, is a free parameter in the framework of the SM. Comparisons of the EW observables with precision measurements can constrain the Higgs mass through the corrections caused by the loop effects which can be computed in a renormalisable quantum field theory.  One can also put limits on this parameter from theoretical considerations of consistency of the SM as a field theory at high scales: the triviality and vacuum stability, all to be discussed in the lectures.

Let us discuss in detail the case of the $t$ quark  which 
is quite interesting. The existence of the $t$ quark and the information on its mass came from a variety of  theoretical and phenomenological observations in flavour physics and physics of the $W/Z$ bosons. As already mentioned the explanation of the experimentaly observed CP violation in terms of the quark mixing matrix requires at least three generations of quarks. This mixing is described by the famous CKM mixing matrix \Refs~\cite{bib:CCKM}--~\cite{bib:KMCKM}.  So in that sense existence of the $t$ and $b$ was indicated by this observation.\footnote{The requirement of anomaly cancellation for the gauge theory of EW interactions to be renormalisable, further indicated existence of an additional generation of leptons, $\tau, \nu_{\tau}$ as well.} Experimental manifestation of $B_{0}$--$\bar B_{0}$ oscillations at the ARGUS experiment~\cite{bib:four}  was a harbinger of the presence of the $t$ quark.
Further indications for the expected mass actually came from precision  measurements of many EW observables, ie. properties of the $Z$ and the $W$ boson and the quantum corrections caused to them  by loops containing top quarks.
  
Experimental observation of the $t$ quark at the Tevatron~\cite{bib:five}, with a mass value consistent with the implications of the EW precision measurements, provided a test of the description, at loop level, of EW interaction in terms of an $SU(2)_{L} \times U(1)_{Y}$ gauge field theory with SSB. 
\begin{figure}[htb]
\begin{center}
\centering\includegraphics*[width=10cm,height=8cm]{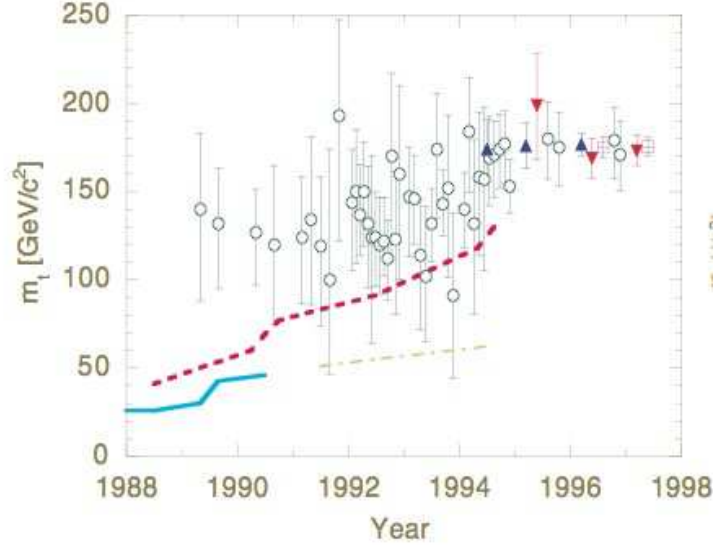}
\caption{Comparison of the limits on the mass of the top quark from direct searches at the $e^{+} e^{-}$ collider (solid line) and the hadronic colliders (red dashed line) with the indirect limits, indicated by open circles, coming from precision EW measurements as a function of time. The dot dashed line is an indirect lower limit obtained from the observed rates of inclusive $W/Z$ production in $p \bar p$ colliders.The solid triangles indicate directly measured values of the  mass of the observed $t$-quark. This is taken from \protect\cite{Quigg}.}
\label{fig:topology}
\end{center}
\end{figure}
Fig.~\ref{fig:topology} shows, by open circles, evolution with time of the values of the top mass extracted {\it indirectly} by comparing the measured  EW parameters with the SM predictions. Also shown are the $95\%$ c.l. upper limits from direct searches from the $e^{+}e^{-}$ experiments (solid line) and from $p \bar p$ experiments (the dashed line). In the last part of the plot the solid triangles show the mass of the top quark measured directly at the Tevatron and the 'indirectly' extracted values of $M_{t}$ at the same time. The remarkable agreement between  directly measured and the 'indirectly' extracted values around the time of the discovery, was a test of the SM at loop level. 

Once this was achieved, the same information could be used to obtain constraints on the Higgs mass, now looking at quantum corrections to the $W,Z$ mass as well as to the $Z$ couplings, caused by loops containing the Higgs boson. Finally finding a Higgs boson in 2012~\cite{bib:one} with a mass consistent with these constraints was the biggest success of the SM \footnote{
Knowledge of QCD, the part of the SM which we are not  discussing in these lectures, was essential in making  precision predictions for the Higgs signal and hence to this mass determination!}
\begin{figure}[htb]
\centering\includegraphics*[width=10cm,height=8cm]{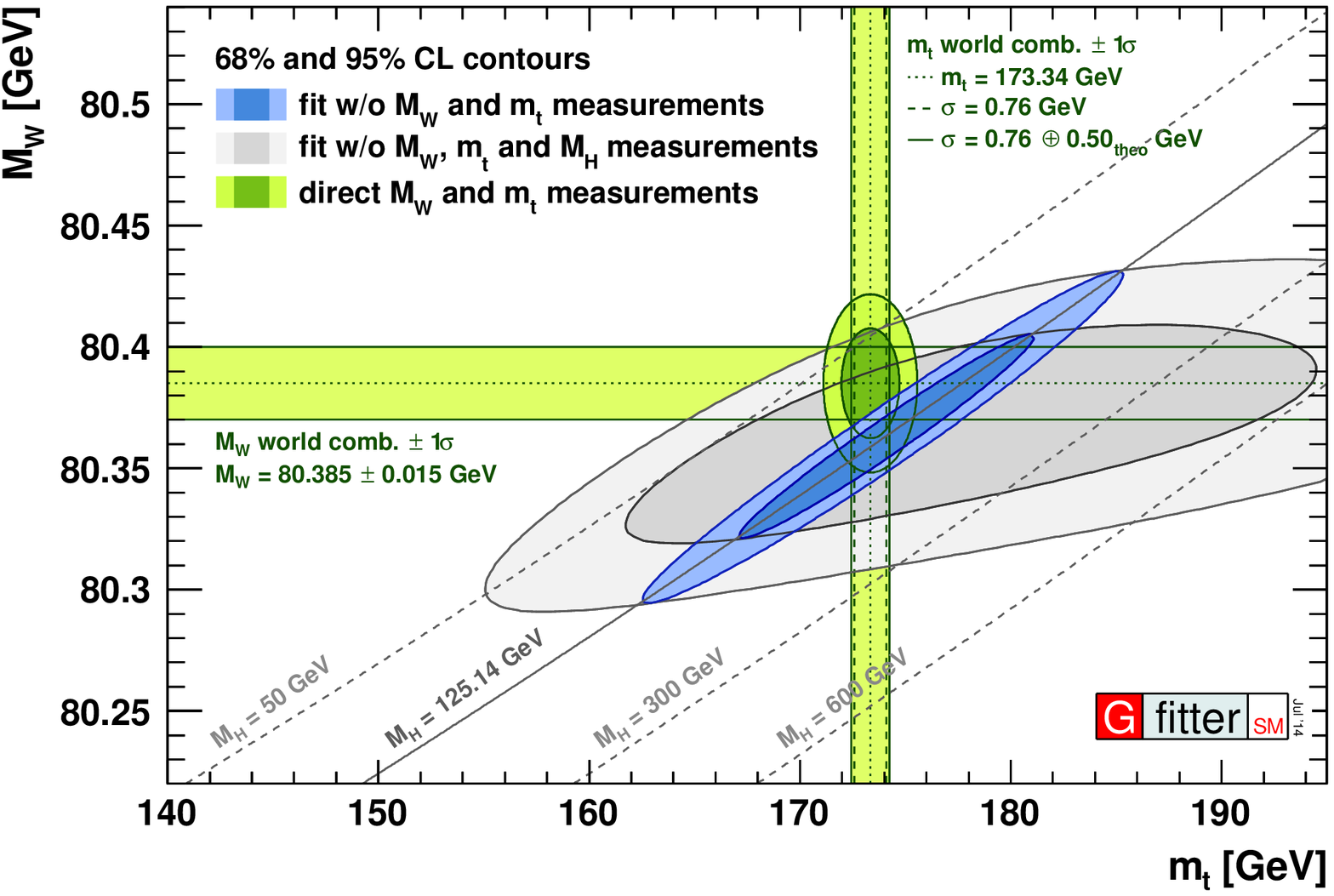}
\caption{Testing of the SM at loop level from  mass measurements. $M_{W}$ -- $M_{t}$ values consistent with the EW precision data with and without using measured value of $M_{t},M_{W}$ and $M_{h}$ as inputs. Taken from ~\cite{bib:gfitter}}
\label{fig:fig2}
\end{figure}
Fig.~\ref{fig:fig2} reproduced from the Gfitter webpage~\cite{bib:gfitter} illustrates this. The various dark and light shaded regions correspond to $68\%$ and $95\%$ c.l. contours in all cases. The green bands between the vertical and horizontal lines indicate experimentally measured values of $M_{t}$ and $M_{W}$. The region shaded in blue (the long and narrow ellipses)  indicates the region allowed in the $M_{t}$--$M_{W}$ plane, by fits of the SM prediction for precision measurements of EW observables where the Higgs mass~\cite{bib:one} information is used. The big elliptical regions, one of them open at one end,  shaded in light and dark grey, are the ones allowed when none of the mass measurements are used as input and one lets the EW precision data choose the best fit values. Consistency of the values obtained in these fits with each other and with the experimental measurements indicated by the small oval with dark and pale green regions, leaves us with no doubt about the correctness of the SM. This tests the correctness of quantum corrections to $M_{W}$ coming from the loops containing the $t$ and $h$;  hence of the quantum field theoretic description of the EW interactions as a gauge theory. 

Alongside this  spectacular testimonial of the correctness of the EW part of the SM, is also the equally impressive demonstration of a highly accurate description of all the CP violating phenomena in terms of the flavour mixing in the quark sector. In the three flavour picture the $3 \times 3$ CKM matrix is unitary. Making detailed fits of theoretical predictions to a large variety of data on meson mixing and decays, to determine the elements of the CKM matrix with high precision,  is an involved exercise as it requires a synthesis of a variety of theoretical tools and high precision data. These elements are parameterised in terms of 
two parameters : $\bar \rho$ -- $\bar \eta$~\cite{bib:PDG}.
\begin{figure}[htb]
\centering\includegraphics[height=8cm,width=12cm]{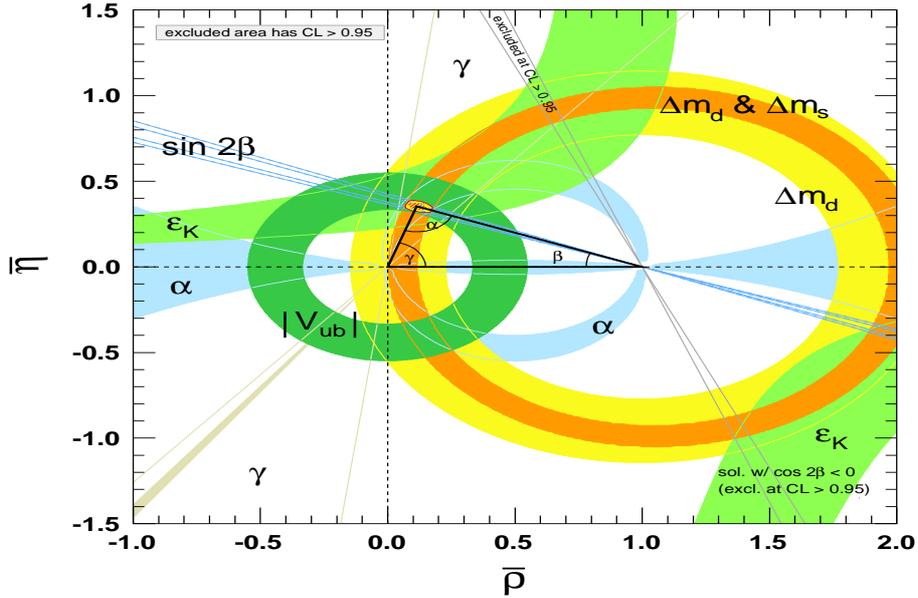}
\caption{Constraints in the $\bar \rho$ -- $\bar \eta$ plane from a variety of measurements around the global fit point. Taken from ~\protect\cite{bib:PDG}.}
\label{fig:utfit}
\end{figure}
 Fig.~\ref{fig:utfit} taken from PDG-2015 shows the constraints in the $\bar \rho$ -- $\bar \eta$ plane from a variety of measurements around the global fit point. Various shaded areas indicate the regions allowed at $95 \% $ c.l. from a  given measurement. The unitarity of the CKM matrix is indicated by the fact that the 'tip' of the unitarity triangle lies in the small intersection region allowed by all the various measurements. Since for many of these observables their relationship with the parameters of the SM is given by loop computations, this success too provides a test of the SM as a quantum gauge field theory.
 
\section{$SU(2)_{L} \times U(1)_{Y}$ gauge theory}
\subsection{Gauge principle}
\label{sec:qed}
Gauge principle is the basis of the theoretical description of  three of the fundamental interactions {\it viz.} strong, weak and electromagnetic, among the quarks, leptons and the force carrying gauge bosons.  QED is the first gauge theory to be established.  We therefore can begin our discussion of gauge theories, by looking at QED: a theory of a Dirac fermion field $\psi(x \equiv \vec x, t)$ of charge  $e$. For a free Dirac fermion of mass $m$ the Lagrangian density consists of the kinetic term supplemented with the mass term and is given by
\[
{\cal L}_{f} = i \bar \psi \gamma^{\mu} \partial_{\mu} \psi - m \bar \psi \psi.
\]
However, this Lagrangian density is not invariant under a local $U(1)$  gauge transformation, 
\begin{equation}
\psi (x) \rightarrow e^{i \alpha} \psi(x), {\mathrm with}~~\alpha = \alpha(x) .
\label{eq:u1trans}
\end{equation}
Note that this non invariance of the Lagrangian density is true only for the local gauge transformation with $\alpha = \alpha (x)$. To construct a gauge invariant Lagrangian density, one needs to introduce a vector field $A_{\mu}$ and generalise the derivative  $\partial_{\mu} \rightarrow \partial_{\mu} + i q_{f} |e| A_{\mu}$ where $q_{f}$ is the charge of the fermion in units of positron charge $|e|$. Thus for the electron, the covariant derivative is 
\begin{equation}
D_{\mu} = \partial_{\mu} - i e A_{\mu}
\label{eq:u1covariant}
\end{equation}
Combining this generalization of the kinetic term for the fermion, with the gauge transformation of the vector field
\begin{equation}
A_{\mu} \rightarrow A_{\mu} + {\frac{1}{e}} \partial_{\mu} \alpha(x),
\label{eq:u1gauge}
\end{equation}
one can show that $(\partial_{\mu} - i e A_{\mu})\psi \rightarrow e^{i \alpha(x)} (\partial_{\mu} - i e A_{\mu}) \psi$. Thus,
under the gauge (phase) transformation of the fermion field, the vector field too has to transform with the same transformation parameter $\alpha(x)$. Note now that the Lagrangian density
\begin{eqnarray}
{\cal L}_{QED} &=&i\bar\psi \gamma^{\mu} D_{\mu} \psi - m \bar \psi \psi-\frac{1}{4} F_{\mu \nu} F^{\mu \nu} \nonumber = i\bar\psi \gamma^{\mu} (\partial_{\mu} - i e A_{\mu}) \psi - m \bar \psi \psi- {\frac{1}{4}} F_{\mu \nu} F^{\mu \nu} \\  &=&i\bar \psi \gamma^{\mu} \partial_{\mu}\psi - m \bar \psi \psi- {\frac{1}{4}} F_{\mu \nu} F^{\mu \nu} + e~\bar\psi \gamma^{\mu} \psi A_{\mu}\nonumber \\ &=&{\cal L}_{f} + ~ {\cal L}_{gauge} + e~\bar\psi \gamma_{\mu} \psi A^{\mu} = {\cal L}_{f} + ~ {\cal L}_{gauge} + {\cal L}_{int},
\label{eq:lqed}
\end{eqnarray}
with $F_{\mu \nu} = \partial_{\mu} A_{\nu} - \partial_{\nu} A_{\mu}$, is gauge invariant. Note that a mass term for the vector field {\it viz.} $M_{A}^{2} A_{\mu} A^{\mu}$ will break this gauge invariance (cf. Eq.~\ref{eq:u1gauge}). Further, this Lagrangian density is just the sum of three Lagrangian densities: ${\cal L}_{f}$ for the free fermion field $\psi$  of mass $m$ as given by the first two terms in the third line of Eq.~\ref{eq:lqed}, ${\cal L}_{gauge}$ for free {\it massless} gauge field $A_{\mu}$   given by the third term and the interaction term ${\cal L}_{int}$ being given by the last one. Note that the form of the interaction of the fermion with the gauge field is {\it completely fixed} by the form of the covariant derivative $D_{\mu}$. Further, the mass term
\[
m \bar \psi \psi = m (\bar\psi_{L} \psi_{R} + \bar \psi_{R} \psi_{L})
\]
will not be invariant under an $U(1)$ local gauge transformation similar to that given by Eq.~\ref{eq:u1trans} if, for example, the left and right chiral fermions have different $U(1)$ charges. This will be the case with $U(1)_{Y}$ gauge group of the Standard Model, as we will see very soon.

\noindent Note also the interaction term given by: 
\begin{equation}
{\cal L}^{QED}_{int} = e \bar \psi \gamma^{\mu} \psi A_{\mu} = e J^{\mu, em} A_{\mu}.
\label{eq:intqed}
\end{equation}
The current $J^{em}_{\mu}$ of Eq.~\ref{eq:intqed} is the vector bilinear constructed out of the fermion fields $\psi$ and $\bar \psi$. 
As opposed to this, the weak current $J_{12}^{\pm \mu}$ defined in section~\ref{pregauge} contains a linear combination of both the vector and axial vector bilinears. This phenomenologically ascertained form of the weak current therefore pointed already towards a gauge theory of weak interactions albeit with parity violation. The form of this chirality conserving current indicated existence of two charged vector bosons which however couple only to left chiral fermions. Thus the $V$--$A$ form of the current-current interaction already gives indications about the representation of this gauge group, to which different types of fermions should belong since, as seen above, in a gauge theory it is this representation that decides the interaction of the fermions with the vector gauge bosons. The similarity and the differences in the nature of the weak and electromagnetic current and description of electromagnetic interactions in terms of a $U(1)$ gauge theory, paved the way towards an unified description of electromagnetic and weak interactions as the  electro-weak gauge theory based on the gauge group $SU(2)_{L}\times U(1)_{Y}$.

Before we formally write down the complete Lagrangian density  for
the EW part of the SM, let us discuss the generalisation  of the above discussion to non ablian gauge transformations. To that end let us begin by summarising some of the relevant observations for QED, which we have stated above. The local phase transformations given by Eq.~\ref{eq:u1trans} form  an  unitary group and is called $U(1)$. The Lagrangian density of matter fields is invariant under this $U(1)$ transformation only if there exists a vector field which simultaneously transforms with the same transformation parameter and the matter field interacts with this vector field in a specific manner.  We consider now, a generalisation of this simple symmetry transformation of Eq.~\ref{eq:u1trans} to a case where matrix valued analogues of this simple phase transformations act on a set of fields and again the elements of the matrices can depend on the space time coordinates of the point : $\vec x, t$.  Again, invariance of the matter Lagrangian density under this local transformation requires a set of spin $1$, vector fields which transform under the local gauge transformation according to a generalisation of Eq.~\ref{eq:u1gauge} in addition to a modification of the kinetic term of the matter fields by replacing $\partial_{\mu}$ by the covariant derivative $D_{\mu}$ as done above. Thus there exists now a multiplet of gauge bosons. Another curious property of the Lagrangian density involving these gauge fields is that even in absence of matter fields and interactions, the equations of motion are non linear. This in turn means that the associated spin $1$ particles interact with each other in the absence of matter.  Further,  unlike the phase transformations of the QED, these  matrix valued transformations  do not commute with each other. Hence these generalized gauge theories are also called non-abelian gauge  theories. 

Lagrangian density of a free, massless non-Abelian gauge theory is given by
\begin{equation}
{\cal L}_{non abelian} = -\frac{1}{4} F^{a}_{\mu \nu} F^{a,\mu \nu}
\label{eq:nabelian-density} 
\end{equation}
with 
\begin{equation}
F^{a}_{\mu \nu} = \partial_{\mu} W^{a}_{\nu} - \partial_{\nu} W^{a}_{\mu} + g f^{abc} W^{c}_{\mu} W^{c}_{\nu}
\label{eq:fieldtensor}
\end{equation}
Here $f^{abc}$ are structure constants which are specific to each gauge group defined by,
\begin{equation}
[T^{a}, T^{b}] = i f^{abc} T^{c},
\label{eq:commutator}
\end{equation}
$T^{a}$ being the generators of the gauge transformation.  $f^{abc}$ are called the structure constants. $T^{a}$ are called 
generators because,  in general if $\Phi$ represents a matter field (spin $\Half$ or spin $0$)  transforming according to a representation $T_{IJ}$ of the gauge group  then  
\begin{equation}
\Phi_{I} \rightarrow exp^{-ig (T^{a})_{IJ}\alpha^{a}(x)}~~ \Phi_{J},
\label{eq:gaugetrans}
\end{equation}
where $g$ is the coupling constant.  The repetition of index $a$ indicates sum over all the generators of the transformation. 
The covariant derivative is then given by
\begin{equation}
D_{\mu} \Phi_{I} = \partial_{\mu} \Phi_{I} - i g V^{a}_{\mu} (T^{a})_{IJ} \Phi_{J},
\label{eq:covariant}   
\end{equation}
where $V_{\mu}^{a}$ denote the associated spin $1$ vector fields. The kinetic term for the matter fields, defined in terms of the $D_{\mu}$ along with the one for massless gauge fields given by Eq.~\ref{eq:nabelian-density}, are both invariant under the gauge transformation if the gauge field also transforms as
\begin{equation}
V^{a}_{\mu} \rightarrow V^{a}_{\mu} + {\frac{1}{g}} \partial_{\mu} \alpha^{a} + f^{abc} V^{a}_{\mu} \alpha^{c}.
\label{eq:spin1transf}
\end{equation}  
Again the couplings of the matter particles with the gauge bosons 
$V_{\mu}^{a}$, are then given by the kinetic term written down using the covariant derivative given by Eq.~\ref{eq:covariant},
just like we did in Eqs.~\ref{eq:lqed} and \ref{eq:intqed}. We can then write down currents $J_{\mu}^{V}$ analogous to $J_{\mu}^{em}$ of Eq.~\ref{eq:intqed}. This is completely determined once we specify the gauge group, \ie $T^{a}$, the representation of the gauge group to which the matter particles belong and the coupling $g$. 

When $a=1$, \ie  when there exists only one gauge boson, these gauge transformations and covariant derivative  given by Equations~\ref{eq:gaugetrans}--\ref{eq:spin1transf} reduce to those for simple phase transformation corresponding to the $U(1)$ case, viz., Equations ~\ref{eq:u1trans}--\ref{eq:u1gauge}.
For the case where $a$ is different from 1,  because of the commutator relation, the normalisation of the charge $g$ is fixed for all the representations. For $U(1)$ gauge transformation on the other hand the normalisation of the charge can be different for different representations.  For future reference, let us also note here that for the $SU(2)$ gauge group we have
\[
T^{a}= \frac{\tau^{a}}{2}~~ {\mathrm and}~~  f^{abc} = \epsilon^{abc},~~~~a =1-3
\]
where $\tau^{a}, a =1-3$ are the Pauli matrices and $\epsilon^{abc}$ is the constant, completely antisymmetric tensor.
Hence, for $SU(2)$ the index $a$ takes values $1$--$3$ in Eq.~\ref{eq:gaugetrans}.

\subsection{GSW model}
\label{sec:GSW}
Let us first write down the gauge boson and matter particle content for the GSW model along the interactions among all these. The gauge group for the GSW model is $SU(2)_{L} \times U(1)_{Y}$. The subscript $L$ means that the gauge transformations corresponding to this gauge group are non trivial ONLY for the left chiral(handed)\footnote{The word handedness and chirality can be used interchangeably for massless fermions.} fermions and the right chiral fermions remain unchanged under it. The direct product means that these two groups are independent, \ie  the left handed fermions belonging to a given representation of $SU(2)_{L}$ will all have the same value of the charge under $U(1)_{Y}$. Thus ONLY the left chiral fermions belong to the nontrivial representation of the $SU(2)_{L}$ group and the right chiral fermions are singlets under the $SU(2)_{L}$ gauge group. Therefore these have NO interactions with the gauge bosons corresponding to the $SU(2)_{L}$ gauge group.

\subsubsection{Particle content and Currents of the GSW model}
For the $SU(2)$ group, each representation is labelled by two quantum numbers $T_{L}$ and $T_{3L}$, where $T_{L}$ takes integral or half integral values: $0,1/2,1,3/2...$ \etc and  for a given $T_{L}$, $T_{3L}$ takes values from $- T_{L}$ to $+ T_{L}$ in steps of $1$. Thus number of fields belonging to representation labelled by $T_{L}$ is then $2 T_{L} + 1$. For singlet representation $T_{L} = 0$ and for the doublet it is $1/2$. Thus a doublet of $SU(2)_{L}$ contains two members with $T_{3L} = \pm 1/2$.
The gauge bosons belong to the $T=1$ representation (called the adjoint representation) and hence they are three in number called
$W_{\mu}^{a}, a=1-3$. The $U(1)_{Y}$ gauge group  has only one generator like the QED case discussed above. We denote the corresponding single gauge boson $B_{\mu}$. The corresponding current is $J_{\mu}^{Y}$ and the charge is called ``hypercharge''. The electromagnetic charge of a charged fermion is independent of its chirality. On the other hand, the two left chiral fermions of different electromagnetic charges have to have the same $U(1)_{Y}$ charge. Thus it is clear that the $U(1)_{Y}$ can not be identified with $U(1)_{em}$, \ie the hypercharge is different from the electromagnetic charge. Thus $U(1)_{em}$ arises out of a linear combination of $U(1)_{Y}$ and a $U(1)$ subgroup of$SU(2)_{L}$. 

First let us discuss the physics in terms of $W_{\mu}^{a}, a=1-3$ and $B_{\mu}$.
The gauge groups, the corresponding spin-$1$ huge bosons and the couplings are indicated in Table~\ref{tab:ggroup}. 
\begin{table}[hbt] 
\begin{center}
\begin{tabular}{|c|c|c|}
\hline
&&\\
Gauge Group& Gauge Boson Fields& Coupling\\
&&\\
\hline
&&\\
$SU(2)_{L}$&$W_{\mu}^{a}$, $a =1,2,3$& $g_{2}$\\
&&\\
\hline
&&\\
$U(1)_{Y}$& $B_{\mu}$ & $g_{1}$\\
&&\\
\hline
\end{tabular}
\caption{Gauge group, gauge bosons and couplings for the GSW model}
\label{tab:ggroup}
\end{center}
\end{table}
As we will see in a minute, if the left handed fermions belong to the doublet representation of $SU(2)_{L}$, the corresponding charge changing gauge current $J_{\mu}^{W}$ we would construct from the covariant derivative, has the same form as the  $J_{\mu}^{CC}$ of  Eq.~\ref{eq:cccurrent}, of the $V-A$ current Lagrangian describing the charge changing weak interactions.
Let $\frac{Y}{2}$ denote the charge of the fermion under the $U(1)_{Y}$ gauge group. The  corresponding transformation is given by
\begin{equation}
\psi \rightarrow e^{-i(g_{1}Y/2) \alpha_{Y}(x)}~~\psi
\label{eq:covarY}
\end{equation}
whereas, for a $SU(2)_{L}$ doublet the  gauge transformation is given by
\begin{equation}
\Psi = \left( \begin{array}{c} f_{1}\\ f_{2}\\ \end{array} \right) \rightarrow \Psi' = e^{- ig_{2} (\tau^{a}/2) \alpha^{a}(x)} \Psi.
\label{eq:doubtransf}
\end{equation}
$f_{1}$ and $f_{2}$ are the $T_{3L} = \pm 1/2$ members of this doublet $\Psi$ respectively. $\tau^{a}/2$ are the generators $T^{a}$ for the 2-dimensional fundamental representation. 

The fermion content of the GSW model can then be written as shown in Table~\ref{tab:fermsu2}.
\begin{table}[hbt]
\begin{center}
\begin{tabular}{|c|c|}
\hline
&\\
Quarks & Leptons \\
&\\
\hline
&\\
$\left( \begin{array}{c} u\\ d\\ \end{array} \right)_{L}$
\hspace{0.2cm}
$\left( \begin{array}{c} c\\ s\\ \end{array} \right)_{L}$
\hspace{0.2cm}
$\left( \begin{array}{c} t\\ b\\ \end{array} \right)_{L}$
&
$\left( \begin{array}{c} \nu_{e}\\ e^{-}\\ \end{array} \right)_{L}$
\hspace{0.2cm}
$\left( \begin{array}{c} \nu_{\mu}\\ \mu^{-}\\ \end{array} \right)_{L}$
\hspace{0.2cm}
$\left( \begin{array}{c} \nu_{\tau}\\ \tau^{-}\\ \end{array} \right)_{L}$
\\
&\\
$u_{R}$, $c_{R}$, $t_{R}$& $e_{R}$, $\mu_{R}$, $\tau_{R}$\\
$d_{R}$, $s_{R}$,$b_{R}$&\\
&\\
+anti-quarks & + anti-leptons \\ \hline
\end{tabular}
\caption{The fermions and the representation of $SU(2)_{L}$ to which they belong.}
\label{tab:fermsu2}
\end{center}
\end{table}
All the left chiral fermions belong to the doublet representation, with the up-type quarks and neutrinos having $T_{3L} = 1/2$ and d-type quarks and negatively charged leptons having $T_{3L} = -1/2$. Note that according to this there are no right handed neutrinos in the particle spectrum of the SM. The colour gauge group $SU(3)_{c}$ commutes with the electroweak gauge group : $SU(2)_{L} \times U(1)_{Y}$. Hence the electroweak interactions of a quark are independent of its colour. Therefore we suppress here the colour index.

\begin{figure}[hbt]
\begin{center}
\includegraphics*[scale=0.7]{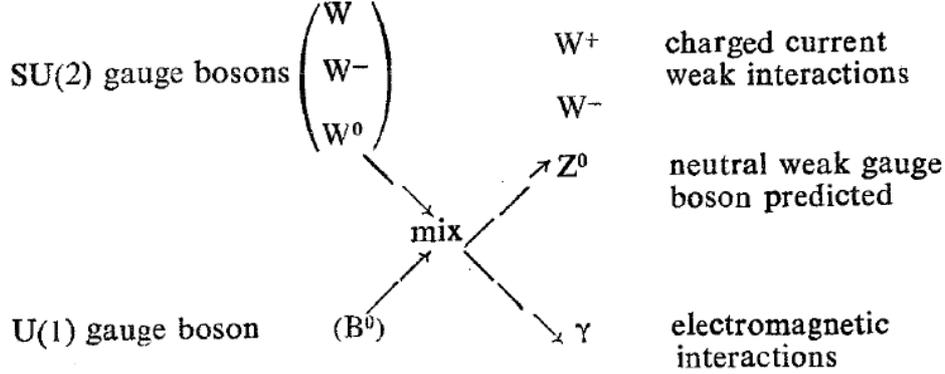}
\caption{A schematic description of mixing  between the $W^{3}_{\mu}$ and $B_{\mu}$. This is taken from~\protect\cite{bib:horizons}.}
\label{fig:W-B mixing}
\end{center}
\end{figure}

As already discussed $U(1)_{em}$ is a linear combination of $U(1)_{Y}$ and a $U(1)$ subgroup of $SU(2)$. This is really the essence of Electro-Weak  unification and is embodied  in Glashow's observation: 
\begin{equation}
Q_{f} = T_{3L} +  Y/2.
\label{eq:qt3ly}
\end{equation}
Here $Q_{f}$ is the electromagnetic charge in units of $|e|$, where e is electron charge, $T_{3L}$  and $Y/2$ denote  the $SU(2)_{L}$  and $U(1)_{Y}$ charges respectively.  Writing the electromagnetic charge as a linear combination  of $T_{3L}$ and the hyper-charge $Y$, embodies the fact that the carrier of electromagnetic  interactions, the photon  $A_{\mu}$ will appear as a linear combination of the neutral vector boson $W_{\mu}^{3}$ and the $U(1)_{Y}$ gauge boson $B_{\mu}$. We can discuss this mixing without making any explicit reference to the Higgs sector. This is what we will do first and then summarise the details of the SSB.  Note that the three  gauge boson fields $W^1_{\mu},W^2_{\mu},W^3_{\mu}$ : all couple only to left handed fermions and $B_{\mu}$ couples to both the left handed and right handed fermions. $B_{\mu}$ and $W^3_{\mu}$ mix, giving one zero mass eigenstate $\gamma$. One then identifies the  other one with a new  neutral vector boson called $Z$. 
One can schematically represent this as shown in the diagram in \Fref{fig:W-B mixing}. Note here that one can discuss this simply at the level of currents which give interactions among matter and gauge bosons in terms of the gauge principle enunciated in Section~\ref{sec:qed}, without making any reference to a specific model which will generate these mixing and masses.
The essence of this mixing is to define two fields  $A_{\mu}$ and $Z_{\mu}$ as a linear combination of $B_{\mu}$ and $W_{\mu}^{3}$ as:
\begin{equation}
A_\mu = cos \theta_W B_\mu  + \sin \theta_W W^3_\mu, \qquad
Z_\mu = -\sin \theta_W B_\mu + \cos \theta_W W^3_\mu
\label{eq:AZWW3}
\end{equation}
Here, $\theta_{W}$, called the `weak mixing angle',  is just an arbitrary parameter denoting the mixing between the $W^{3}_{\mu}$ and $B_{\mu}$. To see how the electric charge $e$ is related to $g_{1},g_{2}$ and $\sin \theta_{W}$, let us construct the currents $J_{\mu}^{W}$ and 
$J_{\mu}^{Y}$ the way electromagnetic current was constructed in Section~\ref{sec:qed}. To do this we need to know the $Y$ values for the different fermion fields written in Table~\ref{tab:fermsu2}. Let us consider a single generation of leptons: $e^{-},\nu_{e}$.  Eq.~\ref{eq:qt3ly} means that the lepton doublet 
${\cal L}^{1}_{L} = \left( \begin{array}{c} \nu_{e}\\ e^{-}\\ \end{array} \right)_{L}$ has $Y=-1$ and $e_{1R}=e_{R}$ which is an $SU(2)_{L}$ singlet has to have $Y=-2$.  Let us indicate the three lepton doublets written in the last three rows of the Table~\ref{tab:fermsu2} by ${\cal L}^{i}$ with $i=1,3$ respectively. We also use ${\cal Q}^{i}_{L}$ with $i=1,3$ to indicate the doublets $\left( \begin{array}{c} u^{i}\\ d^{i}\\ \end{array} \right)_{L}$
where $u^{1} = u, d^{1} = d$ \etc , as written in the first three rows of the same table. For the quark doublets ${\cal Q}^{i}$  the hypercharge $Y$ has value $1/3$. For all the right handed quarks the hypercharge is twice the quark charge and $Y = 2 Q_{q}$, since the value of $T_{3L}$ is zero for all the right handed fields.

Following the discussions in Section~\ref{sec:qed}, let us start from the kinetic part of the Lagrangian for all the fermions in Table~\ref{tab:fermsu2}, to construct the physical currents of the GSW model. For the quarks it is simplest when written  in the gauge eigenstate basis $u'^{i},d'^{i}, i=1,3$. The kinetic term for a fermion field $\psi$ is given by
\begin{equation}
{\cal L}_{\rm fermion kin} = i \bar \psi_L\,\dsl\,\psi_L + i \bar \psi_R\,\dsl\,\psi_R,
\end{equation}
For the $SU(2)_{L} \times U(1)_{Y}$ gauge theory the $\dsl$ is to be replaced by the covariant derivative. This can be  written in terms of the hyper charges  for the fermions given in the earlier paragraph. For a fermion $f$ which is a member of the doublet $\Psi$ this is given by:
\begin{equation}
\partial_\mu \Psi_{L} \to D_\mu \Psi_{L} = \partial_\mu \Psi_{L} - i\frac{g_1 Y_{\Psi}}{2} B_\mu \Psi_{L} -ig_2 W_\mu^a\, \frac{\tau^a}{2} \Psi_{L}.
\label{eq:SMcovarl}
\end{equation}
where  $\Psi_{L} = {\cal L}^{i}_{L},{\cal Q}_{L}$ and $Y_{\Psi}$ is the hypercharge for the doublet $\Psi$. For the case of $SU(2)_{L}$ singlets the covariant derivative is given by
\begin{equation}
D_\mu f_{R} = \partial_\mu f_{R} - i{\frac{g_1 Y_{f_{R}}}{2}} B_\mu f_{R}.
\label{eq:SMcovarr}
\end{equation}
The kinetic terms for all the fermions can be written as:
\begin{equation}
{\cal L}_{\rm fermkin} = 
\sum_{i=1}^{3} \left[ i{\cal L}^{i}_{L}\,\Dsl\,{\cal L}^{i}_{L} + ie^i_R\,\Dsl\,e^i_R +
i{\cal Q}^{'i}_{L}\,\Dsl\,{\cal Q}^{'i}_{L} + iu^{'i}_R\,\Dsl\,u^{'i}_R+ id'^i_R\,\Dsl\,d'^i_R\right].
\label{eq:fermkin}
\end{equation}
Since there are no right handed neutrinos in the strictest version of the SM, for the lepton sector the mass basis and interaction basis are the same. Using the expressions for the covariant derivative $D_{\mu}$ of Eqs.~\ref{eq:SMcovarl}, \ref{eq:SMcovarr}, along with Eq.~\ref{eq:CKM},  we find the interaction Lagrangian to be
\begin{equation}
\Delta {\cal L}_{\rm int} = \Half g_1 J^{\mu\,Y} B_\mu + g_2 \Big(\frac{1}{2\sqrt2}(J^{\mu\,+} W^+_\mu + J^{\mu\,-} W^-_\mu) + J^{\mu\,3} W^3_\mu\Big) 
\label{eq:wjpm}
\end{equation}
where:
\begin{eqnarray}
J^{\mu +} &= & 2\big({\bar\nu}^i_L\,\gamma^{\mu} e^i_L+ {\bar u}^i_L\,\gamma^{\mu} {\bf V}_{ij} d^j_L\big)~~, J^{\mu -} = (J^{\mu +})^{\dag}, \nonumber\\
J^{\mu Y} &=& - {\bar\nu}^i_L\,\gamma^{\mu} \nu^i_L -{\bar e}^i_L\,\gamma^\mu e^i_L -2 {\bar e}^i_R\,\gamma^{\mu} e^i_R
 + {\frac{1}{3}} {\bar u}^i_L\gamma^\mu u^i_L + {\frac{1}{3}} {\bar d}'^i_L \gamma^\mu d'^i_L +{\frac{4}{3}}{\bar u}^i_R\gamma^{\mu} u^i_R - {\frac{2}{3}} {\bar d}'^i_R \gamma^\mu d'^i_R,\nonumber \\
J^{\mu 3} &=& \Half{\bar\nu}^i_L\,\gamma^{\mu}\nu^i_L
- \Half{\bar e}^i_L\,\gamma^{\mu} e^i_L
+ \Half{\bar u}^i_L\,\gamma^{\mu} u^i_L 
- \Half{\bar d}'^i_L\,\gamma^{\mu} d'^i_L,\nonumber \\
W^\pm_\mu &=& \frac{1}{\sqrt2}(W^1_\mu\mp i W^2_\mu).
\label{eq:currents}
\end{eqnarray}
The couplings must now be rewritten so that one linear combination of $B_\mu,W^3_\mu$ couples to the electromagnetic current and an orthogonal one couples to $J^{\mu\,3}$. For this purpose we may ignore the terms in $\Delta {\cal L}$ depending on $W^\pm$. For the remaining part, we may think of the physical fields $A_\mu,Z_\mu$ as the result of a rotation in the $B_\mu,W^3_\mu$ plane, as already discussed in Eq.~\ref{eq:AZWW3}. We write the inverse rotation:
\begin{equation}
W^3_\mu = \cos\theta_W Z_\mu + \sin\theta_W A_\mu,\qquad
B_\mu = - \sin\theta_W Z_\mu+ \cos\theta_W A_\mu 
\label{weakmixing}
\end{equation}
Inserting into the Lagrangian Eq.~\ref{eq:wjpm}, we find:
\begin{equation}
\Delta {\cal L} (B_\mu,W^3_\mu) = 
\left[\Half g_1\cos\theta_W\,J^{\mu\,Y} 
+g_2\sin\theta_W\,J^{\mu\,3}\right]A_\mu\\
 +\left[-\Half g_1\sin\theta_W\,J^{\mu\,Y} 
+g_2\cos\theta_W\,J^{\mu\,3}\right]Z_\mu 
\label{eq:currcoup}
\end{equation}
The expression in the first square bracket in Eq.~\ref{eq:currcoup} must be equal to $e J^{\mu\,{\rm em}}A_\mu$ where $e$ is the unit of electric charge and $J^{\mu\,{\rm em}}$ is given by an expression for all the charged fermions according to 
Eq.\ref{eq:intqed} and can be written as
\begin{equation}
J_\mu^{\rm em} = -{\bar e}^i_L \gamma_\mu e^i_L -{\bar e}^i_R \gamma_\mu e^i_R  + {\frac{2}{3}}\Big( {\bar u}^i_L\gamma_\mu u^i_L + {\bar u}^i_R\gamma_\mu u^i_R \Big)
-{\frac{1}{3}} \Big( {\bar d}'^i_L\gamma_\mu d'^i_L + {\bar d}'^i_R\gamma_\mu d'^i_R \Big).
\label{eq:jmuem}
\end{equation}
This can happen {\it only if}
\begin{equation}
e=g_1\cos\theta_W = g_2\sin\theta_W
\label{eq:egrel}
\end{equation}
It follows that:
\begin{equation}
\tan\theta_W = \frac{g_1}{g_2},\qquad e=\frac{g_1g_2}{\sqrt{g_1^2+g_2^2}}
\label{eq:thetawrels}
\end{equation}
Inserting this into Eq.~\ref{eq:currcoup} we learn that the coupling of the $Z$-boson is:
\begin{equation}
\frac{1}{\sqrt{g_1^2+g_2^2}}\left(-\Half g_1^2 J^{\mu\,Y}+g_2^2 J^{\mu\,3}\right)\,Z_\mu \equiv g_{z} J^{\mu \rm NC} Z_{\mu}
\end{equation}
Thus the weak neutral current is given by:
\begin{equation}
g_z J_\mu^{\rm NC}=\frac{1}{\sqrt{g_1^2+g_2^2}}\left(-\Half g_1^2 J_{\mu}^{Y}+g_2^2 J_{\mu}^{3}\right)
\label{neutcurr}
\end{equation}
where $g_z$ is the coupling constant we associate to the $Z$-boson. This is a convention, because only the combination $g_z J_\mu^{\rm NC}$ appears in formulae. For convenience we choose:
\begin{equation}
g_z = \frac{g_2}{\cos\theta_W} = \sqrt{g_1^2+g_2^2}
\label{eq:gz}
\end{equation}
With this, the weak neutral current is:
\begin{eqnarray}
J_{\mu}^{\rm Z} = J_\mu^{\rm NC} &=& -\Half \frac{g_1^2}{g_1^2+g_2^2}J_\mu^Y + 
\frac{g_2^2}{g_1^2+g_2^2} J_\mu^3\nonumber \\
&=& -\Half \sin^2\theta_W J^Y_{\mu} + \cos^2\theta_W J^3_{\mu}
= J_{\mu}^3 - \sin^2\theta_W J_{\mu}^{\rm em}
\label{eq:weaknc}
\end{eqnarray}
where we have written two different forms that are both useful. 

Taking a look at the first of Eqs.~\ref{eq:currents} show us that the charged currents $J^{\mu \pm}$ involve only the left chiral fermions and have the so called  V(ector)$-$A(xial vector) structure. $J_{\mu}^{\rm em}$ given by Eq.~\ref{eq:jmuem} has pure vector nature. Eqs.~\ref{eq:currents} and~\ref{eq:weaknc} clearly show that, unlike the $W^\pm$ bosons, the $Z$-boson does {\em not} have V$-$A couplings with the fermions. It must be kept in mind that when coupling it to $Z_\mu$, this current should be multiplied by ${\displaystyle g_z=\frac{g_2}{\cos\theta_W}}$. Note that the expression of the current will remain the same even when it is written in terms of the mass eigenstates $d^{i}$ of instead of $d'^{i}$.

The weak neutral current can also be written in terms of the $T_{3}$ and $Y$ of the various fermions and also as a combination of $V$ and $A$ currents as follows.
\begin{eqnarray}
J_{\mu}^{\rm Z} = \sum_{f} J_{\mu}^{\rm Z,f} &=& \sum_{f} \left[\bar f \gamma_{\mu} f_{L} g_{L}^{f} + \bar f \gamma_{\mu} f_{R} g_{R}^{f}\right]\nonumber\\
&=& \left[ {\frac{1}{2}}g_{V}^{f} \bar f \gamma_{\mu} f - 
{\frac{1}{2}} g_{A}^{f} \bar f \gamma_{\mu} \gamma_{5} f \right].
\label{eq:jmuNCZ}
\end{eqnarray}
Here the sum is over all fermions $f^{i} = u^{i},d^{i},e^{i}, \nu^{i}, i =1-3$. The couplings $g_{L}^{f}, g_{R}^{f}, g_{V}^{f}, g_{A}^{f}$ can be read off from Eqs.~\ref{eq:currents} and \ref{eq:weaknc} to be
\begin{equation}
\begin{array}{ll}
g_{L}^{f}  = T_{3}(f_{L}) - \sin^{2}\theta_{W}~Q_{f}, &g_{V}^{f} = T_{3}(f_{L}) + T_{3}(f_{R})  - 2~Q_{f} \sin^{2} \theta_{W} \\[2mm]
g_{R}^{f}  = T_{3}(f_{R}) - \sin^2\theta_{W}~Q_{f}, &g_{A}^{f} = T_{3}(f_{L}) - T_{3}(f_{R})
\end{array}
\label{eq:glrva} 
\end{equation}
In the above equation, we have written down $T_{3} (f_{R}) $ explicitly, which in the GSW model is zero, with a view to generalize the expressions for the weak neutral current, should the fermions belong to other representations of $SU(2)_{L} \times U(1)_{Y}$, other than the one in the GSW model. Recall that $Q_{\rm f}$ is the electromagnetic charge of the fermion in units of positron charge.  

Note now that the form for the neutral current of Eq.~\ref{eq:jmuNCZ} is exactly the same, for all the fermions of a given electrical charge and  given values of the$SU(2)_{L}$ quantum numbers. Since in the GSW model, all the quarks or leptons of a {\it given electric charge and handedness} belong to the same representation of $SU(2)$ the weak neutral current automatically conserves 'flavour', be it the leptonic one or the quark one. This is indeed quite reassuring since the experiments had shown that while 'flavour' changing charged weak current (Eq.~\ref{eq:currents}) exist,  decays caused by 'flavour' changing weak neutral current, FCNC mentioned before, are either forbidden or suppressed by orders of magnitude. Their absence at the tree level is automatically guaranteed in the GSW model, just by the particle content. The values of $g_{A}^{f}, g_{V}^{f}, g_{L}^{f}, g_{R}^{f}$ for the fermions of the GSW model are given in the Table~\ref{tab:04gagv}.

\begin{table}[h]
\begin{center}
\begin{tabular}{|c|c|c|c|c|}
\hline
&&&&\\
$f$& $\nu$ & $e^{-}$&$u$&$d$\\
&&&&\\
\hline
&&&&\\
$g_{L}^{f}$&$\frac12$&$-\frac12 + \sin^2 \theta_W$&$\frac12 - \frac23 \sin^2 \theta_{W}$& $-\frac12 + \frac13 \sin^2 \theta_W$\\
&&&&\\
\hline
$g_{R}^{f}$&$0$&$\sin^{2}\theta_{W}$&$- \frac23 \sin^2 \theta_{W}$& $\frac13 \sin^2 \theta_W$\\
&&&&\\
\hline
&&&&\\
$g_{A}^{f}$&$\frac12$&$-\frac12$&$\frac12 $& $-\frac12 $\\
&&&&\\
\hline
&&&&\\
$g_{V}^{f}$&$\frac12$&$-\frac12 + 2 \sin^{2}\theta_{W}$&$\frac12 - \frac43 \sin^2 \theta_{W}$& $-\frac12 + \frac23 \sin^2 \theta_W$\\
&&&&\\
\hline
\end{tabular}
\end{center}
\caption{The values of axial and vector neutral current couplings $g_{A}^{f}, g_{V}^{f}$ for the fermions of the GSW model. Also
given are the neutral current couplings $g_{L}^{f}, g_{R}^{f}$ for the left and right handed fermion fields.} 
\label{tab:04gagv}
\end{table}
Thus we see that in the GSW model, the weak neutral current couplings are completely determined by $g_{2}$ and $\sin \theta_{W}$. The weak neutral current involving $\nu^{i}$ is pure left handed just like the corresponding charged current, where as for the charged fermions the $V$-$A$ mixture depends on the electromagnetic charge of the fermion because the relative weight of $L$ and $R$  currents is decided by the hypercharge $Y$. While the strength of the axial current is completely decided by the $T_{3}$ value of $f^{i}_{L}$, the vector coupling depends on the weak mixing angle $\theta_{W}$. As we will see later, the experimentally determined value of $\sin^{2} \theta_{W} \sim 0.25$. As a result the weak neutral current coupling of the charged lepton ($e,\mu,\tau$)  is in fact close to zero.

\begin{figure}[htb]
\includegraphics*[width=8cm]{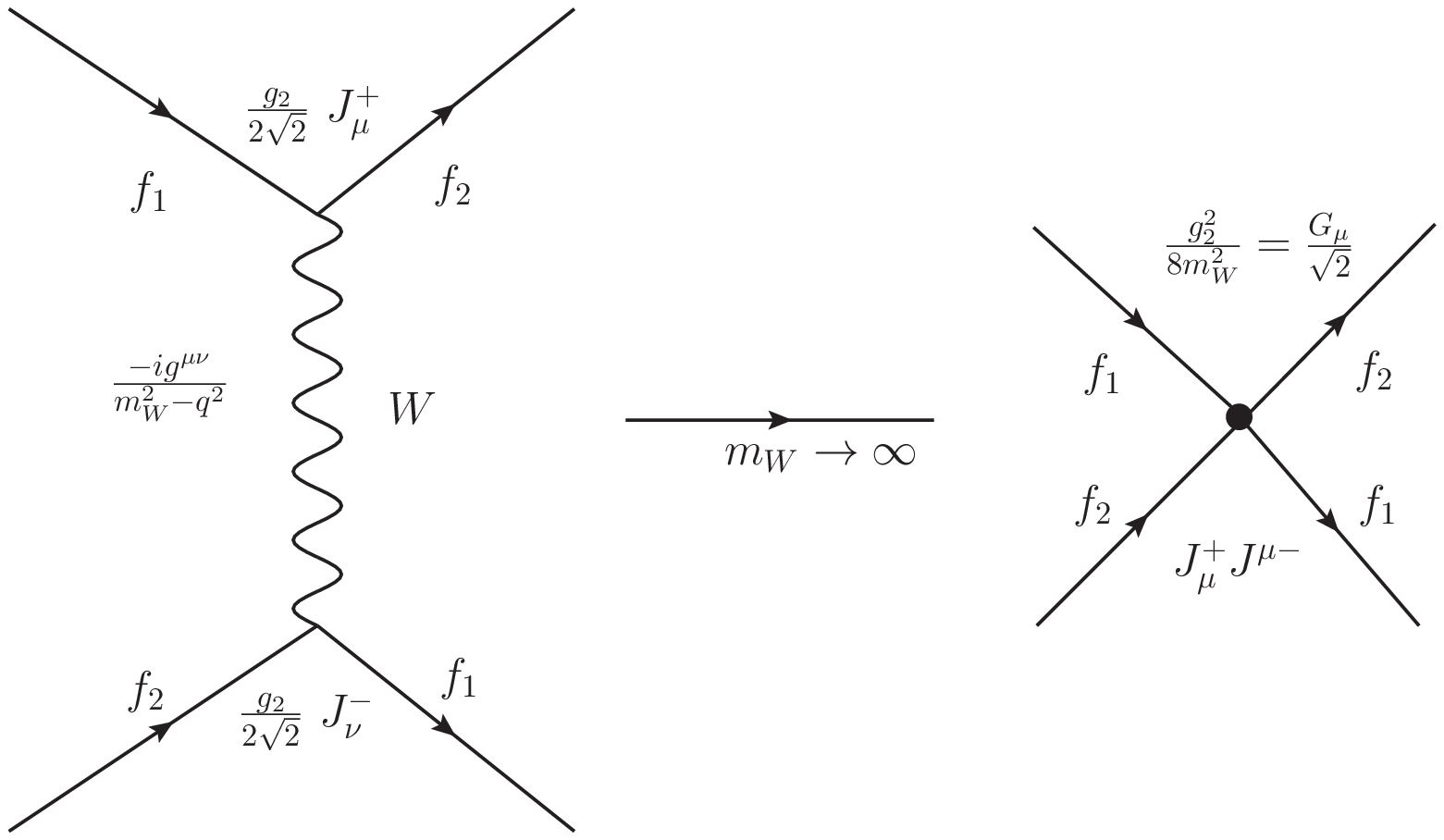}
\hspace{1cm}
\includegraphics*[width=8cm]{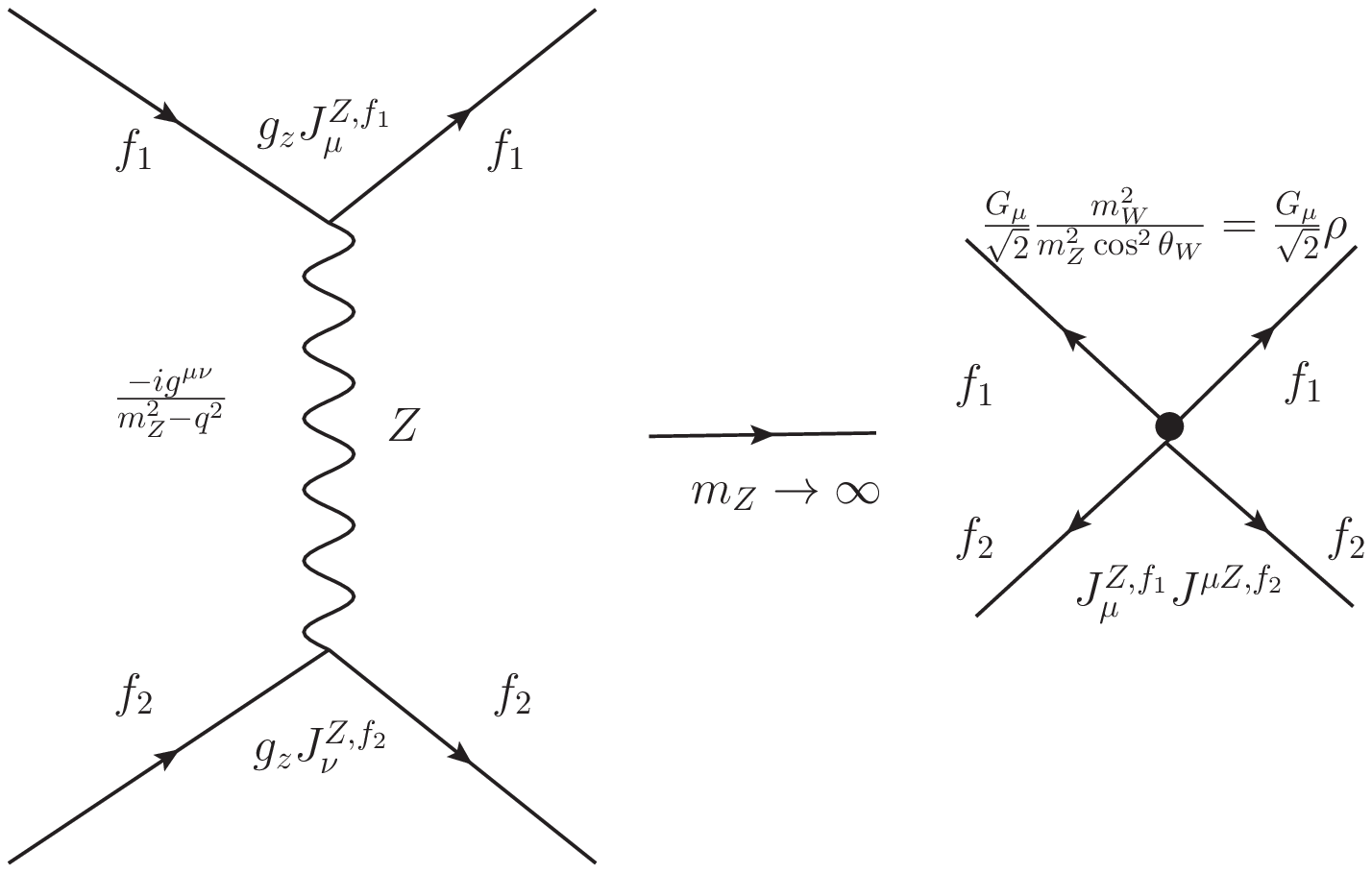}
\caption{Effective current current interactions for charged and neutral current processes in the left and right panel respectively.}
\label{fig:CCNC}
\end{figure}

The interaction of all the quarks and leptons with the electroweak gauge bosons is encoded in the currents $J_{\mu}^{\rm em}$, $J_{\mu}^{\pm}$ and $J_{\mu}^{\rm Z}$ given by \Eq[b]\ref{eq:jmuem}, first of \Eq[b]s~\ref{eq:currents} and Eq.\ref{eq:jmuNCZ}.  In low energy reactions, the appropriate way to adjudge the strength of processes mediated by the weak neutral current is to derive  the current-current form of the interaction Lagrangian starting from Eq.~\ref{eq:jmuNCZ}. This is done by  considering the matrix element of a four fermion scattering process and taking the limit in which the mass of the exchanged gauge boson is infinite. Let us consider the scattering process $f_{1} + f_{2} \rightarrow f_{1} + f_{2}$ through the exchange of a  massive $W^{\pm}$ (\ie via charged current:CC) as indicated in the left panel of Fig.~\ref{fig:CCNC}.
The effective current-current Lagrangian for the scattering process of Fig.\ref{fig:CCNC} can then be written as
\begin{equation}
{\cal L}_{\rm eff}^{\rm CC} = - \frac{g_{2} ^{2}}{8 M_{W}^{2}}
 J_{\mu}^{+} J^{- \mu} = - \frac {G_{\mu}}{\sqrt{2}} J_{\mu}^{+}
J^{- \mu}
\label{eq:LCC}
\end{equation}
with $J_{\mu}^{\pm}$ as given by Eq.~\ref{eq:currents}. On comparing with the current-current interactions of the pre gauge theory days, one then gets:
\begin{equation}
\frac{G_\mu}{\sqrt{2}} = \frac{g_2^2}{8 M_W^2} = \frac{e^2}{8 M_W^2 \sin^2 \theta_W},
\label{eq:Gmumw}
\end{equation}
where $G_{\mu} V_{ud} = G_{F}$. It can be noted here that since $|\sin \theta_{W}| < 1$, the experimentally measured value of $G_{\mu}$ and $e$, tells us that $M_{W} > 37.43\UGeV$. For the limiting value of $\sin \theta_{W} \sim 1$ we get $M_{W} \sim 100 \UGeV$. 

One can similarly write down the effective neutral current interaction effective Lagrangian under the approximation that the $Z$ boson mass is large, by considering the four-fermion scattering process shown in the right panel of Fig.~\ref{fig:CCNC}. This is given by
\begin{equation}
{\cal L}_{\rm eff}^{\rm NC} = - {\frac{g_z^{2}}{2}} \left( \sum_{f} J_{\mu}^{\rm Z,f}\right) \left( \sum_{f} J^{\mu, \rm Z,f} \right)
\label{eq:LNC}
\end{equation}
If one calculates the matrix elements for scattering process $\nu_{e} + e^{-} \rightarrow  \nu_{e} + e^{-}$ taking place via the interaction of Eq.~\ref{eq:LCC} and Eq.~\ref{eq:LNC} respectively,  {\it viz.},  ${\cal M}_{CC}$ and ${\cal M}_{NC}$,  it can be seen that their ratio is given in terms of  $M_{Z}, M_{W}$ and $\sin \theta_{W}$ as:
\begin{equation}
{\frac{{\cal M}_{NC}}{{\cal M}_{CC}}} = {\frac {M_{W}^{2}}{M_{Z}^{2} \cos^{2}\theta_{W}}} \equiv \rho .
\label{eq:rhodef}
\end{equation}
Note further, that this effective Lagrangian involves couplings $g_{2},g_{1}$ and $M_{W},M_{Z}$.  More directly we can use the two measured couplings $G_{\mu}$ and $\alpha_{\rm em}$ along with $\rho$ and one arbitrary parameter of the model the weak mixing angle $\sin \theta_{W}$. $M_{W}, M_{Z}$ are then given in terms of these and we have traded $g_{1},g_{2}$ for $G_{\mu}$ and $\alpha_{\rm em}$.  We will come back to this later in our discussion of the experimental validation of the SM. 

Note also that in these discussions we have completely sidestepped the issue of how the non-zero masses for the gauge bosons and the fermions written can be made consistent with gauge invariance. In case of the gauge bosons the loss of gauge invariance also means loss of renormalisability and hence consequently of the ability to make any predictions. So one of the problems to be addressed is how to generate the mass terms below in a gauge invariant manner.
\begin{equation}
{\cal L}_{mass}  = {\frac{1}{2}} M_{Z}^{2} Z_{\mu} Z^{\mu} +  M_{W}^{2} W_{\mu}^{+} W^{- \mu}  
+ \sum_{i} m_{i} \left[\bar  \psi_{iL} \psi_{iR} + \bar \psi_{iR} \psi_{iL}\right]. 
\label{eq:lmass}
\end{equation}
It should be noted that the sum in Eq.~\ref{eq:lmass} is over all the fermions except the neutrinos which are assumed to be massless here in this discussion.  

\subsubsection{SSB and generation of $W/Z$ masses.}
\label{sec:WZmasses}
Before we move on to discuss more about the novel phenomenon of the existence of the weak neutral current, which was but the first step in testing and establishing the GSW model, let us first look at the issue of how nonzero masses for the gauge bosons and all the fermions can be generated in a gauge invariant manner. This is achieved~\cite{bib:two} through the famous SSB mechanism~\cite{bib:six}. 

One starts with the $SU(2)_{L} \times U(1)_{Y}$ gauge invariant Lagrangian, for the nonabelian gauge  fields $W_{\mu}^{i},~~i =1,3$ and  the abelian gauge field $B_{\mu}$, analogous to Eqs.~\ref{eq:nabelian-density} and \ref{eq:lqed} respectively.
\begin{eqnarray*}
{\cal L}_{massless}  &=& {\cal L}_{gauge} + {\cal L}_{fermikin}\\
&=& -{\frac{1}{4}} B_{\mu \nu} B^{\mu \nu} - {\frac{1}{4}} F^{a}_{\mu \nu} F^{a,\mu \nu} + {\cal L}_{fermikin}.
\end{eqnarray*}
Here $B_{\mu \nu} = \partial_{\mu} B_{\nu} - \partial_{\nu} B_{\mu}$ and $F^{a}_{\mu \nu} = \partial_{\mu} W^{a}_{\nu} - \partial_{\nu} W^{a}_{\mu} + g f^{abc} W^{c}_{\mu} W^{c}_{\nu}$
with $f^{abc} = \epsilon^{abc}$. Further,  ${\cal L}_{fermkin}$ is given by Eq.~\ref{eq:fermkin}.

The considerations of SSB begin by considering a complex scalar field $\Phi$, which is a colour singlet and an $SU(2)_{L}$ doublet with hypercharge $Y_{\phi} = 1$, given by
\[
\Phi = \left( \begin{array}{c} \phi_{1}\\ \phi_{2}\\ \end{array} \right) \equiv  \left( \begin{array}{c} \phi^{+}\\ \phi^{0}\\ \end{array} \right)
\]
where $\phi_{i} = Re (\phi_{i}) + i Im (\phi_{i})$ and similarly for $\phi^{+}, \phi^{0}$. Thus we have four real scalar fields and the Lagrangian we consider is, 
\begin{equation}
{\cal L}_{\Phi} = (D_{\mu} \Phi)^{\dag} D^{\mu} \Phi - V(\Phi)
= (D_{\mu}\Phi)^{\dag} (D_{\mu} \Phi) + \mu^{2} \Phi^{\dag} \Phi - \lambda (\Phi^{\dag} \Phi)^{2},
\label{eq:Lscalar}
\end{equation}
with $\mu^{2} > 0$.  Note that compared to the Lagrangian for a free complex scalar field, this has the wrong sign for the quadratic term. So $\mu$ is not the mass and we can not interpret the excitations of the field $\Phi$ as propagating degrees of freedom. But it is precisely this wrong sign that is required for the spontaneous symmetry breaking to occur.
\begin{figure}[hbt]
\begin{center}
\includegraphics*[width=8cm,height=6cm]{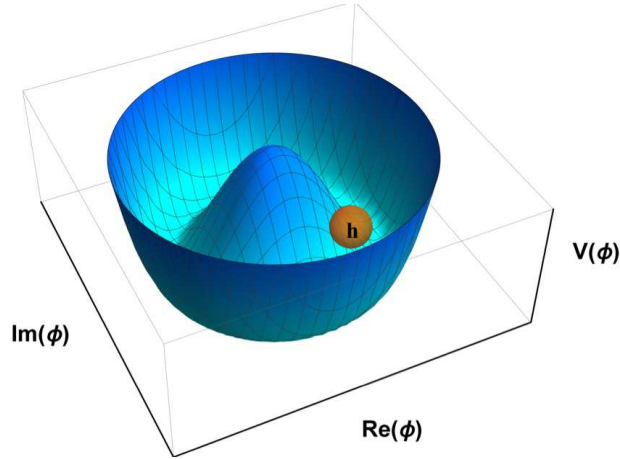}
\caption{A sketch of the mexican hat potential}
\label{fig:mexicanhat}
\end{center}
\end{figure}

Let us look at \Figure[b]~\ref{fig:mexicanhat} which shows a sketch of a similar potential, but for a single complex scalar field $\phi$:
$V(\phi) = - \mu^{2} \phi^{\dag} \phi + \lambda |\phi^{\dag} \phi|^{2}$.  This shows clearly that classically the point $\Re e \phi = \Im m \phi = 0$ is in fact a maximum and there exist a continuum of minima where the field is nonzero, all related to each other by the symmetry transformation of the Lagrangian, which is a $U(1)$ transformation for the case shown in Fig.~\ref{fig:mexicanhat}. SSB occurs when the quantum field configuration is such that the field has a nonzero vacuum expectation value corresponding  to one of these minima, thus breaking the symmetry. The system is then described by the fluctuations of the fields around this minimum.

For the $V(\Phi)$ of Eq.~\ref{eq:Lscalar} the minimum occurs for
\begin{equation}
\Phi^\dag \Phi = \frac{\mu^2}{2\lambda}\equiv\frac{v^2}{2}.
\label{eq:vmin}
\end{equation}
The $SU(2)$ symmetry is broken when the vacuum field configuration chooses a particular direction in the $\phi^{1}, \phi^{2}$ space.  The choice of the representation of the Higgs field decides pattern of symmetry breaking. For the case of $SU(2)_{L}\times U(1)_{Y}$ case under consideration, the unbroken symmetry should correspond to the $U(1)_{\rm em}$ invariance since the $\gamma$ is massless. Glashow's partial symmetry breaking with $Q = T_{3L} + Y/2$ aids in deciding how to implement and helps us decide  which of the four scalar fields can acquire a nonzero vev. The charge operator should annihilate the vacuum and hence only the electrically neutral, real scalar field can have a nonzero vev. The required symmetry  breaking pattern is guaranteed (with the choice $Y_{\Phi} = 1$) by
\begin{equation}
 <0|\Phi |0> = <\Phi>_{0} = \left( \begin{array}{c} 0 \\ v/\sqrt{2}\\ \end{array}\right)
\label{eq:vev}
\end{equation}
As follows from Eq.~\ref{eq:vmin},  $v = \sqrt{\frac{\mu^{2}}{\lambda}}$.  
Since $\Phi$ is a $SU(2)_{L}$  doublet clearly this choice for the vev  means that the vacuum configuration breaks the symmetry and chooses a particular minimum from amongst the continuum of minima, similar to the situation depicted in the picture in Fig.~\ref{fig:mexicanhat}. Since the electromagnetic charge still annihilates the vacuum, the  symmetry breaking pattern is 
$SU(2)_{L} \times U(1)_{Y} \rightarrow U(1)_{em}$

One can rewrite the field $\Phi$  using the following parameterisation in terms of $\theta_{a}, a =1,3$ and $h$ all of which have vacuum expectation value to be $0$.
\begin{equation}
\Phi(x) = \frac{1}{\sqrt{2}} \left( \begin{array}{c} \theta_{2} + i \theta_{1}\\ v + h(x)  - i\theta_{3} \\ \end{array}\right).
\label{eq:phipara}
\end{equation}
If $\theta_{a}(x), h(x)$ are small then  we get
\begin{equation}
\Phi(x) = \exp (i \theta_{a} \tau^{a}/v) \left (\begin{array} {c} 0 \\ v/\sqrt{2} + h(x)/\sqrt{2} \\ \end{array} \right).
\end{equation}
This is then an expansion of the field $\Phi$ in terms of the fluctuations around the minimum. One recognizes the factor outside as that for a gauge transformation for a $SU(2)_{L}$ doublet. Comparing this expression with Eq.~\ref{eq:gaugetrans} we see immediately that by doing a gauge transformation $\Phi' = -\exp (i \theta_{a} \tau^{a}/v) \Phi$ we get, 
\begin{equation} 
\Phi'(x) = \left( \begin{array}{c} 0 \\ v/\sqrt{2} + h(x)/\sqrt{2} \\ \end{array}\right)
\label{eq:phiugauge}
\end{equation}
This gauge is called the Unitary gauge.
\Eq[b]~\ref{eq:vev}  also means that the vev is zero for field $h$. The three scalar degrees of freedom $\theta_{i}$ in fact have disappeared from the spectrum in this gauge. Indeed these three correspond to three Goldstone Bosons corresponding to the three generators of the symmetry group that are broken spontaneously. 

Let us now evaluate ${\cal L}_{\Phi}$ of Eq.~\ref{eq:Lscalar} in the unitary gauge using $\Phi'$ from Eq.~\ref{eq:phiugauge}. We use 
\begin{equation}
D_\mu\Phi = \partial_\mu \Phi - i\frac{g_1}{2} B_\mu\Phi - ig_2 W_\mu^a\,\frac{\tau^a}{2}\,\Phi.
\label{eq:covarSM}
\end{equation}
The covariant derivative term in Eq.~\ref{eq:Lscalar} gives rise to terms quadratic in the gauge boson fields which are given as below:
\begin{eqnarray}
\Bigg|\bigg(\frac{g_1}{2}B_\mu+g_2 \frac{\tau^a}{2}W_\mu^a\bigg)\left(\begin{array}{c}
0 \\ \frac{v}{\sqrt2} \end{array}\right)\Bigg|^2
&=& \frac{g_2^2v^2}{8}\Big(W_\mu^a W^{a\,\mu}\Big)
+\frac{g_1^2v^2}{8}B_\mu B^\mu \nonumber\\
&&\quad -\frac{g_1g_2 v^2}{4}W_\mu^3 B^\mu \nonumber\\
&=& \frac{g_2^2v^2}{4}W_\mu^+ W^{-\,\mu}
+\frac{v^2}{8}\Big(g_1 B_\mu - g_2W_\mu^3\Big)^2\nonumber\\
& =& \frac{g_2^2v^2}{4}W_\mu^+ W^{-\,\mu}
+\frac{(g_1^2+g_2^2)v^2}{8}Z_\mu Z^\mu
\label{eq:WZmassterm}
\end{eqnarray}
This then tells us directly that three of the four gauge bosons
become massive: the $W^{\pm}$ and one linear combination of $B_{\mu}, W_{\mu}^{3}$ which we call $Z_{\mu}$ and the
orthogonal linear combination remains massless. This also tells us
\begin{equation}
M_W^2  = \frac{g_2^2 v^2}{4},~~~~ M_Z^2 = \frac{(g_1^2+g_2^2)v^2}{4} = \frac{M_{W}^{2}}{\cos^{2} \theta_{W}}.
\label{eq:WZmasses}
\end{equation}
Identifying $g_{1} B_{\mu} - g_{2} W_{\mu}^{3}$ with $Z_{\mu}$ with proper normalisation we see that expression for $Z_{\mu}$ is the same as that given in Eq.~\ref{eq:AZWW3} and $\tan \theta_{W}$ same as that in Eq.~\ref{eq:thetawrels}.  
 
The new thing compared to the earlier discussion of the GSW model,  is that now one has a model for generating masses for the gauge bosons from the gauge invariant kinetic term of the scalar field.  The combination $A_{\mu}$ remains massless as it must. The fact that the same linear combination which has mass zero also has the couplings to fermions that a photon field $A_{\mu}$ must have (cf. Eqs.~\ref{eq:currcoup},\ref{eq:jmuem}) means that the SSB has achieved the desired symmetry breaking pattern. Further, in the earlier discussion $M_{W}, M_{Z}$ were unknowns, put in by hand; but now we find that the two are related to each other.

Another fact worth noticing is that the value of the vev $v$ gets determined in terms of measured value of $G_{\mu}$. Using  the expression for $M_{W}$ in Eq.~\ref{eq:WZmasses} and that for $G_{\mu}$ in 
Eq.~\ref{eq:Gmumw}, we get
\begin{equation}
v = \left (\frac{1}{\sqrt{2} G_{\mu}}\right)^{1/2} \simeq 246
 \UGeV.
\label{eq:vevgmu}
\end{equation}
Using the expression for $g_{2}$ in terms of $e$ and $\sin \theta_{W}$ and Eq.~\ref{eq:WZmasses}, one can then see that,
\begin{equation}
M_{W} = \sqrt{\frac{\pi}{\sqrt{2} G_{\mu}} \frac{\alpha_{\rm em} }{\sin^{2} \theta_{W}}} = \frac{37.3}{\sin \theta_{W}} \UGeV;~~\qquad\qquad M_{Z} = \frac{37.3}{\sin \theta_{W} \cos\theta_{W}} \UGeV.
\label{eq:mwsinthw}
\end{equation}
This is the promised  reduction in the number of free parameters. Now everything in the GSW model is predicted in terms of the two known constants $\alpha_{\rm em}, G_{\mu}$ and one free parameter $\sin^{2}\theta_{W}$. An accurate determination of $G_{\mu}$ is possible via life time of the muon,  $\tau_{\mu}$. Since $|\sin\theta_{W}| < 1$ this also means we have an automatic lower limit on the masses of the $W,Z$ bosons of $37.3$ GeV.

We further notice from Eq.~\ref{eq:WZmasses} that the ratio $\rho$ defined in Eq.~\ref{eq:rhodef} is {\it predicted} to be unity in the GSW model and we have   
\[
\rho = \frac {M_{W}^{2}}{M_{Z}^{2} \cos^{2}\theta_{W}} = 1 .
\]
Noting, in addition, from Table~\ref{tab:04gagv} that $g_{A}^{f}, g_{A}^{f}$ are numbers of ${\cal O} (1)$, we can then conclude from Eqs.~\ref{eq:LCC}--\ref{eq:LNC} that one should expect the $\nu$ induced scattering processes via neutral current interactions to  happen at rates similar to those via charged current interactions. This conclusion is of course independent of the actual values of $M_{W}, M_{Z}$ with the proviso that the energies are much smaller compared to these masses. Thus the GSW model not only predicted the existence of a weak neutral gauge boson and weak neutral current processes mediated by it, but it also predicted their strength  to be ${\cal O} (G_{\mu})$.

The experiments with the bubble chamber Gargamelle  at CERN, found evidence for the processes induced by neutral current interactions as predicted by the GSW model. The energies involved were smaller than the lower bound on the $W/Z$ masses implied by Eq.~\ref{eq:mwsinthw}. Hence one can use the effective lagrangian description of Eqs.~\ref{eq:LCC} and \ref{eq:LNC}. In addition, measurements of cross-sections for neutral current processes further showed the ratio $\rho$ to be close to 1. Thus these provided both the qualitative and quantitative support for the GSW model. This was  before $W,Z$ were  experimentally discovered and their masses measured.

It was further seen that the model prediction of $\rho =1$ is true  even with additional Higgs fields as long as the scalars responsible for the SSB belong to the doublet representation. This can be understood in terms of an accidental symmetry that the scalar  potential $V(\Phi )$ seems to have for this choice of the representation of the Higgs field. We shall discuss later this symmetry called the Custodial Symmetry.

After working out the remaining terms also in terms of the field $\Phi'$ in the unitary gauge we get,
\begin{eqnarray}
{\cal L}_{\Phi'}^{U} &=&  \left[M_{W}^{2} W_{\mu}^{+} W^{-\mu} +  \frac{1}{2} M_{Z}^{2} Z_{\mu} Z^{\mu}\right]\left(1 + \frac{h}{v}\right)^{2} + \frac{1}{2} (\partial_{\mu} h)^{2} + \mu^{2} h^{2} - \lambda v h^{3} - \lambda/4 h^{4} \nonumber \\
&=& {\cal L}_{VVh} + {\cal L}_{h}.
\label{eq:LhV}
\end{eqnarray}
The first two terms are the mass terms for the $W,Z$ as well as the term describing the interaction between a pair of gauge bosons and the $h$. The form of this term makes it very clear that the strength of the $VVh$ coupling is simply proportional to the mass of the corresponding gauge boson. This proportionality between the mass and the coupling is the most critical prediction of the SSB.

The remaining terms describe now a real, scalar field which is a propagating degree of freedom with mass $M_{h} = \sqrt {2 \mu^{2}}$. Since $v = \sqrt{\mu^{2}/\lambda}$, the mass of the Higgs boson is given in terms of self coupling $\lambda$. This being an arbitrary parameter of the Higgs potential, not fixed by any condition, $M_{h}$ too is a free parameter of the SM, with no prediction for it. We will come back to this later when we look at theoretical constraints on the Higgs mass!

In  the unitary gauge now the propagating degrees of freedom are the three  massive gauge bosons $W^{\pm}, Z$ , one massless gauge boson $\gamma$  and ONE propagating massive scalar.
A massless vector boson has two degrees of freedom corresponding to the two degrees of polarisation it can have whereas a massive gauge boson has three degrees of freedom as it can also have longitudinal polarisation. Out of the four scalar degrees of freedom  only one, $h$, is left in the particle spectrum and the other three provide the remaining degrees of freedom corresponding to the longitudinal polarisation necessary for the three gauge bosons to be massive. The total number of bosonic degrees of freedom before SSB are twelve: eight corresponding to four massless gauge boson fields $W_{\mu}^{a,a=1,3},B_{\mu}$ and the four scalars in $\Phi$. After the SSB one has again twelve bosonic degrees of freedom : nine corresponding to the three massive gauge bosons $W^{\pm}, Z$, two corresponding to the massless photon $\gamma$  and  one corresponding to the massive neutral scalar $h$. In the unitary gauge the particle spectrum contains only the physical fields and the Goldstone boson fields $\theta_{a}, a=1,3$ of Eq.~\ref{eq:phipara}, are absent from the spectrum.  The same is depicted somewhat pictorially below:

\begin{table}[htb]
\begin{center}
\begin{tabular}{|cc|c|cc|}
\hline
&&&&\\
${\cal L}_{gauge}^{massless}$ & + ${\cal L}_{\Phi}$ &~~~~~~  & ${\cal L}_{gauge}^{massive}$ & +~~$ {\cal L}_{h}$\\
&&&&\\
4 massless & 4 scalar &$\xrightarrow{SSB, Unitary gauge}$ & 3 massive, 1 massless & 1 physical \\
 gauge bosons& fields&~~~~~~~& gauge bosons &scalar\\
 &&&&\\
8 d.o.f.&4 d.o.f.&~~~~~~~~&11 d.o.f&1 d.o.f.\\
\hline
\end{tabular} 
\caption{Bosonic degrees of freedom before and after the SSB.}
\label{tab:dof}
\end{center}
\end{table}
\subsubsection{SSB and generation of lepton masses}
It was really Weinberg's  genius that he saw that exactly the same  mechanism can be used effectively to give masses to {\it all} the fermions.  He did so by  postulating a gauge invariant term for interaction between the fermionic matter fields and the Higgs field! For the electron, it can be written as  
\begin{equation}
{\cal L}_{yukawa}^{e} = -f^{*e} \bar{\cal L'}_{1L} \Phi e'_{1R} +  h.c.
\end{equation}
The 'prime' on the lepton fields are to indicate that these the interaction eigenstates. One can also see clearly that this is a singlet under $SU(2)_{L}$ and $U(1)_{Y}$). Using $\Phi'$ of Eq.~\ref{eq:phiugauge}, we get  
we get 
\begin{equation}
{\cal L}_{yukawa}^{e,U} = - \frac {f^{*e} v}{\sqrt{2}} (\bar e'_{L} e'_{R}) (1 + h/v) + h.c.
\end{equation}
The first term in the bracket is clearly the mass term. Hence we have
\begin{equation}
m_{e} = + f^{*e} v /\sqrt{2} \qquad \qquad e' = e
\end{equation}
Second term in the bracket also then tells us that the $hee$ coupling is just $m_{e}$. One can do the same for all the charged leptons. Thus the {\it gauge invariant} Lagrangian ${\cal L}_{yukawa}^{i}$, gives rise to the mass term for the leptons. 

The original paper by Weinberg~\cite{bib:three} talked {\it only} of leptons.  With some extra work the procedure works for the  case of quarks as well. The most general Yukawa interaction can be written as, 
\begin{equation}
{\cal L}_{yukawa}^{q} = -f^{*d}_{ij} \bar{\cal Q}^{'i}_{L} \Phi d{'i}_{R} - f^{*u}_{ij} \bar {\cal Q}^{'i}_{L} \tilde \Phi u^{'j}_{R} 
+ h.c.
\label{eq:Yukawaq}
\end{equation}
where $\tilde \Phi = i \sigma_{2} \Phi^{*}$. We want the  $\cal L$ to be invariant under $SU(2)_{L} \times U(1)_{Y}$ transformations. The  $SU(2)_{L}$ invariance is guaranteed by construction. Recall,  for the right handed quark fields the hyper charges are $Y = - \frac{2}{3}$ and $\frac{4}{3}$ for the down-type and up-type quarks respectively whereas $\bar {\cal Q}^{i'}$ has $Y =  - \frac{1}{3}$. As a result, the second term involving up-type quarks in ${\cal L}_{yukawa}^{q}$ is invariant {\it ONLY} if the hypercharge of the scalar doublet has $Y = -1$.  The most economical choice for such a field is then $\tilde \Phi$.  Again the $'$ for the quark fields indicate that these are interaction eigenstates. In the unitary gauge, using $\Phi'$ of Eq.~\ref{eq:phiugauge} we get,
\begin{equation}
{\cal L}_{yukawa}^{q,U} = -\frac{f^{*d}_{ij}}{\sqrt{2}} v  ~ \bar d'^{i}_{L} (1 + h/v) d'^{j}_{jR}  
- \frac{f^{*u}_{ij}}{\sqrt{2}} v ~\bar u'^{i}_{L}(1+h/v)u'^{j}_{R} +  h.c.
\end{equation}
We see that after the SSB, the $SU(2)_{L} \times U(1)_{Y}$ gauge invariant Lagrangian ${\cal L}_{yukawa}^{q}$ of Eq.~\ref{eq:Yukawaq} contains mass terms for both the up-type and down-type quarks. These are matrices in the generation space and are given by;
\begin{equation}
m_{ij}^{d} = \frac{f^{*d}_{ij}}{\sqrt{2}} v  \qquad , \qquad m_{ij}^{u} = \frac{f^{*u}_{ij}}{\sqrt{2}} v .
\end{equation}
Since in general $f^{*d}_{ij}, f^{*u}_{ij}$ are completely arbitrary matrices in the generation space, these mass matrices are not  diagonal in the basis $d^{'i},u^{'i}$, in the most general case. The states $d^{'i},u^{'i}, i= 1-3$ are therefore clearly not mass eigenstates.  $d^{i},u^{i}, i=1,3$ are thus linear combinations of $d^{'i},u^{'i}, i =1,3$. In the most general case, after diagonalisation of both the $m^{d}, m^{u}$ matrices given above, we can write the weak charged current in terms of the mass eigenstates $u^{i}, d^{i}$ as indicated in Eqs.~\ref{eq:wjpm} and~\ref{eq:currents}. An alert reader might have wondered why one does not have such a mixing matrices for the charged leptons. This has to do with the fact that the mixing matrix ${\bf V}$ given in Eq.~\ref{eq:CKM}, arises from a mismatch in the matrices which diagonalise the $d$ and $u$ mass matrices, and will be different from each other in the most general case. However, for the charged lepton case, the neutrinos being massless, the corresponding mismatch between matrices diagonalising the charged lepton and neutrino mass matrices,  can not have any physical implications. 

\subsubsection{Flavour changing neutral currents}
An alert reader might wonder why we emphasize the issue of FCNC so much. To appreciate this, we have to discuss briefly one more puzzle that the weak decays of the $K$ mesons had presented to  the theorists during the development of a theory of weak interactions. Let us  consider the leptonic decay of $K^{+} \rightarrow l^{+} \nu_{l}$. The big difference in  the measured branching ratios for the leptonic decays $l \nu_{l}$,  $(63.55 \pm 0.11) \%  $ and $(1.581 \pm 0.007)\times 10^{-5}$ for $l = \mu, e$ respectively, can be understood in terms of the $V-A$ structure of the leptonic current  in first of the equations in Eq.~\ref{eq:currents}. The $K^{\pm}$ were known to have a non-leptonic decay as well, with a branching ratio of about $25 \% $. On the other hand, the $K^{0}$ mesons were found to decay  only in the non-leptonic final states. For example, even today only an upper limit of $9 \times 10^{-9}$ is available for the branching ratio for $K_{S}^{0} \rightarrow \mu^{+}\mu^{-}$, meaning thereby that this decay is not yet seen.
This big difference in the leptonic branching ratios for the $K^{\pm}$ on the one hand  and $K_{S}^{0}$ on the other,  was  interpreted as suppression of strangeness changing weak neutral current as compared to the strangeness changing, weak charged  current. However, there was no 'understanding' as to why this should be so. So after the postulation of weak neutral currents in the GSW model, it was an obvious question to ask whether the model provides a 'natural' understanding of the observed fact of suppression of the flavour changing weak neutral currents.

\begin{figure}[htb]
\begin{center}
\includegraphics*[width=5.5cm]{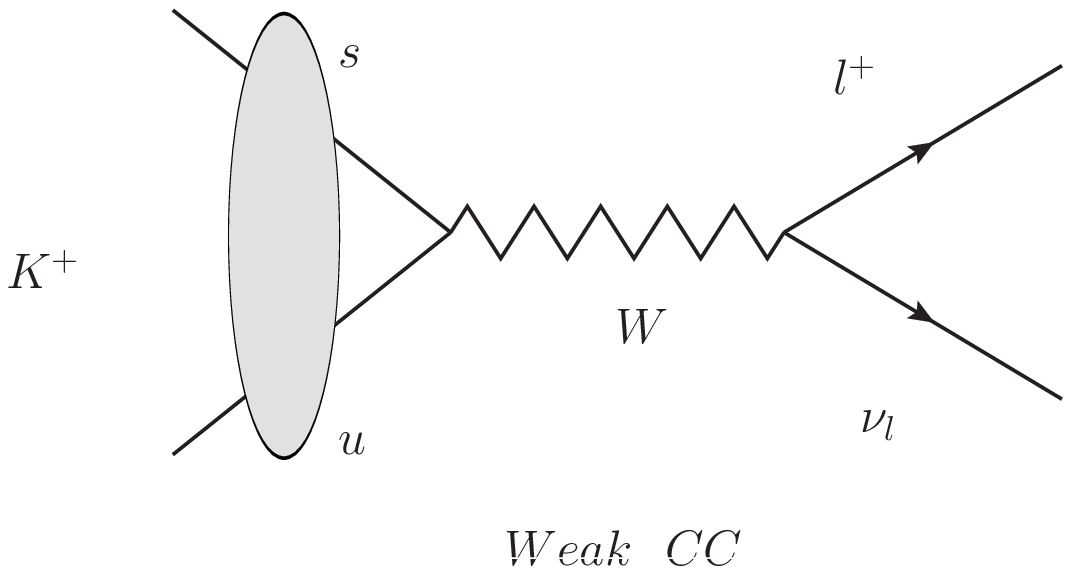}
\hspace{2cm}
\includegraphics*[width=5.5cm]{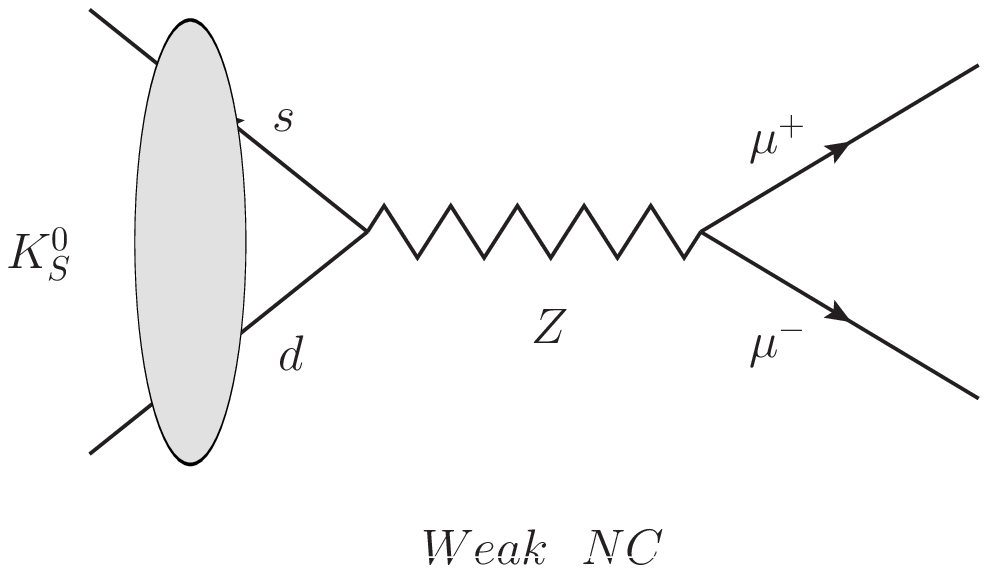}
\caption{Leptonic decay of  mesons via currents. The blob indicates that the quarks are bound in the $K$ mesons.}
\label{fig:Kdecay}
\end{center}
\end{figure}

Weak decays of hadrons can be understood (and calculated)  in the framework of the quark model and $W^{\pm}$ bosons. The left panel of Fig.~\ref{fig:Kdecay} shows the diagram which needs to be computed for (say) the $\Delta S = 1$ weak decay, $K^{+} \rightarrow \nu_{l} l^{+}$  taking place via charged current.
The hadronic decays of the $K^{\pm}$ mesons can then be understood in terms of hadronic decays of the $W^{\pm}$. Both the non-leptonic and leptonic decays of the $K^{\pm}$ thus happen at the weak rate; amplitude being proportional to $G_{\mu}$, the relative branching ratios being controlled by those of the $W^{\pm}$ which are known in the GSW model.

The existence of the weak neutral $Z$ boson, in principle, could have given rise to weak leptonic decay of $K_{S}^{0}$ mediated by the  $Z$ as depicted in the right hand panel of Fig.~\ref{fig:Kdecay} with rates similar to the charged weak current processes, should a $u-d-Z$ vertex exist.  This too would be then a $\Delta S = 1$ process. The happy instance of absence of such a term in the $J_{\mu}^{Z}$ of Eq.~\ref{eq:jmuNCZ}, explains the absence of  pure leptonic decays of $K_{S}^{0}$ via the weak neutral current at the tree level. This is then consistent with the experimentally observed suppression of such decays.  As has been already mentioned, absence of this current is due to the fact that the fermions of the SM with a given electromagnetic charge and handedness, belong to the {\it same} representation of the EW gauge group. Thus, the observed suppression of the FCNC decays, in fact indicated the {\it need} of the existence of the $c$ quark with $Q = \frac23$, which is a $T_{3} = \frac{1}{2}$ member of the $SU(2)$ doublet along with the $s$ quark. The mere presence of a $c$-quark in the spectrum is enough to achieve this absence of the FCN. Further, this result is independent of the masses of the quarks involved.

Even though such a decay is forbidden at the {\it tree level} by the absence of FCNC couplings in Eq.~\ref{eq:jmuNCZ}, it can take place through loop processes at a higher order in $G_{\mu}$ through the charged current (CC) interactions.  In a renormalisable gauge theory such as the GSW model, one should be able to  compute the rate at which it is predicted to occur. This can  then be compared  with the observed suppression of less than one part in $10^{9}$.  

\begin{figure}[hbt]
\begin{center}
\includegraphics*[width=5.5cm,height=5cm]{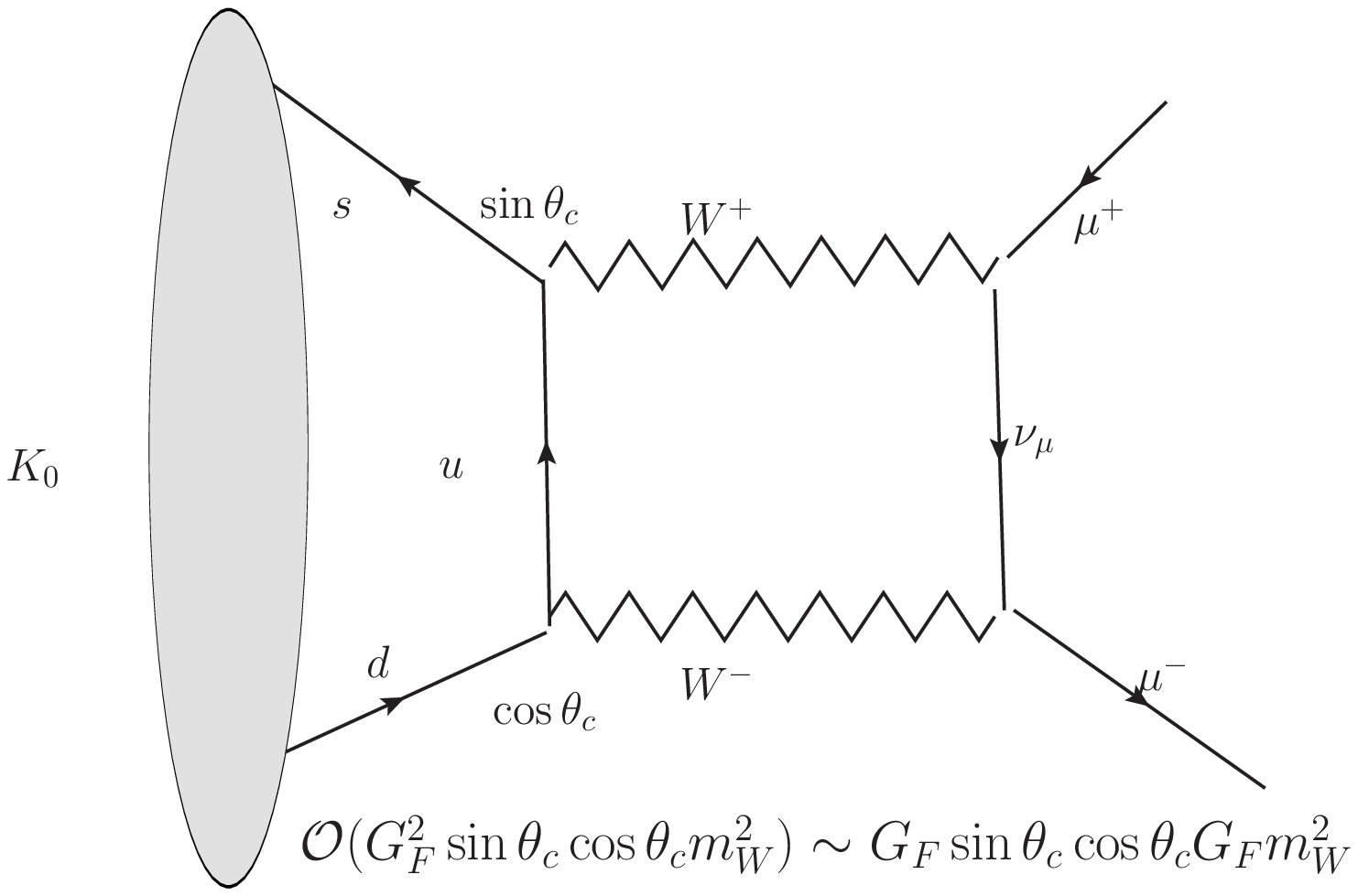}
\hspace{2.0cm}
\includegraphics*[width=5.5cm,height=5cm]
{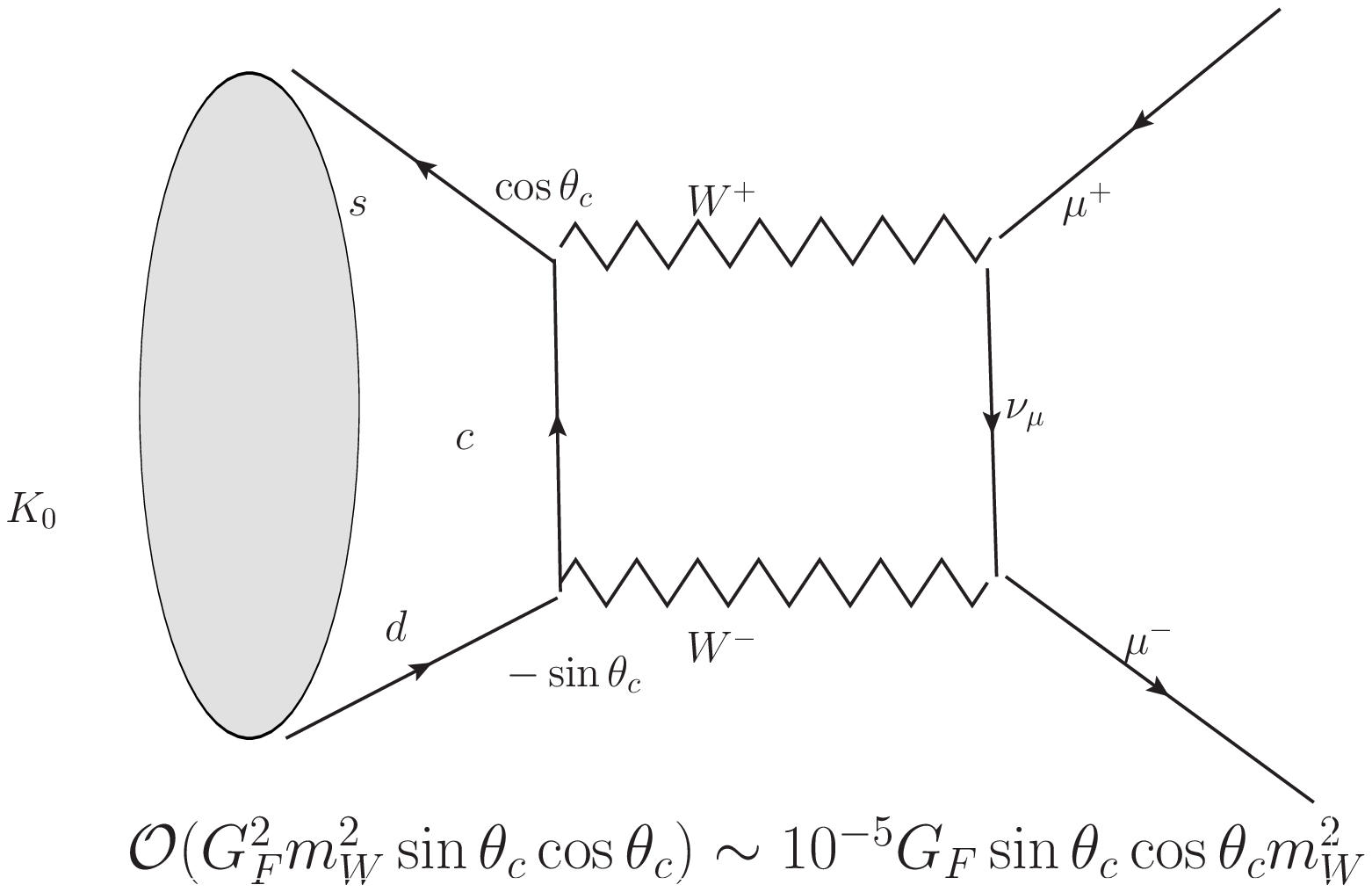}
\caption{One loop diagrams giving rise to $K_{S}^{0} \rightarrow \mu^{+} \mu^{-}$}
\label{fig:kmu+mu-}
\end{center}
\end{figure}

Fig.~\ref{fig:kmu+mu-} depicts two of the possible four box diagrams which would give rise to this decay at the loop level, in a world with only four quarks $u,d,s$ and $c$. The difference between the left and the right panel is in identity of the charge $+\frac{2}{3}$ quark which is exchanged in the $t$-channel.  There will also be two more  diagrams where the $W$'s form the vertical legs of the box.
One calculates these loops explicitly in a gauge theory with SSB as it is renormalisable. In a world with only $3$ quarks, one would compute only the diagram in the left panel where the $u$ quark is exchanged in the $t$ channel and the amplitude of a second diagram where it is $W$ which is exchanged in the $t$ channel and $u$ forms the horizontal leg of the box.
Recall we already know that for a unified theory $M_{W} > 37.3 $ GeV. The loop amplitude, can then be computed in the approximation $m_{u}^{2} << M_{W}^{2}$.  The amplitude of the box will be proportional to $G_{\mu}^{2} \sin \theta_{c} \cos \theta_{c} \times {\rm loop ~~function}$, modulo the wave function factors which will describe how the $\bar s$ and $d$ quarks are held together to form a $\bar K_{S}^{0}$ . One then gets 
\begin{equation}
{\cal M}_{\mu \mu}^{loop} (K^{0} \rightarrow \mu^{+} \mu^{-})
\propto \frac{g_{2}^{2}}{M_{W}^{4}} \cos\theta_{c} \sin\theta_{c} ~~g_{2}^{2} \times M_{W}^{2} (1 + {\cal O} (m_{u}^{2}/M_{W}^{2})
\label{eq:3quark}
\end{equation}
The factors of $\sin \theta_{c}, \cos \theta_{c}$ that appear at various vertices in these diagrams are a reflection of Cabibo mixing. In the limit where all the masses can be neglected, the loop function can only involve $M_{W}^{2}$, which is what explicit computations will yield. The $M_{W}^{4}$ in the denominator comes from the $W$-propagators. Remembering the relation between $G_{\mu}$ and $M_{W}^{2}$ (Eq.~\ref{eq:Gmumw}), we then find that the amplitude can be written as:
\begin{equation}
{\cal M}_{\mu \mu}^{loop} ( K^{0} \rightarrow \mu^{+} \mu^{-} ) \sim G_{\mu}^{2} \cos \theta_{c} \sin \theta_{c} M_{W}^{2}.
\label{eq:loop-leading}
\end{equation}
\begin{figure}[hbt]
\begin{center}
\includegraphics*[width = 6cm]{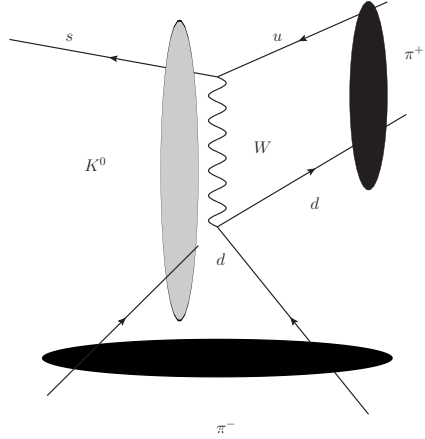}
\caption{One loop diagrams giving rise to $K_{S}^{0} \rightarrow \pi^{0} \pi^{0}$}
\label{fig:k0pipi}
\end{center}
\end{figure}
Let us compare then  the order of magnitude for this amplitude with the one expected for the non leptonic weak decay  $K_{S}^{0} \rightarrow \pi^{0} \pi^{0}$. The latter takes place not through a loop diagram but via the weak charged current at tree level and  occurs at ${\cal O} (G_{\mu})$. A possible digram is shown in Fig.~\ref{fig:k0pipi}.
Amplitude for this decay will be proportional to $G_{\mu} \sin \theta_{c} \cos \theta_{c}$, modulo the aforementioned wave function factors describing $q \bar q'$  bound state. If it were not for the factor of $M_{W}^{2}$ ($ M_{W} > 37.3 \UGeV^{2})$,~the additional factor of $G_{\mu}$ present in the loop amplitude of Eq.~\ref{eq:loop-leading}, could have suppressed the ${\cal M}_{\mu \mu}^{loop}$ by a factor $10^{-5}$ compared to the charged current induced, tree level amplitude for $K^{0} \rightarrow \pi^{0} \pi^{0}$. Thus the rate for the $\mu^{+} \mu^{-}$ decay  could have been suppressed to the experimentally observed low level as compared to the $\pi^{0} \pi^{0}$ decay.  However, the factor $M_{W}^{2}$ removes this suppression of the ${\cal M}_{\mu \mu}^{loop} ( K^{0} \rightarrow \mu^{+} \mu^{-} )$. As a result, in the three quark picture,  the amplitude for the $\mu^{+}\mu^{-}$ decay is suppressed though not {\it hugely} compared to the $\pi^{0} \pi^{0}$ decay which in turn  occurs at the usual weak rate. This then is in contradiction with the  experimentally observed branching ratio of about $\sim 31 \%$ for the $\pi^{0} \pi^{0}$ final state and the  observed upper limit on the branching ratio for the $\mu^{+} \mu^{-}$ channel of $10^{-9}$.

When one adds to the loop amplitude of Eq.~\ref{eq:3quark} the contribution coming from $c$ loop as well, something interesting happens. Due to the relative negative sign of the term containing $\sin \theta_{c}$, we note that the amplitudes from the two box diagrams in the left and right panel of Fig.~\ref{fig:kmu+mu-}, will cancel each other exactly in the case where the masses of the $u$ and $c$ quarks are equal. The large term independent of the mass of the quark in the loop thus cancels between these two diagrams! The non leading terms dependent on the mass of the quark in the loop, will give zero when $m_{u} = m_{c}$ and will be proportional to $m_{c}^{2} - m_{u}^{2}$. So the factor with mass dimension two, in the amplitude ${\cal M}_{\mu\mu}$ is no longer the large $M_{W}^{2}$, but $m_{c}^{2} - m_{u}^{2} \sim m_{c}^{2}$.
Thus, in the four quark picture, the observed suppression happens due to the very existence of the charm quark and is guaranteed here by the {\it orthogonality} of the quark mixing matrix.  Further, any deviation from zero for  the  branching ratio will then depend on the difference in the masses of the quarks being exchanged in the loops and in fact can give {\it indirect} information on these, in the framework of a gauge theory when the various parameter values
$g_{1},g_{2},v$  and mixing angles are known. However, particularly in the case of $K^{0} \rightarrow \mu^{+} \mu^{-}$  no firm constraint on the charm mass can be drawn due to the existence of additional contributions to this process which do not come from the weak charged current interactions along with  some accidental cancellations.

A similar suppression of FCNC is also observed experimentally in the the $K^{0}$-$\bar K^{0}$ mixing which is a $\Delta S = 2$ transition. In principle, this could occur at higher order in the CC weak interactions which are strangeness changing with $\Delta S =1$. The $K_{L}$--$K_{S}$ mass difference is $\Delta m_{K} = |m_{K_{L}} - m_{K_{S}}| = (3.484 \pm 0.006) \times 10^{-12}$ MeV, with  $\frac{\Delta m_{K}}{m_{K^{0}}} \simeq 8.5 \times 10^{-15}$. Recall here that the strength of weak interactions is given by $G_{\mu} \sim \frac{1.01 \times 10^{-5}}{m_{p}^{2}}$.  The strength of the $\Delta S =2$ transition which causes the $K^{0}$--$\bar K^{0}$ oscillations and gives rise to the $K_{L}$--$K_{S}$ mass difference, is thus clearly weaker than that expected from just two insertions of the CC weak interaction and is thus suppressed perhaps even further. In the early days of gauge theory it was not clear whether the $K_{0}$--$\bar K_{0}$ mixing is caused by a new interaction {\it weaker} than the weak  or whether it can be understood as a higher order effect of the  $|\Delta S| =1$ weak charged current interaction. 

\begin{figure}[hbt]
\begin{center}
\includegraphics*[width = 6 cm,height=5cm]{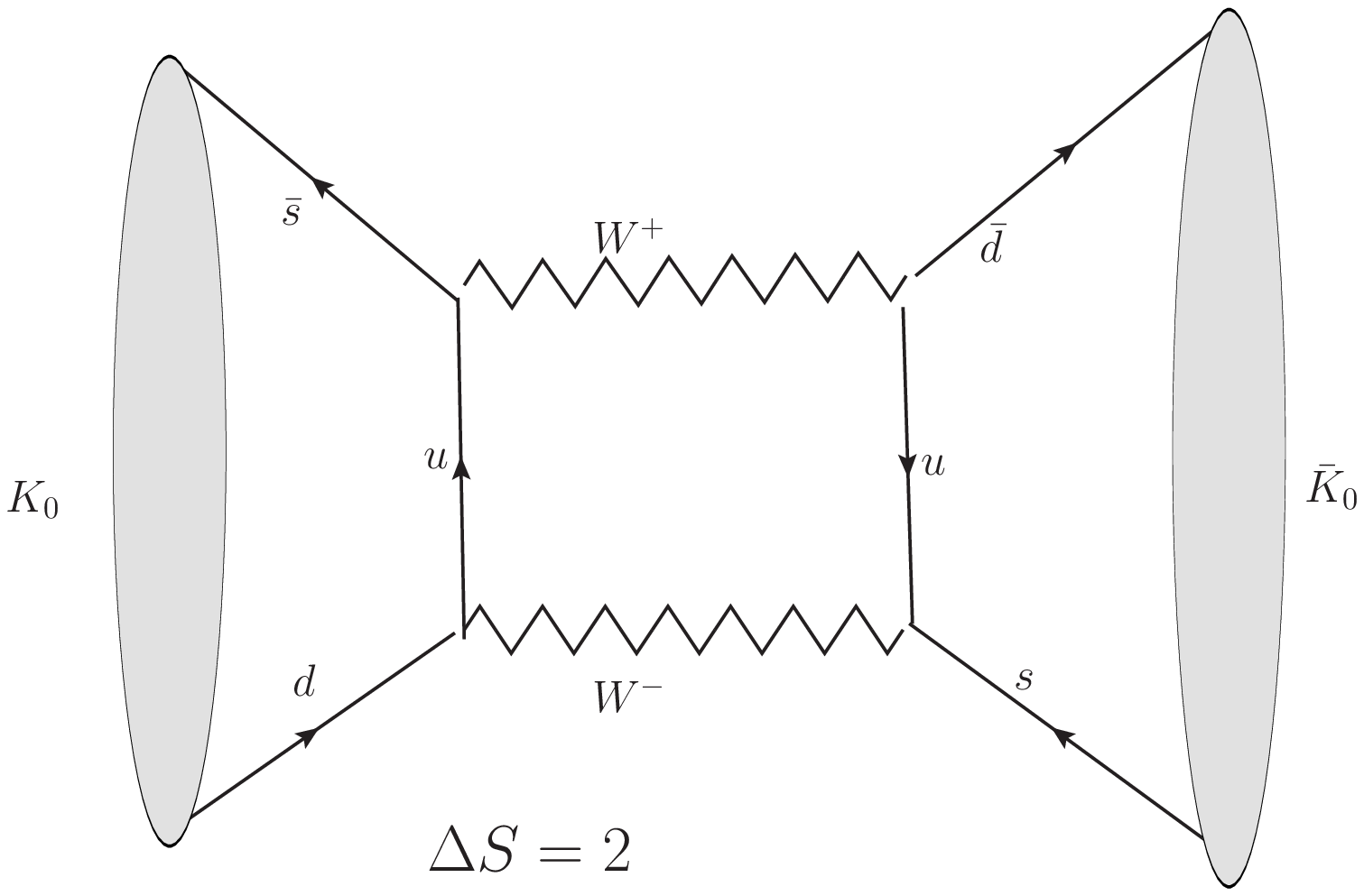} 
\hspace{1cm}
\includegraphics*[width = 6 cm, height=5cm]{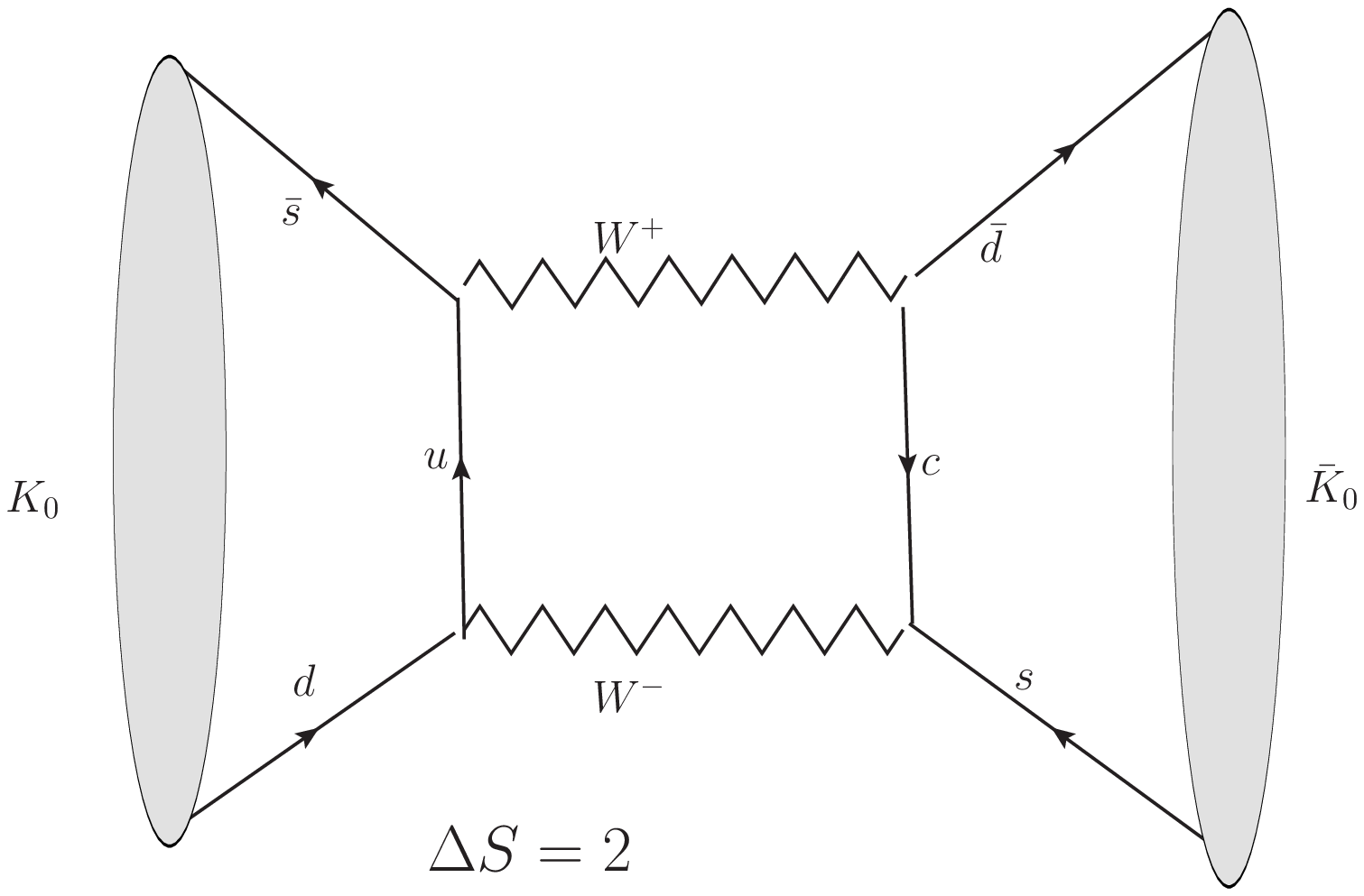}
\caption{One loop diagrams giving rise to $K^{0}-\bar K^{0}$ mixing.}
\label{fig:koko-massdiff}
\end{center}
\end{figure}

In a gauge theory one can compute the expected value of this mixing  in terms of loop diagrams very similar to those shown in Fig.~\ref{fig:kmu+mu-}, where at the right hand end of the box the $\nu_{\mu}$ is replaced by a $u$ or $c$-quark line and the $\mu^{+}, \mu^{-}$ lines are replaced by the $\bar d$ and  $s$ quark line which are bound in a $\bar K^{0}$ meson. Again, we show only  two of these diagrams  contributing to it and that too in the 4-quark picture,  in Fig.~\ref{fig:koko-massdiff}.
Again one can make very similar observations as before. 
If the model had only three quarks $u,d,s$ then only the digram involving the $u$ quarks would have contributed and it is very clear that the predicted $K_{0}$--$\bar K_{0}$ mass difference will not be proportional to $G_{\mu}^{2}$ in the limit that $u,d,s$ quark masses are much smaller than $M_{W}$. As a result this contribution would have been much bigger than the experimental measurement mentioned above. On the other hand, in the four quark picture, if the masses of $u$ and $c$ quarks were equal the contribution from the two diagrams will just cancel each other due to the factors of $\cos \theta_{c}$ and $\sin \theta_{c}$ appearing with appropriate signs and will be zero in this limit of $m_{c} = m_{u}$.  Further, the actual value of the predicted mass difference will now depend on $m_{c},m_{u}$ as well as experimentally measured values of $G_{\mu}, \theta_{c}$ etc. The observed mass difference could then be interpreted as an upper limit on the mass difference $m_{c} - m_{u}$ and further as a limit on $m_{c}$ of about a few $\UGeV$ neglecting $m_{u}$. This is perhaps the first example of prediction of the 'scale' of new physics (in this case the charm quark) through virtual effects on quantities measured at energies much below the scale.

There are two parts to this calculation. One is evaluation of the transition amplitude indicated by the box diagram drawn involving the $W$'s and the quark lines, and the other is conversion of that amplitude into mass difference between the mesons. This requires evaluation of the matrix element between the meson states of the effective Lagrangian which in turn has been extracted from the transition amplitude at the quark level. One can relate the former to the meson wave function factor encoded in the  decay constant $f_{K}$ which in turn can be extracted from the  measured life times of the kaons. The loop calculation, yields  a result for the mass difference $\Delta M_{K} = |M_{K_{0}} - M_{\bar K_{0}}|$,
\begin{equation}
\frac {\Delta M_K}{M_K} = {\frac{2}{3}} \frac{G_\mu^2}{4 \pi^{2}} m_c^2 \cos^2 \theta_c \sin^2 \theta_c  f_K^2
\label{eq:dmk4qrk}
\end{equation}
In principle, the large mass of the $t$ quark means that this could change substantially in the six-quark picture. A calculation of the mass difference in the six-quark case can be shown to be
\begin{equation}
\frac {\Delta M_K}{M_K} = \frac{2}{3} \frac{G_\mu^2}{4 \pi^{2}} m_c^2 \cos^2 \theta_c \sin^2 \theta_c  f_K^2 X
\label{eq:dm6qrk}
\end{equation}
with
\begin{equation}
X = (\sin^{2}\theta_{c} \cos^{2} \theta_{c})^{-1} \Re e \left[(V_{cs} V_{cd}^{*})^{2} + \frac{m_{t}^{2}}{m_{c}^{2}} (V_{ts} V_{td}^{*})^{2} + V_{cs} V_{cd}^{*} V_{ts} V_{td}^{*} \frac{2 m_{c}^{2}}{m_{t}^{2} - m_{c}^{2}} \ln \left(\frac{m_{c}^{2}}{m_{t}^{2}}\right) \right] 
\label{eq:dm6qrkX}
\end{equation}
For the four-quark case the CKM matrix is just a $2 \times 2$ matrix and hence Eq.~\ref{eq:dm6qrk} just reduces to Eq.~\ref{eq:dmk4qrk}. For the six quark case, indeed $X$ in Eq.~\ref{eq:dm6qrkX} contains terms $\propto m_{t}^{2}$. These terms
can, in principle, dominate the mass difference $\Delta M_{K}$.
However, since the elements of the CKM matrix which  connect the third generation with the first and the second generation, $V_{td}, V_{ts}$,  are extremely small, the dominant contribution
to $\frac {\Delta M_K}{M_K}$  is still given by Eq.~\ref{eq:dmk4qrk}. 

In fact, even without calculating the loop one could try to estimate the size of expected value of $\Delta m_{K}$ assuming that the $\Delta S =2$ transitions are caused by an interaction with strength proportional to $ G_{K}^{2} = G_{\mu}^{2} \sin^{2} \theta_{c}$. Since $G_{\mu}$ has mass dimension $- 2$, we need to add appropriate factors of the only mass available at the meson level, viz. $m_{K}$. Thus the expected mass difference is 
\begin{equation}
{\frac{\Delta M_{K}}{M_{K}}} = G_{\mu}^{2} \times m_{K}^{4} = (1.01 \times 10^{-5})^{2} \times \left({m_{K} \over m_{p}}\right)^{4} \sin^{2} \theta_{C}  \simeq {\cal O} (10^{-14}) 
\end{equation}
which is indeed the right order of magnitude. This thus means that 
this amplitude must be $\propto G_{\mu}^{2} \sin^2{\theta_{c}}$ and can NOT be $\sim {\cal O} (G_{\mu})$. 

Thus one sees that the suppression of FCNC that has been observed experimentally is 'understood' neatly, both at the tree and loop level in a gauge theory, in terms of the chosen particle spectrum of the SM. At the tree level case it is just guaranteed by the representation of the group to which quarks of a given electromagnetic charge and handedness belong where as at the loop level it is the orthogonality of the mixing matrix. I.e,  the mere presence of charm quark in the spectrum is sufficient to achieve both. The latter observation is the celebrated GIM mechanism~\cite{bib:GIM}. In the six quark case, it is not the orthogonality of the mixing matrix but the Unitarity of ${\bf V}$ matrix that guarantees the GIM cancellation.  Further, the actual observed suppression can give a hint about the masses of the quarks involved. In fact, the first 'prediction'~\cite{bib:QGL} for the charm mass around a scale $\LTS$ a few $\UGeV$ was made, using the GIM idea by comparing the observed $\Delta M$, with the one calculated theoretically. The uncertainties in the upper limit were mainly due to the gaps in the theoretical understanding of strong interactions at the time.  As explained above, while in principle this 'prediction' could have had 'large' corrections, for the values of the mixing matrix elements realised in nature, the prediction was correct.

\subsubsection{Anomaly cancellation}
As we have seen above, the GSW model contains both the vector and the axial vector currents. This causes a problem when we try to renormalise the theory and do loop computations.  The gauge invariance of axial vector currents of the type
\[
J_{\mu}^{5} = \bar \psi \gamma_{\mu} \gamma_{5} \psi', 
\]
($\psi'=\psi$ for neutral currents) is not preserved by dimensional regularization due to the presence of $\gamma_{5}$ in the current. This means that even though, 
\[
\partial_{\mu} J^{\mu}_{5} = 0
\] 
classically, at loop level due to the non invariance of the regulator,$\partial_{\mu} J^{\mu}_{5} \ne 0$ and the RHS develops a nonzero term on the RHS. Hence, this axial gauge current is no longer conserved. The current is said to be `anomalous'. As we know from Noether's theorem if the current is not conserved, it means gauge invariance is broken. Gauge symmetry along with Higgs mechanism is needed to have a consistent quantum theory with massive gauge bosons. Thus if the theory has an anomalous current (or has anomaly) the theory may not make sense at quantum level. It was shown by Adler and Bell-Jakciw, that there is only one type of loop diagram with a logarithmic divergence which  can make $\partial_{\mu} J^{\mu}_{5}$ non- vanishing and poses a danger to the conservation of the axial gauge current. This is a triangle diagram with a fermion loop and two gauge boson legs and one current insertion;  equivalently one can also consider a fermion loop with three gauge boson legs.  In the GSW model with its $SU(2)_{L}$ gauge bosons which have couplings only to left chiral fermions and the $U(1)_{Y}$ gauge bosons which have unequal couplings to the left and right chiral fermions, these triangle diagrams are in general not zero. Further, one can show that the anomalous contribution is independent of the mass of the fermions in the internal loop. 

\begin{figure}[hbt]
\begin{center}
\includegraphics*[width=5cm,height=4cm]{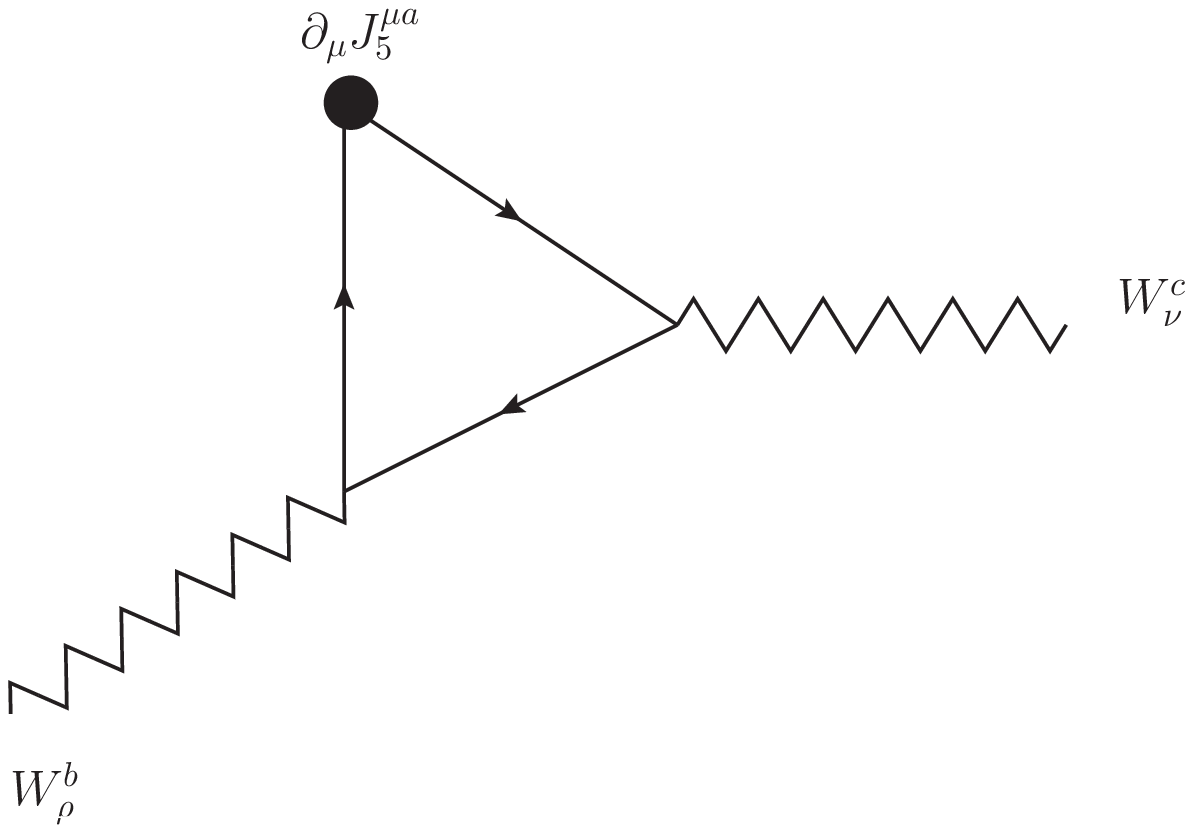}
\includegraphics*[width=5cm,height=4cm]{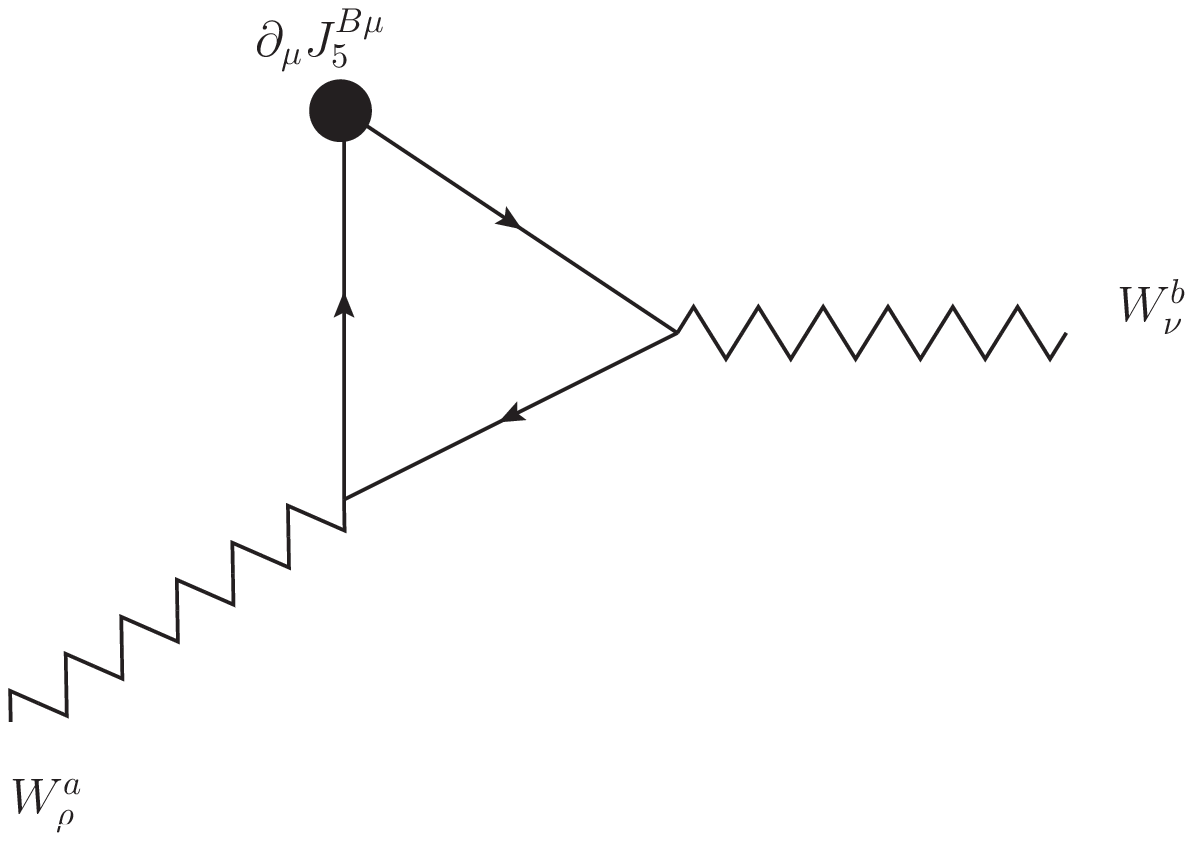}
\includegraphics*[width=5cm,height=4cm]{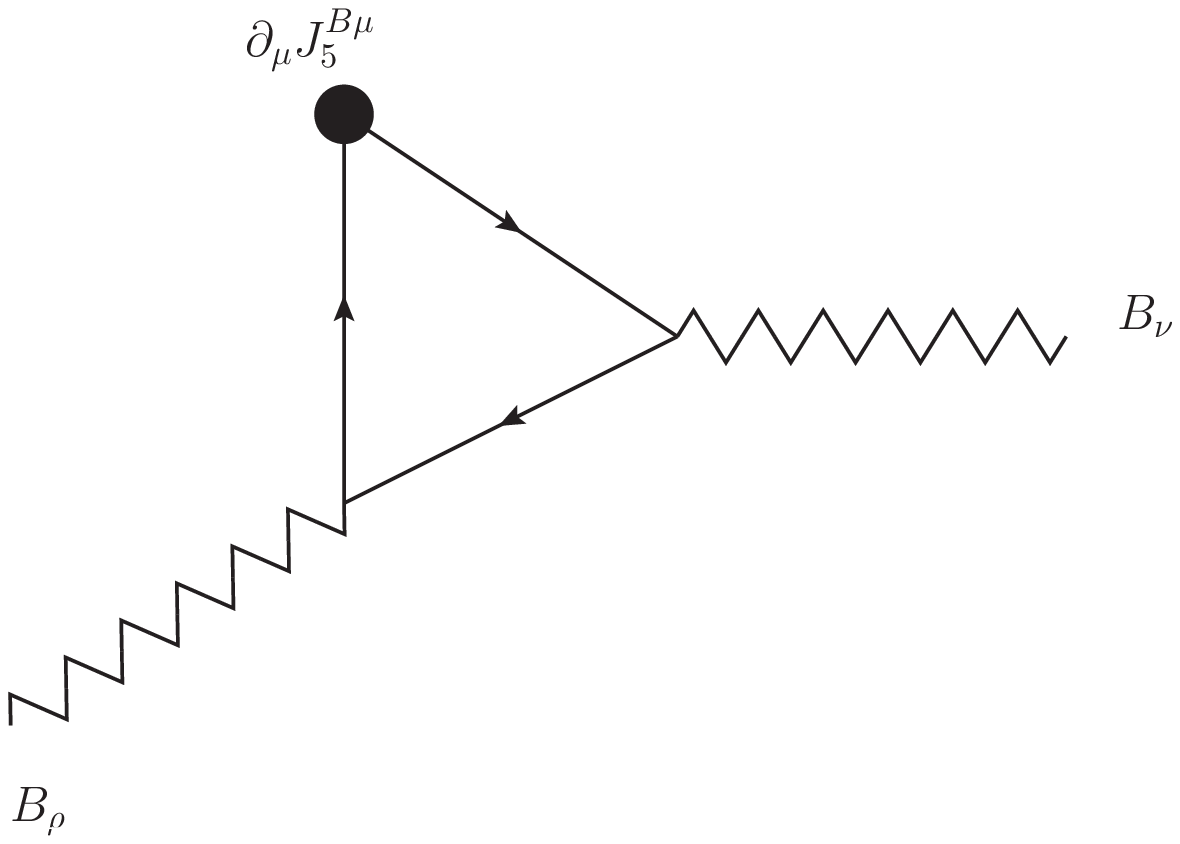}
\caption{Triangle diagrams with anomalies.}
\label{fig:anomalies}
\end{center}
\end{figure}

There are in fact four types of triangle diagrams we need to consider out of which three are shown in Fig.~\ref{fig:anomalies}.
Consider the  diagram in the left most panel which contains matrix element of a pure $V-A$ current insertion along with two $SU(2)_{L}$ gauge boson legs. Only left handed fermions contribute to this anomaly and it can be shown that
\begin{equation}
\partial_{\mu} J^{\mu a}_{5} \sim tr \tau^{a} \{\tau^{b}, \tau^{c}\} \epsilon^{\alpha \rho \beta \nu} F_{\alpha \rho}^{b} F_{\beta \nu}^{c} .
\end{equation}
Here the 'tr' refers to the trace over representation matrices and indicates the sum over all the fermions in the representation. Since $\{\tau^{b},\tau^{c}\} = 2 \delta^{bc}$ and $\tau^{a}$ are traceless matrices this anomaly is zero identically. In fact, the diagram with just  one $SU(2)_{L}$ V--A current insertion not shown here will also give zero contribution  to the anomaly due to the traceless property of $\tau^{a}, a =1,3$ matrices. The central diagram also gets contribution only from the left chiral fermions and is given by
\begin{equation}
\partial_{\mu} J^{\mu}_{5} \sim tr (Y_{L}) \epsilon^{\alpha \rho \beta \nu} F_{\alpha \rho}^{a} F_{\beta \nu}^{a} 
\label{eq:su2-u1-anomaly}
\end{equation}
The notation $tr (Y_{L})$ indicates that only the left chiral fermions contribute to this quantity and sum is to be taken over one $SU(2)_{L}$ representation. The contribution of the rightmost diagram in Fig.~\ref{fig:anomalies} is given by 
\begin{equation}
\partial_{\mu} J^{\mu}_{5} \sim tr \left(Y_L^{3} - Y_{R}^{3}\right)  \epsilon^{\alpha \rho \beta \nu} B_{\alpha \rho} B_{\beta \nu}.
\label{eq:u1-anomaly}
\end{equation}
We see that for a single lepton generation the anomaly of Eq.~\ref{eq:su2-u1-anomaly} is proportional to $ 2 \times Y_{L} = -2$. 
Summing over all the lepton doublets it will have a value $-6$.
However, one notices that, for a single quark generation it is 
$2 \times {1/3}$. The three colours add another factor of 3. Thus we find,
\[ 
tr (Y_{L})|_{l} + tr (Y_{L})|_{q} = - 2 + 3 \times 2 \times 1/3 = 0.
\]
Thus this anomaly vanishes identically for the particle content of the left chiral fermions in the GSW model. Further, we also notice that while $(2 Y_{L}^{l})^{3} - (Y_{R}^{e})^{3} = -2 + 8 = 6$ is not zero, it is again compensated by the value for the quark doublets which is $3 \times (-2/27 - \left(\frac{4}{3} \right)^{3} + \left(\frac{2}{3}\right)^{3} ) = -6 $. Thus again
\[
tr \left(Y_L^{3} - Y_{R}^{3}\right) |_{l} + tr \left(Y_L^{3} - Y_{R}^{3}\right) _{q} = 6 -6 = 0 .
\] 
Hence contributions to both the anomalies, from loops of fermions of one quark and one lepton doublet of the GSW model, are equal and opposite in sign. This means that the numbers of the lepton and quark doublets have to be exactly equal so that the anomalies do not spoil the gauge invariance of the GSW model and hence the renormalisability. 

\subsubsection{Custodial Symmetry}
\label{sec:custodial}
Let us discuss further the $\rho$ parameter. To that end let us understand in a little more detail the origin of the prediction of unity for $\rho$ defined Eq.~\ref{eq:rhodef}. Let us begin first writing down the most general gauge boson mass terms  that one could generate by spontaneous symmetry breaking. In the $W_{\mu}^{a}, a=1,3$ and $B_{\mu}$ basis this can be written as 
\begin{equation}
\left( 
\begin{array}{c c c c}
M_{W^{1}}^{2}&0&0&0\\
0&M_{W^{2}}^{2}&0&0\\
0&0&M_{W^{3}}^{2}&M_{W B}\\
0&0&M_{W B}&M_{B}^{2}\\
\end{array}
\right)
\label{eq:GB-massmatrix}
\end{equation}
The mass terms  $M_{W^a}, a=1-3, M_B$ and $M_{W^3 B} $ arise from the covariant derivative term $D_{\mu} \Phi^\dag D^\mu \Phi$ (cf. Eq.~\ref{eq:covarSM}), after the field $\Phi$ acquires a non zero vev.
The expressions for  $M_{W^{3} B}$, $M_{W^{a}}, a=1,3$ and $M_{B}$
that one would get as a result by expanding the field around the minimum, will depend on the weak isospin charges $T,T_{3L}$ of the field $\Phi$. Demanding that the EW minimum conserves electromagnetic charge, as it must because $SU(2)_{L} \times U(1)_{Y}$ breaks to $U(1)_{em}$ after $\Phi$ acquires the nonzero vev, implies that the $T_{3L}$ value of field which acquires the nonzero vev will be given by $ Q = 0 = T_{3L} + Y/2$. While various entries in this mass matrix will then depend on the isospin and the hyper charge of the $\Phi$, conservation of the electromagnetic charge will mean that the mass matrix will have a block diagonal form. The same  also implies $m_{W^{1}}^{2} = M_{W^{2}}^{2} = M_{W}^{2}$ where $M_{W}$ is the mass of the $W^{\pm}$ boson, defined via the last of equations in  Eq.~\ref{eq:currents}.
The $W^{3}_{\mu}$ and $B_{\mu}$ will mix. Irrespective of the representation to which the scalar $\Phi$ belongs we are interested in the symmetry breaking patterns where $SU(2)_{L} \times U(1)_{Y}$ breaks to $U(1)_{em}$ on , on $\Phi$ achieving a nonzero vev. Hence one of the eigenvalue of the $2 \times 2$ block diagonal matrix aught to be 0. The value of $M_{Z}$ as well as the $\rho$ parameter will thus depend on the representation of $\Phi$.  In fact, it is possible to  write a general expression for $\rho$. 

For the present, let us continue with this general form of the matrix without committing to a representation for $\Phi$.
Again defining $Z_{\mu}, A_{\mu}$ as in Eq.~\ref{weakmixing},
to be the eigenstates of the above block diagonal mass matrix, it is easy to see 
\begin{eqnarray}
M^{2}_{\gamma} &= M^{2}_{W^{3}} \cos^{2}\theta_{W} + M^{2}_{B} \sin^{2} \theta_{W} + 2 M^{2}_{WB} \sin \theta_{W} \cos \theta_{W}
=0\nonumber \\
M^{2}_{Z} &= M^{2}_{W^{3}} \cos^{2}\theta_{W} + M^{2}_{B} \sin^{2} \theta_{W} - 2 M^{2}_{WB} \sin \theta_{W} \cos \theta_{W}\nonumber\\
0  &= (M^{2}_{W^{3}} - M^{2}_{B}) \sin \theta_{W} \cos \theta_{W} + M^{2}_{BW} (\cos^{2}\theta_{W} - \sin^{2}\theta_{W})\nonumber \\
\label{generalWB}
\end{eqnarray}
This also means 
$M^{2}_{B} + M^{2}_{W^{3}} = M^{2}_{Z} + M^{2}_{\gamma} = M^{2}_{Z}$, as it should be since the trace of a matrix is equal to sum of the eigenvalues. Thus we can eliminate $M^{2}_{B}$ in favor of $M^{2}_{Z}$. Using Eq.~\ref{generalWB}, we can easily see that
\begin{equation}
-M^{2}_{WB} = \frac{M^{2}_{W^{3}} (\sin^{2}\theta_{W}- \cos^{2} \theta_{W}) + M^{2}_{Z} \cos^{2}\theta_{W}} {2 \sin \theta_{W} \cos \theta_{W}}  = \frac{(2 M^{2}_{W^{3}} - M^{2}_{Z})\sin \theta_{W} \cos \theta_{W}}
{\cos^{2}\theta_{W} - \sin^{2}\theta_{W}}
\label{WBcrossterm}
\end{equation}
Thus $\cos \theta_{W}$ can be expressed in terms of $M^{2}_{W^3}$ and $M^{2}_{Z}$. On comparing Eq.~\ref{eq:WZmassterm} with Eq.~\ref{eq:GB-massmatrix}, we see that for the case of the Higgs doublet we would have 
\begin{eqnarray}
M_{W^{1}}^{2} = M_{W^{2}}^{2} = &M_{W^{3}}^{2} = \frac{g_2^2 v^2}{4}, \qquad M_{B}^{2} = \frac{g_1^{2} v^2}{4}, \nonumber\\
M_{W B} = -\frac{g_1 g_2 v^2}{4}
\label{HDWB}
\end{eqnarray}
Using Eq.~\ref{WBcrossterm}, we then get $M_{W} = M_{Z}  \cos \theta_{W}$, precisely  the result of Eq.~\ref{eq:WZmasses}. Thus, we see that the $\rho = 1$ prediction is tied to the equality of $M^{2}_{{W}^{a}}, a=1,3$ terms in Eq.~\ref{eq:GB-massmatrix}. 

In fact a closer inspection of the scalar potential of Eq.~\ref{eq:Lscalar} reveals that this equality of all $m^{2}_{W^{a}}$ is in fact due to an accidental symmetry of the scalar potential for doublet $\Phi$. The doublet $\Phi$ contains, in all, four real fields as $\phi^{+}, \phi^{0}$ are both complex fields. Writing,
\begin{equation}
\Phi = \left( \begin{array}{c} 
\Re e \phi^{+} \\ 
\Im m \phi^{+} \\
\Im m \phi^{0} \\
\Re e \phi^{0}\\
\end{array}
\right)
\label{eq:phicomp}
\end{equation}
we can see that the scalar potential 
\begin{eqnarray}
V ({\Phi}) = - \mu^{2} \left[(\Re e \phi^{+})^{2} + (\Im m \phi^{+})^2 + (\Re e \phi^{0})^{2} + (\Im m \phi^{0})^{2}\right]
    \nonumber\\
    + \lambda \left[(\Re e \phi^{+})^{2} + (\Im m \phi^{+})^2 + (\Re e \phi^{0})^{2} + (\Im m \phi^{0})^{2}\right]^{2}
\end{eqnarray}
has an $O(4)$ symmetry under a rotation of  the vector $\Phi$ of Eq.~\ref{eq:phicomp}. 

Upon SSB, the lowermost component of $\Phi$ acquires a non zero vev $\frac{v}{\sqrt{2}}$, whereas all the three components have zero vev..  Hence the scalar potential loses this $O(4)$ symmetry. However, there is still a left over  $O(3)$ symmetry corresponding to rotations of the first three components of $\Phi$. among each other. It is this left over $O(3)$ symmetry, called the Custodial Symmetry,  which reflects itself in the equality of the masses $M^{2}_{W^{a}}$ for $a=1,3$ in the matrix Eq.~\ref{eq:GB-massmatrix}, yielding $\rho = 1$.

This also means that even though in the original formulation we had discussed the case of just a single Higgs doublet $\Phi$ being involved in the SSB, as long as we use only doublet fields, Eq.~\ref{eq:rhodef} is always guaranteed. Of course the statement is true only at the {\it tree level}. The custodial symmetry,  is isomorphic to 
an $SU(2)$ involving the $W^{a}$. This $SU(2)$ is broken by the different masses of the fermions of a $SU(2)_{L}$ doublet. The value of $\rho$ can change due to contributions coming from loops (as we will discuss in the next section) and also if there exist Higgs belonging to a representation of $SU(2)_{L}$ other than the doublet.

\subsubsection{High energy scattering}
\label{sec:HE}
Recall the discussion around Eq.~\ref{eq:fermics}.  
We saw there how the postulate of {\it massive} vector boson was inspired by the demand to restore unitarity to the $\nu$ induced processes.  For example, the  amplitude (say) for $\nu e \rightarrow \nu e$ scattering calculated in Fermi theory (current-current interactions) violates tree level unitarity for
$\sqrt{s} \LTS 300 \sim G_{\mu}^{-1/2}  \UGeV$. Hence,  one could also take this value as an upper bound  on the mass of the 'massive' $W$ boson.
 
However, theories with massive vector bosons  have problems with gauge invariance and hence renormalisability. The SSB via Higgs mechanism solved the problem by generating these masses in a gauge invariant manner. This then meant that the theory has renormalisability even with massive gauge bosons.  In fact, as we will discuss below, we can see explicitly that gauge invariance also renders nice high energy behaviour to all the scattering amplitudes of the EW theory. 

\begin{figure}[tbh]
\begin{center}
\includegraphics*[width=10cm,height=5cm]{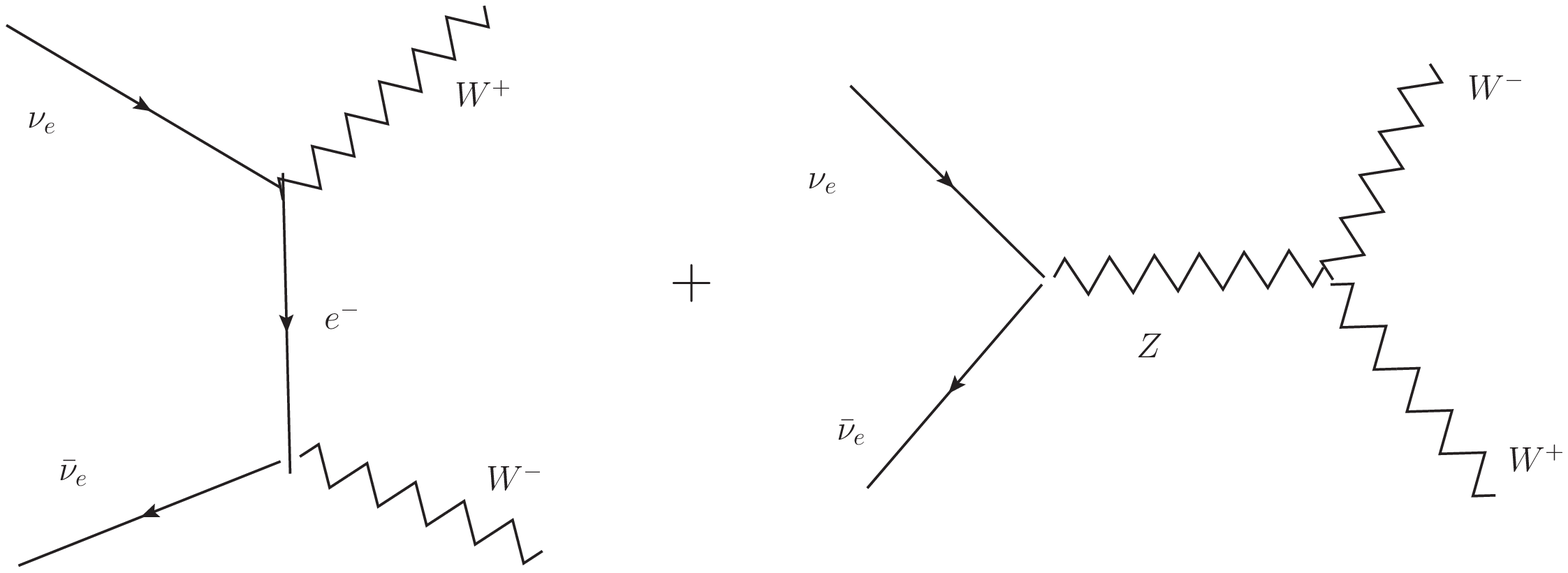}
\caption{Gauge theory restoration of tree level unitarity to the $\nu_{e} \bar \nu_{e} \rightarrow W^{+} W^{-}$ process.}
\label{fig:uniscatter}
\end{center}
\end{figure}

The existence of massive vector gauge bosons restore unitary behavior to processes like (say) $\nu_{\mu} + e^{-} \rightarrow \mu^{-} + \nu_{e}$. But now due to the same non zero mass of the $W$ bosons, amplitudes for processes involving longitudinal $W$'s have a bad high energy behaviour.  For  example, the matrix element for the process  $\nu_e \bar \nu_e \rightarrow W^+W^-$ through a $t$-channel exchange of an $e$, shown in the left panel of Fig.~\ref{fig:uniscatter}, grows too fast  with energy and violates unitarity. One can show that
\begin{equation}
{\cal M} (\nu_{e} \bar \nu_{e} \rightarrow W^{+} W^{-}) ~~ \sim ~~8 \frac{g_{2}^{2}}{M_{W}^{2}} E p' \sin \theta,
\label{eq:wwtch}
\end{equation}
where $E$ is the energy of the incoming $\nu_{e}$ and  $p', \theta$ are the momentum and the angle of scattering of the $W$ boson in the final state. Here we write only the dominant term of the amplitude involving the longitudinal gauge bosons, which is the one with bad high energy behavior. If one does a partial wave analysis of this amplitude, one finds that this amplitude will violate partial wave unitarity, for
${\displaystyle  s ~~\LTS~~ \frac{M_{W}^{2}}{2 g_{2}^{2}}}$
However, what is interesting is that the contribution to the matrix element of the process  $\nu_{e} \bar \nu_{e} \rightarrow W^{+} W^{-}$, from the $s$ channel exchange of a $Z$ boson, shown in the right panel of Fig.~\ref{fig:uniscatter} has exactly the same magnitude as the $t$ channel contribution written above but  opposite in sign. This happens only if the strength and structure of the couplings of the $Z$ with a $\nu$ and $W$ pair is exactly the same as given by  the $SU(2)_{L} \times U(1)$ theory. Thus the violation of unitarity in the amplitude $\nu_{e} \bar \nu_{e} \rightarrow W^{+} W^{-}$ due to the longitudinal gauge boson scattering is cured in a gauge theory. 

In fact, the GSW model contains more such amplitudes which, in principle, could have had bad high energy behaviour but which are rendered safe by the particle content and the coupling structure of the SM. It was demonstrated~\cite{bib:bell} that in the GSW model where the masses are generated through SSB by a Higgs doublet (SM), {\it ALL} such amplitudes satisfy tree level unitarity. In fact the leading  divergence of the  ${\cal M} (W W \rightarrow W W)$  which goes like $s^{2}$ and hence is much worse, is also cured by the $Z$ exchange contribution and the contribution of the quartic coupling among the $W$ bosons which arise from the non abelian gauge invariance of the theory.  Further, the divergent term proportional to $s$ is cancelled by the contribution of the process $W^{+} W^{-} \rightarrow h \rightarrow W^{+} W^{-}$, where the Higgs boson is exchanged in the $s$-channel. Also if one were to calculate high energy behavior of the amplitude  $e^{+} e^{-} \rightarrow W^{+} W^{-}$ obtained by replacing the $\nu_{e}, \bar \nu_{e}$ in the initial state in Fig.~\ref{fig:uniscatter} by $e^{-},e^{+}$, then the same cancellation between the divergent parts of the $t$-channel and $s$-channel amplitudes is seen to take place.
\begin{figure}[hbt]
\begin{center}
\includegraphics*[width=10cm,height=5cm]{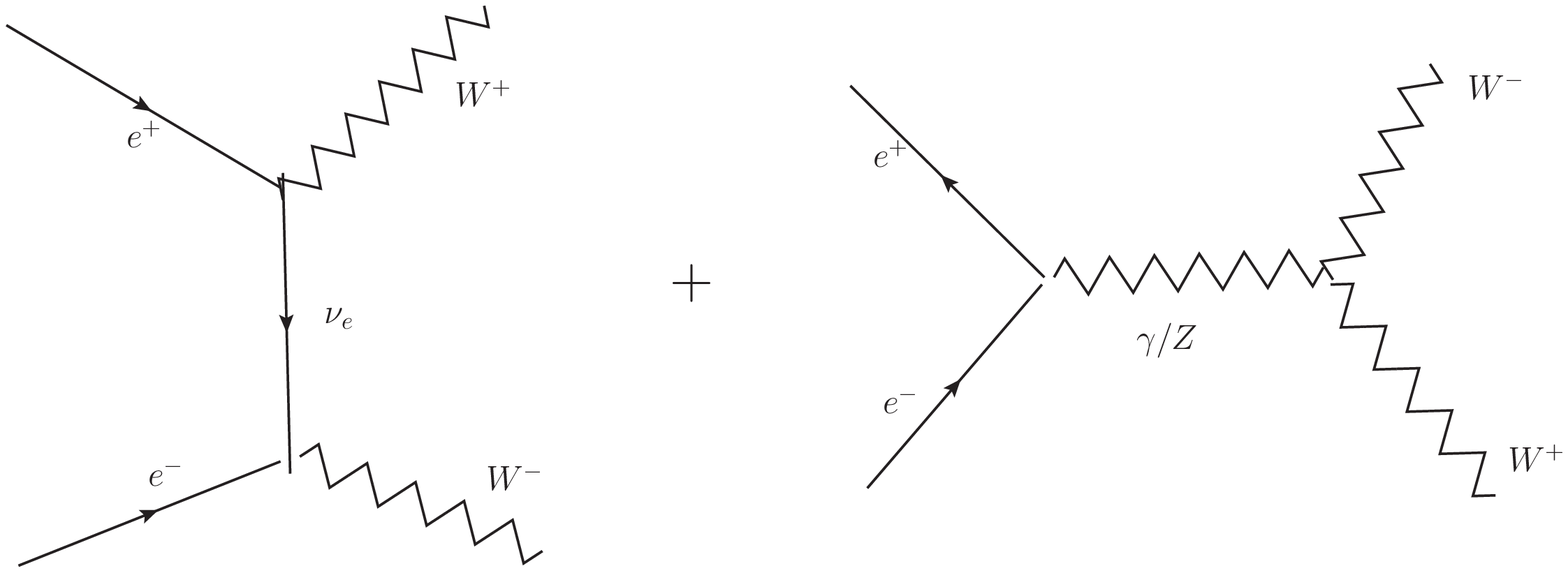}
\hspace{0.2cm}
\includegraphics*[width=5cm,height=5cm]{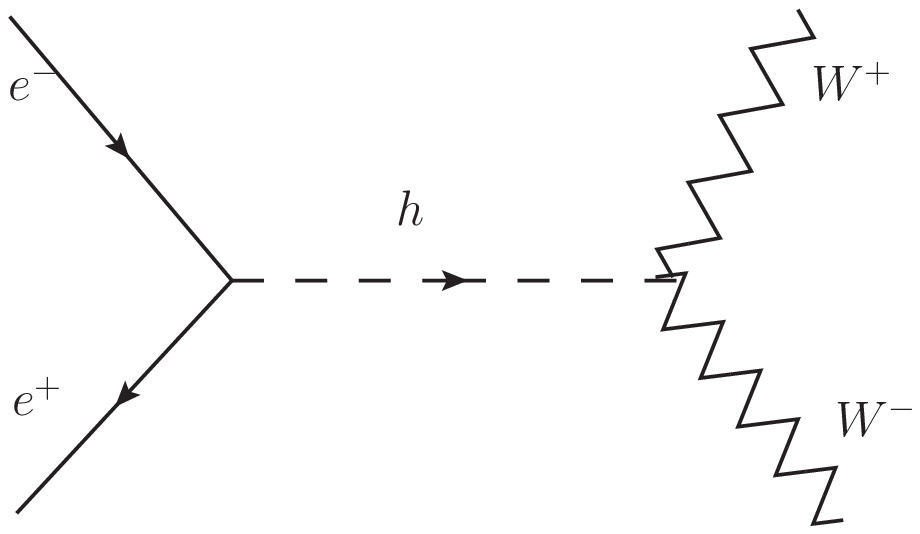}
\caption{$e^{+} e^{-} \rightarrow W^{+} W^{-}$ process in the SM}
\label{fig:eeWW}
\end{center}
\end{figure}

After this observation, a variety of authors~\cite{bib:unitary} investigated the conditions necessary for cancellation of these divergences so that the amplitudes will satisfy tree level unitarity. In fact their analysis indicated that this requires existence of partial wave contributions in the spin $1$ and spin $0$ channel, with the couplings of these particles exchanged in the $s$-channel to be precisely those that are given the SM. 
Recall here that this proportionality of the coupling of the Higgs to the masses of the particles to which it couples is the key prediction of the SSB by Higgs mechanism. The other couplings are of course given by the gauge invariance itself. Thus one could have derived the existence of the Higgs boson as well as the structure of the couplings of the fermions and the gauge bosons to it, without making any reference to the Higgs mechanism and hence the renormalisability. 

The fact that the two different requirements, unitarity and renormalisability, lead us to the same result, indicates that there must be a deep connection between the two. In fact, for the $\nu_{e} \bar \nu_{e} \rightarrow W^{+} W^{-}$ scattering, there is a residual logarithmic violation of unitarity that is left after all the cancellations, which gets cancelled
by the scale dependence of $g_{2}$ which is a loop effect which can be computed reliably only in a renormalisable theory.
     
\subsection{Predictions of GSW model}
Here we  summarize some of  the qualitative and quantitative implications of the $SU(2)_{L} \times U(1)_{Y}$ invariance.  Note that almost all of them are result of the invariance and hence not specific to the actual mechanism of symmetry breaking as long as it preserves the symmetry.
\begin{enumerate}
\item
First and foremost, this is a unification of weak and electromagnetic interaction: \ie   $e,g_1, g_2$ all are of similar order and the apparent difference in strengths of electromagnetic interactions ($\alpha_{\rm em}$ and $G_{\mu}$), is only caused by the large value of the masses of the weak gauge bosons compared to that of the massless photon. The model predicts existence of a new weak gauge boson $Z$ and that of the weak neutral current
(cf. Eq.~\ref{eq:jmuNCZ}) mediated by it, analogous to the weak charged current of Eq.\ref{eq:currents} mediated by the $W$. Further, the strength of this new weak interaction is similar to that of the charged current weak interaction. This is particularly transparent once we use the $\rho=1$ prediction  of the GSW model wherein $W/Z$ masses are generated by SSB using a Higgs doublet. 
\item Further,  FCNC currents are absent at tree level if and only if all the quarks of a given electrical charge belong to the same representation of $SU(2)_{L}$. Thus the experimentally observed absence of FCNC implied existence of the charm quark $c$, in addition to the already known $u,d$ and $s$ quarks. Not only this, one could also `predict' the mass of the $c$ quark from the measured $K_{0}$--$\bar K_{0}$ mass difference.
\item Since $G_{\mu}$ and the electron charge $e$ are measured experimentally, Eq.~\ref{eq:Gmumw} implies that the model has two free parameters, $\sin \theta_{W}$ and $M_{W}$. If $g_{2} = e$, \ie $\sin \theta_{W} =1$, then we get $M_{W} \sim {\cal O} (100) \UGeV$.  However, when the gauge boson masses are generated through the SSB, $M_{W}$ can be expressed in terms of $G_{\mu}, \alpha_{em}$ and $\sin^{2} \theta_{W}$. 
\item The model predicts precise nature of the $WWZ$ coupling, the strength being given by $g_{2}$. 
\item As Table~\ref{tab:04gagv} shows, couplings of all the fermions with the new gauge boson $Z$, are then determined in terms of $\sin\theta_{W}$ once the representations of the two gauge groups to which the fermions belong are specified.   
\item Requirement of anomaly cancellation, necessary for the renormalisability,  predicts that the number of lepton and quark generation seen in nature should be equal. So while the model can not predict how many families of quarks and leptons there should be, it predicts their equality.
\item The conditions of anomaly cancellation and observed  closeness of $\rho$ to unity, then gives strong constraints on new particles that one can be added to the spectrum of the GSW model.
\item As already stated above, generation of gauge bosons masses via SSB provides some more relations among physical quantities and hence reduces the number of free parameters of the model to one, that parameter being $\sin \theta_{W}$.
\end{enumerate}
Thus this model could be easily subjected to experimental tests. This is what we will discuss in the next sections.

\section{Validation and precision testing of the SM.}
\subsection{Early validation.}
Historically, the earliest validation of the correctness of the description of the electromagnetic weak interaction in terms of the EW theory, came from the points 1 and 2 in the list given at the end of the last section. By 1972, the renormalisability of the GSW model was proved explicitly~\cite{bib:'thooft} and the discovery of weak neutral currents had become very urgent. As we have already noted from Table~\ref{tab:04gagv}, the NC couplings are entirely decided by the (anti)fermion charge and $\sin \theta_{W}$.  Neutrino scattering with nuclei offer possibility of studying neutral current interactions of quarks. These typically have higher event rates compared to the pure leptonic scattering processes due to the possibility of using nuclear targets.  However, analysis of these processes requires an understanding and knowledge of the proton structure. Hence the cleanest probe of the neutral current  couplings can come from analysing  pure leptonic reactions. We will discuss both of these below. 

\subsubsection{Discovery of  the Weak Neutral Current.}
To study the properties of the weak neutral current it was necessary first to establish its existence. To that end, it was necessary to predict the characteristics of the events that would result from interactions of $\nu_{\mu}, \bar \nu_{\mu}$ and $\bar \nu_{e}$ beams with electrons, as that would be the cleanest probe.   Let us list different types of elastic scattering processes involving just leptons that can take place through weak charged current and neutral current interactions using the $\nu$ beams and the electron targets. These are 
\begin{enumerate}
\item $\nu_{\mu} + e^{-} \rightarrow \nu_{e} + \mu^{-}$, which can take place only through the CC interaction
\item $\nu_{\mu} + e^{-} \rightarrow \nu_{\mu} + e^{-}$ and  $\bar \nu_{\mu} + e^{-} \rightarrow \bar \nu_{\mu} + e^{-}$, which can take place only through the NC interaction,
\item $\nu_{e} + e^{-} \rightarrow \nu_{e} + e^{-};~~~\bar\nu_{e} + e^{-} \rightarrow \bar \nu_{e} e^{-}$, which can take place both through the NC and CC interactions. 
\end{enumerate}
Calculation of the scattering amplitudes of various NC and CC processes listed above (which are depicted in Fig.~\ref{fig:CCNC}, with appropriate assignments for $f_{i}, i =1,4$)~proceeds using the usual rules of field theory. For the low energies of $\nu$- beams that were available then, the $M_{W}, M_{Z} \rightarrow \infty$ approximation could be used. In situations where both the weak currents (charged and neutral) contribute to a process, the derivation of the effective four fermion interaction in the above limit is a little more involved than our derivations of Eq.~\ref{eq:LNC}, but finally leads to very compact expressions very similar to Eq.~\ref{eq:LNC}. For the $e^{-} \nu_{e}$ scattering mentioned above, for example, the expression resulting from the manipulations is the same as obtained by replacing $g_{A}^{e},g_{V}^{e}$ in Eq.~\ref{eq:LNC} by $g_{V}^{e} +1, g_{A}^{e} +1$. Here, we have used $\rho=1$ prediction of the SM.   

Table~\ref{tab:nccccsec} shows the differential cross-section in terms of the variable {$\displaystyle y = \frac{E_{e}}{E_{\nu}}$} and the integrated cross-section.
\begin{table}[h]
\begin{center}
\begin{tabular}{|c|c|c|}
\hline
&&\\
Process&$d \sigma/dy$&$\sigma$\\
&&\\
\hline
&&\\
$\nu_{\mu} + e^{-} \rightarrow \mu^{-} + \nu_{e}$&$A ~s (g_{L}^{\nu})^{2} (g_{L}^{e})^{2} $& $A~ s~ (g_{L}^{\nu})^{2} (g_{L}^{e})^{2}$\\
&&\\
\hline
&&\\
$\nu_{\mu} + e^{-} \rightarrow \nu_{\mu} + e^{-}$&$A ~s (g_{L}^{\nu})^{2} \left[(g_{L}^{e})^{2} + (1-y)^{2} (g_{R}^{e})^{2}\right]$& $A~ s~ (g_{L}^{\nu})^{2}[ (g_{L}^{e})^{2} + \frac{1}{3} (g_{R}^{e})^{2}$]\\
&&\\
\hline
&&\\
$\bar \nu_{\mu} + e^{-} \rightarrow \bar \nu_{\mu} + e^{-}$&$A~s~ (g_{L}^{\nu})^{2} \left[(g_{R}^{e})^{2} + (1-y)^{2} (g_{L}^{e})^{2}\right]$& $A~ s~(g_{L}^{\nu})^{2} \left[\frac{1}{3} (g_{L}^{e})^{2} +  (g_{R}^{e})^{2}\right]$\\
&&\\
\hline
&&\\
$\nu_{e} + e^{-} \rightarrow  \nu_{e} +e^{-}$&$A~s~ (g_{L}^{\nu})^{2} \left[(g_{L}^{e} + 1 )^{2} + (1-y)^{2} (g_{R}^{e})^{2}\right]$& $A~ s~(g_{L}^{\nu})^{2} \left[\frac{1}{3} (g_{R}^{e})^{2} +  (g_{L}^{e} + 1)^{2}\right]$\\
&&\\
\hline
&&\\
$\bar \nu_{e} + e^{-} \rightarrow  \bar \nu_{e} + e^{-}$&$A~s~ (g_{L}^{\nu})^{2} \left[(g_{R}^{e})^{2} + (1-y)^{2} (g_{L}^{e} +1)^{2}\right]$& $A~ s~(g_{L}^{\nu})^{2} \left[\frac{1}{3} (g_{L}^{e} +1)^{2} +  (g_{R}^{e})^{2}\right]$\\
&&\\
\hline
\end{tabular}
\caption{The differential and total cross-sections for a few $\nu,\bar \nu$ induced CC and NC processes, with $A = 4 G_{\mu}^{2}/\pi$.}
\label{tab:nccccsec}
\end{center}
\end{table}
A few comments are in order. The above expressions use $\rho=1$ as well as the fact that values of $g_{L},g_{R}$ for the $\mu$ and the $e$ are the same.  All the neutrino induced cross-sections are indeed proportional to the square of the centre of mass (com) energy $s$ as we have noted before. The variable $y$ is related to the scattering angle $\theta$ in the com frame. One can see after some manipulations that the angular distribution of the scattered charged lepton is different for the case of $\nu$ and $\bar \nu$. In the first row we have  written the cross-section for the CC process $\nu_{\mu} + e^{-} \rightarrow \mu^{-} + \nu_{e}$ , so that one can indeed see that the size of the expected cross-sections for the NC processes are of the same order of magnitude as the CC process and depend on $\sin \theta_{W}$. Note the last two rows of Table~\ref{tab:nccccsec}. As one changes from the $\nu_{\mu}, \bar \nu_{\mu} $ beams to $\nu_{e},\bar \nu_{e}$ beams the factors of  $(g_{L}^{e})^{2}$ in the total cross-section expressions gets changed to $(g_{L}^{e} + 1)^{2}$. Further, note also the different weights of the $(g^{e}_{L})^{2}$ and $(g^{e}_{R})^{2}$ contributions as one changes from $\nu$ to $\bar \nu$ beams. Both these observations tell us that the contours of constant cross-section for these four processes are ellipses in the $g_{A}$--$g_{V}$ plane with different centers and with major axes of differing orientations. Thus a measurement of these cross-sections can then help us determine $g^{e}_{V}, g^{e}_{A}$, albeit upto sign ambiguities.

Note also from the table that as one changes from $\nu$ to $\bar \nu$,  the terms in the angular distribution proportional to $(g^{e}_{L})^{2}$ and $(g^{e}_{R})^{2}$ get interchanged. This behavior can be understood very easily in terms of the chirality conservation of the gauge interaction and the angular momentum conservation. As a result, one can write the weak NC cross-sections for all the different pairs of fermions rather easily by inspection. In particular, the same table can be used to calculate the cross-section for the weak NC induced processes with nucleon (nuclear) targets as well. The hadronic weak neutral current events arise from the scattering of the $u,d,s$ quarks in the nucleon (nucleus). In the parton model the net rate is then given by the incoherent sum over all the quarks contained in the nucleon (nucelus). Using the information on the momentum distributions of quarks/antiquarks in the nucleon (nucleus), it is also possible to estimate the expected cross-section. Again these too depend only on $\sin^{2}\theta_{W}$ as far as the EW model parameters are concerned. 

At the time of the discovery of weak neutral currents in  hadronic and leptonic production, theoretical estimates were available for the upper limit on the ratio of neutral current to charged current elastic scattering. This was obtained by using experimental knowledge of the form factor of the proton and neutron.  The same was also available for the inelastic process of the inclusive production of hadrons using the language of structure functions of the target nucleus. Two points are worth noting here. While the use of nuclear targets increased the expected rates for NC induced hadron production, establishing that the events are indeed due to weak NC was difficult because of the large neutron induced background. The pure leptonic processes on the other hand, were predicted to be very rare and hence difficult to observe, but could unambiguously prove existence of weak NC as soon as even one event was observed.

Neutral currents were discovered in 1973 in the study of elastic scattering of $\nu_{\mu}$ and $\bar \nu_{\mu}$  off nuclear targets~\cite{bib:firstnc1,bib:firstnc}. The experiment discovered evidence for the neutral current induced hadronic processes
\[ 
\nu_{\mu} + N \rightarrow \nu_{\mu} + {\rm hadrons};~~\bar \nu_{\mu}
+ N \rightarrow \bar\nu_{\mu} + {\rm hadrons}.
\]
as well as pure leptonic processes,
\[
\bar\nu_{\mu} + e^{-} \rightarrow \bar\nu_{\mu} + e^{-},
\]
using the giant bubble chamber Gargamelle. In fact the discovery came in an experiment which had been designed to study the charged current interactions:
\[ 
\nu_{\mu} + N \rightarrow \mu^{-} + {\rm hadrons};~~\bar \nu_{\mu}
+ N \rightarrow \mu^{+} + {\rm hadrons}.
\]
\begin{figure}[hbt]
\begin{center}
\includegraphics*[width=8cm,height=10cm]{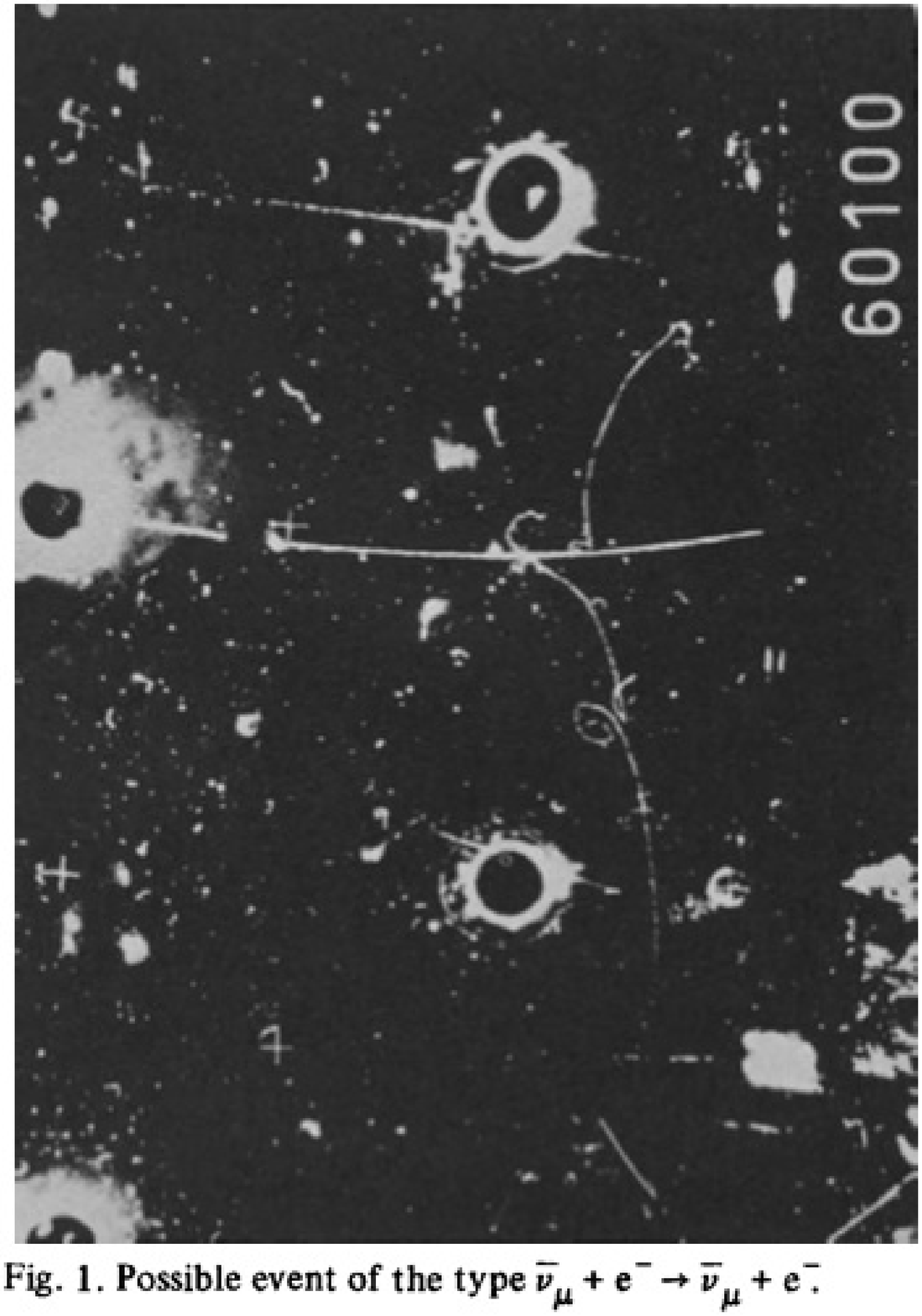}
\caption{Observation of the first  leptonic interaction induced by weak neutral current. The incoming $\bar \nu$ knocks off the $e^{-}$, which then appears as a track accompanied by the shower of $e^{+}e^{-}$ pairs the passage of $e^{-}$ creates. Taken from ~\protect\cite{bib:firstnc}}
\label{fig:wnc}
\end{center}
\end{figure}
Thus the experiment could easily extract  the ratio of the CC to NC events, after the observation of NC in hadronic events. The experiment had seen ${\cal O} \sim 100$ events of different categories (NC and CC) containing hadrons, with 
\[
\frac{NC}{CC}{\Bigg|}_{\nu} = 0.21 \pm 0.03;~~~~~\frac{NC}{CC}{\Bigg|}_{\bar \nu} = 0.45 \pm 0.09
\]
As already mentioned, the same experiment also found evidence for the pure leptonic process, where the $\nu_{\mu}$ was scattered off the atomic electron. Figure~\ref{fig:wnc}, taken from Ref.~\cite{bib:firstnc}, shows the image of the first unambiguous, weak  neutral current event ever observed. The incoming antineutrino, interacts with an atomic electron and knocks it forward. The electron is identified from the characteristic shower created by the electron-positron pairs. This was a considered to be clear evidence for the weak neutral current. The theoretical predictions summarised in Table~\ref{tab:nccccsec} were used to justify the interpretation.  With just one $\bar \nu$ event, the experiment could only quote a range $0.1 < \sin^{2}\theta_{W} < 0.6$ at $90\%$ c.l.  The number of hadronic NC events on the other hand, was big enough to extract a value of $\sin^{2}\theta_{W}$ to be in the range of $0.3$--$0.4$. This was the  first qualitative validation of the prediction of neutral currents.

\subsubsection{Observation of charm with `predicted' mass}
Soon after the observation of the weak neutral current, the charm quark was also discovered with mass very close to that predicted by the analysis of the $\Delta S=2$, $K_{0}$--$\bar K_{0}$ mixing caused by FCNC. We have discussed already details of this prediction in the earlier section. As we understand now, in view of the very large mass of the top quark, it was somewhat 'fortuitous' that the charm quark contribution to the $\Delta S = 2 $ mass difference was the dominant one. Be as it may,  this  was an extremely important second validation of the correctness of the gauge theory of EW interactions based on the gauge group $SU(2)_{L} \times U(1)_{Y}$.
Note that one of the validation came from tree level couplings and the other from loop induced effects. 

\subsubsection{Determination of $\sin^{2}\theta_{W}$ and prediction of $M_{W},M_{Z}$.}
\label{sec:mwmzsnthw} 
\begin{figure}[htb]
\begin{center}
\includegraphics*[width=10cm,height=4cm]{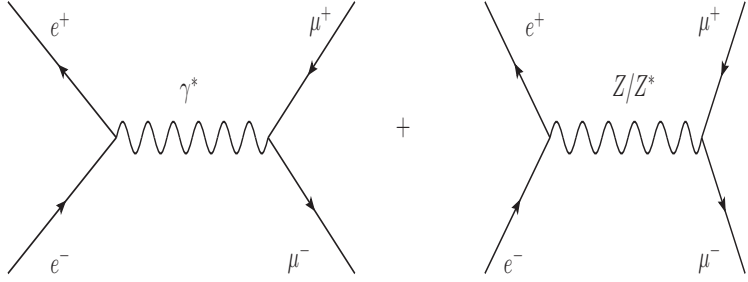}
\caption{Weak neutral current effects in $e^{+} e^{-} \rightarrow \mu^{+} \mu^{-}$.}
\label{fig:eemumu}
\end{center}
\end{figure}

The same leptonic couplings which contribute to the neutral current  scattering processes involving $\nu$'s can also make their presence felt in processes like 
\begin{equation}
e^{+} + e^{-} \rightarrow  \mu^{+} + \mu^{-}.
\label{eq:eemumu}
\end{equation}
This proceeds through a $\gamma^{*}$ exchange in the $s$-channel and a $Z/Z^{*}$ exchange shown in Fig.~\ref{fig:eemumu}. Whether the $Z$ will be on shell or off shell of course depends on the com energy.
The cross-section for this process can be easily computed. Electromagnetic interactions being the same for the left and right chiral fermions, $\gamma^{*}$ exchange diagram gives a forward-backward symmetric contribution whereas both, the square of the amplitude of the $Z$ exchange diagram itself and the interference term, will give contributions which are forward backward asymmetric. Hence the presence of the weak neutral current will manifest itself in the form of a forward-backward asymmetry in (say) $\mu$ production. Both the size and sign of this asymmetry depends on the centre of mass energy of the process $\sqrt{s} = 2 E_{b}$ where $E_{b}$ is the beam energy, relative to the mass of the $Z$ boson. 

\noindent In fact if $\theta$ is the angle made by the outgoing lepton with the incoming lepton, then one can show that
\begin{equation}
\frac{d \sigma (e^{+} e^{-} \rightarrow \mu^{+} \mu^{-})}{d \cos \theta}  =\frac{\pi \alpha_{\rm em}^{2}}{2 s} \left[ A (1 + \cos^{2} \theta ) + B \cos \theta \right]
\label{eq:eemumu1}
\end{equation}
where
\begin{eqnarray}
A &=& 1 + 2 \Re e (\chi) g_{V}^{2}+ |\chi|^{2} (g_{V}^{2} + g_{A}^{2})^{2};~~ B = 4 \Re e (\chi) g_{A}^{2} ~+~8|\chi|^{2} g_{V}^{2}g_{A}^{2}, \nonumber\\
\chi &=& \left( \frac{G_{\mu} M_{Z}^{2}}{2 \sqrt{2} \pi \alpha } \right) \frac{s}{s - M_{Z}^{2} +i M_{Z} \Gamma_{Z}}.
\label{eq:eemumu2}
\end{eqnarray}
Here $g_{V},g_{A}$ denote the (common)  vector and the axial vector NC couplings for the $e$ and the $\mu$, $\Gamma_{Z}$ is the width of the  $Z$. In the chosen normalisation, deviation of $A$ from $1$ and that of $B$ from zero is then indication of the contribution of the weak NC to the process. Both $A$ and $B$ contain terms linear in $\Re e|\chi|$ and $g_{V}^{2}$ or $g_{A}^{2}$. Hence, even for small values of $|\chi|$, both the total cross-section and the angular distribution can be used to probe the weak NC contribution. $B$ is zero without the $Z$ contribution. It is however  nonzero for both, the interference terms containing $\Re e (\chi)$ and the square of the $Z$-exchange diagram alone, containing $|\chi|^{2}$. Hence the angular distribution contains an asymmetric term at all $s$. If we analyze these expressions we find that the results for this asymmetry are very different for $\sqrt {s} \ll M_{Z}$ and $\sqrt{s} = M_{Z}$.

The forward-backward asymmetry, $A_{FB}^{\mu}$ defined as the ratio of the difference of cross-sections  with the $\mu^{-}$ in the forward  and the backward hemisphere and  the total cross-section, is then expected to be nonzero due to the $Z$ contribution and the interference term.
It is clear that this is also the same as charge asymmetry between the muons in the forward hemisphere. Thus one has two asymmetries
$A_{FB}^{\mu}$ and$A_{C}^{\mu}$:
\begin{equation}
A_{FB}^{\mu} = \frac{\sigma(\cos \theta^{\mu} > 0) - \sigma(\cos\theta^{\mu} < 0)} {\sigma(\cos \theta^{\mu} > 0 ) + \sigma(\cos\theta^{\mu} <0)};\qquad\qquad A_{C}^{\mu} = \frac{\sigma(\mu^{-})  - \sigma(\mu^{+})} {\sigma(\mu^{-}) + \sigma({\mu}^{+})}.
\label{eq:asymm}
\end{equation}
and these are equal. The reason for the equality of these two asymmetries is the $CP$ invariance of the gauge Lagrangian, even if the $Z$ has parity violating interactions.  Using Eqs.~\ref{eq:eemumu1} and \ref{eq:eemumu2} one can calculate the $A_{FB}^{\mu}$, which in general depends on $s$. For two different values of $s$ of interest, it can be shown that:
\begin{equation}
A_{FB}^{\mu} {\bigg |}_{s \ll M_{Z}^{2}} = -\frac{3}{\sqrt{2}}\frac{G_{\mu} s}{e^{2}} g_{A}^{2} \frac{1}{1 -  \frac{4 G_{\mu} s}{\sqrt{2} e^{2}}g_{V}^{2}};\qquad\qquad A_{FB}^{\mu}{\bigg|}_{s=M_{Z}^{2}} \sim \frac{g_{A}^{2} g_{V}^{2}}{\left(g_{A}^{2} + g_{V}^{2}\right)^{2}}.
\label{eq:FBasym}
\end{equation}
In the first case $M_{Z}$ drops out as we have made an approximation where $s \ll M_{Z}^{2}$ .
In the second case in Eq.~\ref{eq:FBasym} , while writing the value for  $\sqrt{s} = M_{Z}$, we have used the fact  that $M_{Z}/\Gamma_{Z} \gg 1$ and hence the dependency on the precise value of 
 $M_{Z}$ drops out.  The small width is guaranteed by the weak nature of the NC couplings of the Z with the fermions. The factor in the denominator of $\chi$ gives a characteristic resonant shape to the cross-section for the process $e^{+}e^{-} \rightarrow \mu^{+} \mu^{-}$, the interference term being negative causing the cross-section to reduce below the value expected for the $\gamma$  exchange alone and to start rising again as $\sqrt{s}$ approaches $M_{Z}$.  For $\sqrt{s} \ll M_{Z}$ the value of $A$ differs from $1$, the value expected in QED, by  $\displaystyle{\left(- \frac {G_{\mu} s}{\sqrt{2} \pi \alpha_{\rm em}}  g_{V}^{2}\right)} $. Further the coefficient of the asymmetric, linear term in $\cos \theta$ is given by the same expression with the replacement of $g_{V}^{2}$ by $g_{A}^{2}$. Thus it is possible to get information on {\it both} $g_{V}^{2}$ and $g_{A}^{2}$ from  measurements of $A$ and $B$  even with beam energies that are much lower than $M_{Z}$. Since $G_{\mu} \sim 10^{-5}/M_{p}^{2}$, the effects can become substantial only when  $s \sim  {\cal O} (10^{4} \UGeV^{2})$. Indeed the first hints of weak NC in this process were obtained in $e^{+}e^{-}$ collisions with  $\sqrt{s} \sim 35 \UGeV$. It is worth noting at this point that the calculation of cross-section for quark (and hence hadron) production via $\gamma/Z$ exchange proceeds exactly in the same manner, except the expressions will involve $g_{A}^{q},g_{V}^{q}$ in addition to $g_{V}^{e}, g_{A}^{e}$ in Eqs.~\ref{eq:eemumu1} and \ref{eq:eemumu2}. All the observations about $e^{+} e^{-} \rightarrow \gamma/Z \rightarrow \mu^{+} \mu^{-}$ then apply for the $e^{+} e^{-} \rightarrow \gamma/Z \rightarrow q \bar q \rightarrow {\rm hadrons}$ as well.

Note that just like the various cross-sections in Table~\ref{tab:nccccsec}, the asymmetries of Eqs.~\ref{eq:asymm} and \ref{eq:FBasym} too, depend only on one unknown quantity, {\it viz.}, $\sin^{2}\theta_{W}$ through the vector and axial vector NC  couplings of the charged lepton. The above expressions tell us therefore, that a study of 
the leptonic scattering processes given in the Table~\ref{tab:nccccsec} along with the  energy dependence of the FB asymmetry and that of the cross-section for the reaction given in Eqs.~\ref{eq:eemumu1} and \ref{eq:eemumu2}, can provide information about $\sin^{2}\theta_{W}$ much before reaching the beam energies  close to $M_{Z}$.  If all the measurements of the leptonic cross-sections as well as the asymmetries yielded a unique value of $\sin \theta_{W}$, which is the only free parameter of the model, this can then provide a quantitative validation of the GSW model. It is interesting to note that the energy dependence of the cross-section $\sigma (e^+ e^- \rightarrow \mu^+ \mu^-)$ can also provide indirect information about $M_{Z}$, much before the energy values close to $M_{Z}$ are reached.

Note that production of hadrons by weak NC processes while being very useful for validation of the weak NC due to the large rates possible with nuclear targets, also needed knowledge  about the nuclear structure functions to interpret the data. Both the theoretical and experimental understanding of this structure at that time was somewhat rudimentary.  Hence the validation of the SM would be much more unambiguous, if one would  extract  $\sin^{2} \theta_{W}$ using pure leptonic processes alone, {\it viz.} the $\nu$-charged lepton scattering  and $e^{+} e^{-}$ collisions.

\begin{figure}[hbt]
\includegraphics*[width=7cm,height=5cm]{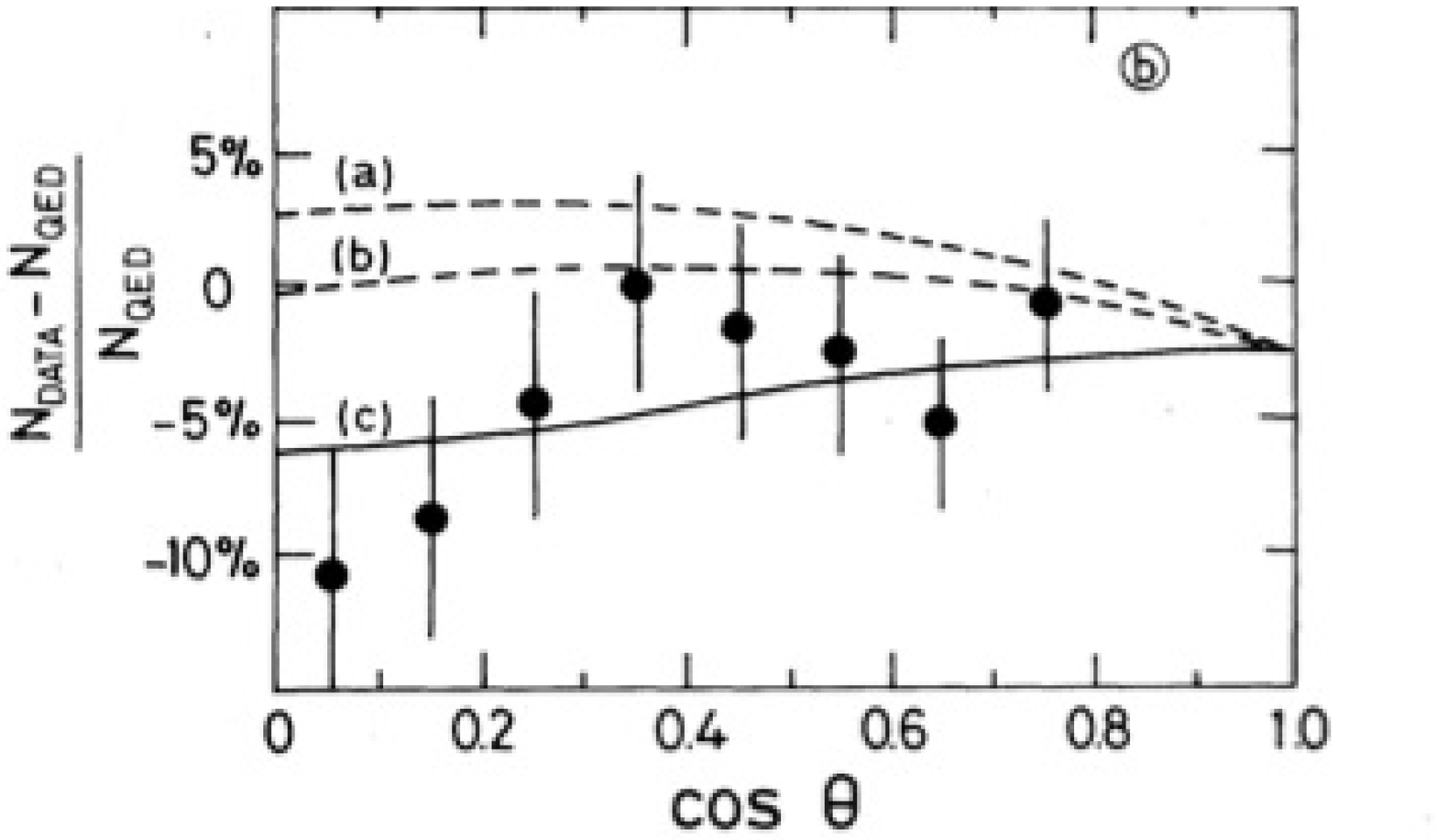}
\includegraphics*[width=10cm,height=7cm]{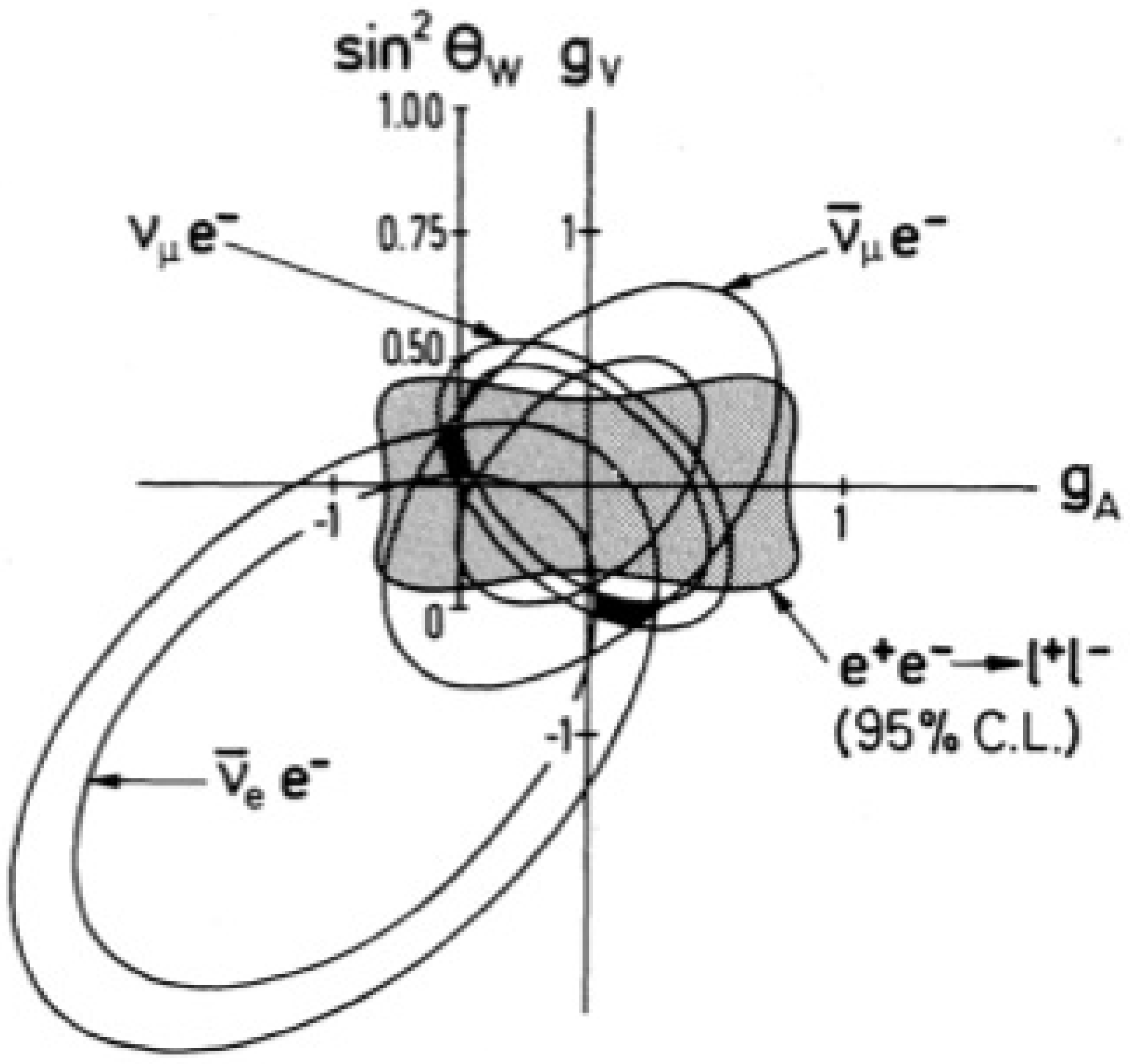}
\caption{Quantitative validation of the weak NC. Details of the data taken from \protect\cite{bib:petradata} are discussed in the text. The left panel shows evidence of asymmetric angular distributions expected from the weak NC contribution. The right panel indicates regions in the $g_{V},g_{A}$ plane and hence values of $\sin^{2}\theta_{W}$ extracted from the leptonic data at different level of confidence.}
\label{fig:earlync}
\end{figure}

Fig.~\ref{fig:earlync} shows compilation of such extraction of 
$g_{V}, g_{A}$ and hence $\sin^{2}\theta_{W}$ from pure leptonic processes.  These results were among the early quantitative validation of the SM.  As explained above the leptonic processes were better suited for a clean and unambiguous extraction of $\sin^{2}\theta_{W}$.   Further, the com energies of the early $\nu$ experiments were limited to  $s <  200 \UGeV^{2}$, whereas the $e^{+}e^{-}$ experiments at PETRA at DESY(Hamburg) had $s ~~\LTS~~ 1400 \UGeV^{2}$. The $e^{+} e^{-}$ experiments could also probe the NC couplings of the quarks as well, by studying the hadron production along with the $\mu^{+}\mu^{-}$ pair production. Thus the information about the weak neutral processes at the $e^{+}e^{-}$ colliders was a value addition to the analysis, even though the beam energies were much below than those required to produce an 'on-shell' $Z$ boson.
The left panel shows results on the deviation from the QED expectations of the angular distribution for the $\mu$ \ie evidence for both : a nonzero value of $B$ and value of $A$ different from $1$.  It was indeed comparable to the deviation of few percents to be expected at these energies as was argued above. The plot shows comparisons with predictions of the GSW model (cf. Eq.~\ref{eq:FBasym}) for different values $\sin^{2}
\theta_{W}$ showing clear sensitivity to the same. Indeed this as well as measurements of  $\mu$ charge asymmetry defined in the Eq.~\ref{eq:asymm} for a limited region in the forward hemisphere and the  cross-section measurement were used to delineate a region in the $g_{A}$--$g_{V}$ plane that was allowed by the data at $95 \%$ c.l. This is indicated by the grey shaded region in the right panel of the Fig.~\ref{fig:earlync}. Superimposed on this grey area are also the regions in the same plane allowed by measurements of $\bar \nu_{e} e^{-}, \bar \nu_{\mu} e^{-}$ and $\nu_{\mu} e^{-}$ scattering. We notice from Table~\ref{tab:nccccsec} that all the cross-section expressions define different ellipses in the $g_{A}$--$g_{V}$ plane. The area between two ellipses is the region allowed at $68\%$  c.l. by  the measurement of the cross-section for that particular neutrino scattering reaction.  

We see from the right panel that if one uses just the elastic $\nu$-charged lepton scattering data, there is a two fold ambiguity in the values of $g_{A},g_{V}$ that are consistent with the totality of the available data. This is indicated by the two dark black regions. This ambiguity is removed on using the $e^{+}  e^{-}
\rightarrow l^{+} l^{-}$ data.  The solution with negative $g_{A}$ and positive $g_{V}$, corresponding to the dark region  in the upper left corner of the grey shaded square region, is chosen
uniquely, after we add determination of $g_{V}, g_{A}$ from the $e^{+} e^{-}$ measurements. This dark region in the upper left corner corresponds to 
\begin{equation}
\sin^2 \theta_{W} = 0.234 \pm 0.011.
\label{eq:sinwnu}
\end{equation}
This was the unique value of $\sin^{2} \theta_{W}$ consistent with all the 'leptonic' NC measurements mentioned before. One could also use only  the $e^{+}e^{-}$ data. Combining all the $e^{+} e^{-}\rightarrow l^{+} l^{-}$ measurements with those for $e^{+} e^{-} \rightarrow q^{+} q^{-}$, $\sin^{2} \theta_{W}$ was determined to be 
\begin{equation}
\sin^{2}\theta_{W} = 0.27 \pm 0.08.
\label{eq:sinwee}
\end{equation}
Clearly the two determinations are consistent with each other.  
These measurements thus conclusively proved existence of the weak NC  as predicted by the GSW model. One could then use the value of $\sin \theta_{W}$ so determined, to further make predictions for the $W,Z$ masses as well as their phenomenology.

The weak neutral couplings of the electron can also be probed by
\begin{figure}[hbt]
\begin{center}
\includegraphics*[width=14cm,height=6cm]{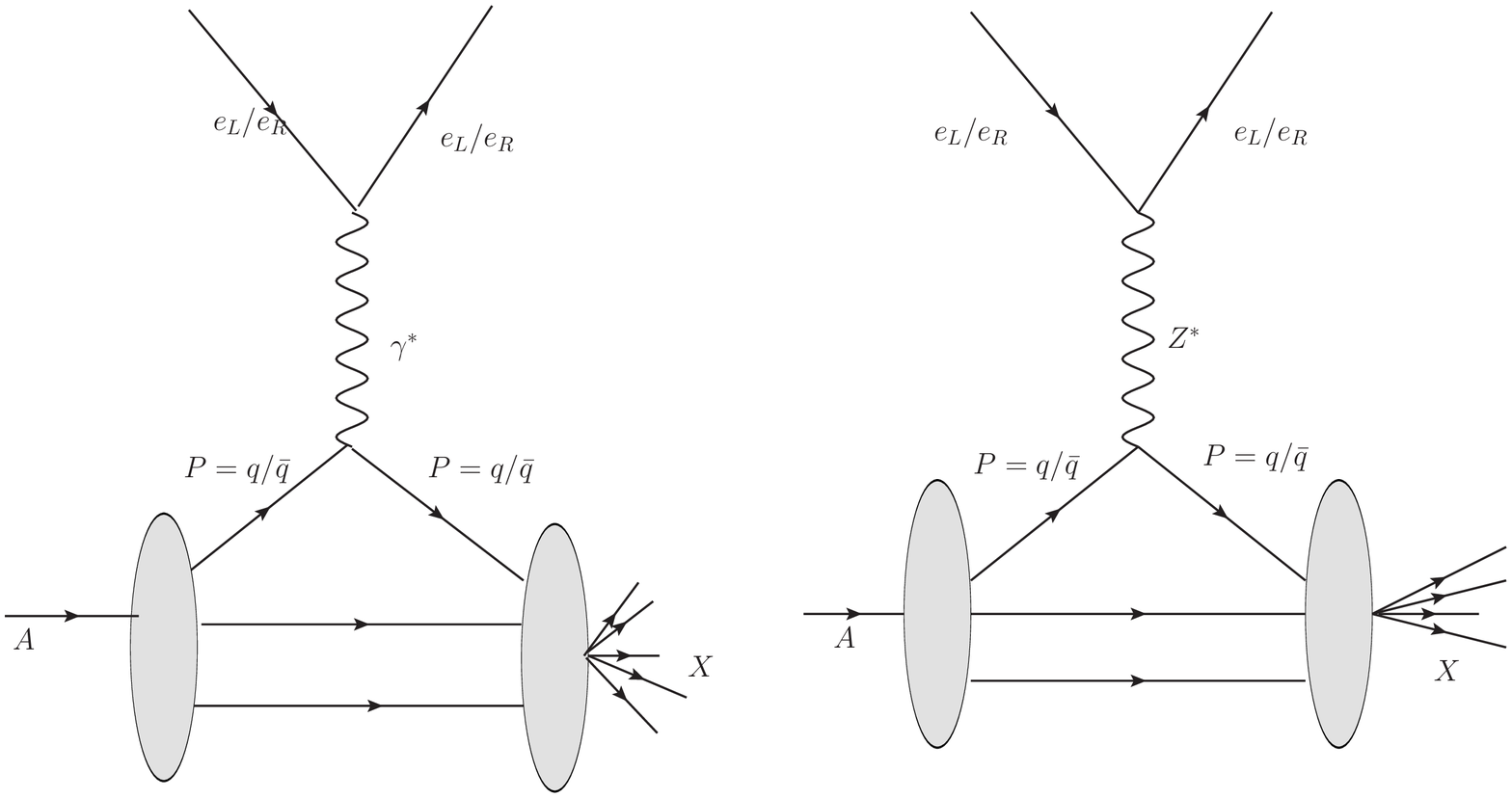}
\caption{Weak NC contributions to the deep inelastic scattering with polarised $e^{-}$ beams.}
\label{fig:DIS}
\end{center}
\end{figure}
studying interference between the $t$-channel $\gamma^{*}$ and $Z$ exchange in the Deep Inelastic Scattering (DIS) processes indicated in Fig.~\ref{fig:DIS}.
This is very similar to the $e^{+} e^{-} \rightarrow l^{+} l^{-}$ case. However, in this case one needs to have longitudinally  polarised electron beams, to be able to see the effect experimentally. The diagram with $\gamma^{*}$ exchange will give a symmetric result for both left and right polarised $e^{-}$ but the $Z$ treats them differently. Recall here the  different values of $g_{L}^{e}$ and $ g_{R}^{e}$ in Table~\ref{tab:04gagv}. Thus there will be a  polarization asymmetry in the cross-section. At lower energies and hence smaller values of the  invariant mass $- Q^{2}$ of the exchanged  $\gamma^{*}/Z^{*}$, it is the interference term between the two diagrams which dominates the size of the observed polarisation asymmetry and hence the evidence for parity violation. The interference effect can be shown to be $\sim G_{\mu} s$ in this case as well and is linear in $g_{V}^{e}$. As mentioned before, for the value of $\sin \theta_{W}$ realized in nature the vector coupling of the electron is very small. Hence an asymmetry which is linear in this small parameter, provides a more sensitive probe of $g_{V}^{e}$ than the one provided by the asymmetry $A_{FB}^{\mu}$ of Eq.~\ref{eq:asymm}.  Measurements of this asymmetry also yielded a value of  $\sin^{2} \theta_{W}$ consistent with the determination from the pure leptonic probes.
 
Finally the best determination of $\sin^{2} \theta_{W}$ came from high statistics data on $\nu$-induced Deep Inelastic Scattering and {\it polarised} $e$- Deuterium scattering (both not discussed here at all) and the value was~\cite{bib:summarync}:
\begin{equation}
\sin^{2}\theta_{W} = 0.224 \pm 0.015, \rho = 0.0992 \pm 0.017;~~~~ \sin^{2} \theta_{W} = 0.229 \pm 0.009 ~~ {\rm assuming }~~ \rho =1 .
\label{eq:rhothwnu}
\end{equation}
In the first case both $\rho$ and $\sin^{2} \theta_{W}$ were taken to be unknown and fitted to the data and in the second case $\rho$ was fixed at $1$. Thus $\rho$ was determined to be $\sim 1$ as expected in the GSW model. Assuming this, around 1981 one could then predict using Eq.~\ref{eq:mwsinthw}: 
\begin{equation}
M_{W} \simeq 78.15 \pm 1.5 \UGeV; \qquad \qquad M_{Z} \simeq 89 \pm 1.3 \UGeV.
\label{eq:wzpred}
\end{equation}
This then sets the goal posts to design  experiments which could produce $W,Z$ directly and study them. In principle, the predictions above receive radiative corrections. A more  accurate prediction would require, for example, discussion of radiative corrections to the couplings involved in the relations given by Eq.~\ref{eq:egrel}. We will come to that in the next subsection.

So the take home message of the above discussion is that the early
$\nu$ experiments as well scattering experiments with polarised electron beams and nuclear targets, along with the $e^{+}  e^{-} \rightarrow l^{+}  l^{-}$ experiments, tested the structure of the NC couplings of the leptons AND those of the quarks predicted by the GSW model. The experiments conclusively proved that all the measurements were consistent with  a {\it unique} value of the one undetermined parameter of the model $\sin^{2} \theta_{W}$. This then also predicted a narrow range of possible masses for both the $W$ and the $Z$ bosons. Inter alia, these measurements also established $\rho \simeq 1$, consistent with the GSW prediction again. Thus at this stage, apart from the direct verification of the tree level $ZWW$ coupling which must exist in this gauge theory, all the other tree level predictions of the model seemed to have been tested. 

Given the knowledge of the quark content of the $p$  available from the DIS experiments, it was also possible to predict the rate of production of these bosons in the process
\[
p + \bar p \rightarrow  W + X \rightarrow l + \nu_{l} + X;
\qquad \qquad p + \bar p \rightarrow  Z + X \rightarrow l^{+} + l^{-} + X.
\]
In fact the CERN super proton synchrotron (SPS) was converted into $S p {\bar p}S$, to collide protons on antiprotons, so as to have enough energy to produce the $W,Z$ in the $p \bar p$ collisions. The observation of the $W$ and the $Z$ bosons in the UA-1 and UA-2 experiments~\cite{bib:ua1,bib:ua2}, with mass values and production rates which agreed with these predictions, was a very important step in confirming the correctness of the GSW model. Later data  confirmed  the $V$--$A$ coupling of the $W$ bosons to fermions from the angular distribution of the events, even though the original observation had only a handful  of these: $6$ in UA-1 and $4$ for UA-2.

The masses of the $W$ and the $Z$ measured in the UA2 experiment~\cite{bib:ua2}, for example, were
\[
M_{W} = 80 +10 -6 \UGeV;~~~M_{Z} = 91.9 \pm 1.3 \pm 1.4 \UGeV.
\]
The larger errors for $M_{W}$ reflect the uncertainties in the measurement of 'missing' transverse momentum due to the $\nu$ which evades detection. For $M_{Z}$, the first number indicates the statistical error and the second systematic. The use of final state containing leptons allowed for much more accurate determination of the invariant mass in the case of the $Z$ boson.  These masses were certainly consistent with the predictions: see, for example, Eq.~\ref{eq:wzpred}. 
One can in principle extract $\rho$ AND $\sin^{2} \theta_{W}$ from this 'direct' measurement of masses (in particular the accurate measurement of $M_{Z}$) and compare these with the values obtained from the  earlier 'indirect' information from $\nu$ scattering, for further tests of the SM. This already used the more accurate predictions using energy dependence of the couplings as well EW corrections to the weak processes used to extract $\sin^{2} \theta_{W}$. We will discuss this in the context of precision testing of the SM.

\subsection{Direct Evidence for the $ZWW$ coupling.}
\begin{figure}[hbt]
\begin{center}
\includegraphics[height=8cm,width = 10cm]{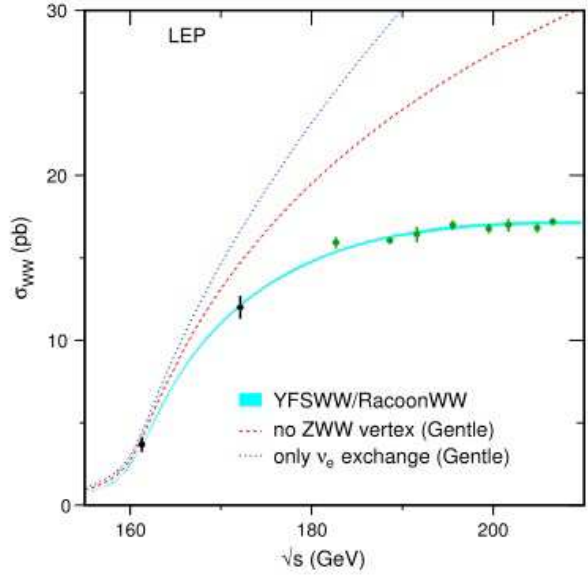}
\caption {Energy dependence of the $W^{+} W^{-}$ cross-section at LEP-II. Taken from \protect\cite{bib:lep2data}.}
\label{fig:lepeeww}
\end{center}
\end{figure}

Before moving on to the discussion of calculation and validation of loop effects in the precision measurements of the EW observables, we need to discuss the validation of the existence of another  tree level coupling of the gauge bosons, {\it viz.}, the triple gauge boson $ZW^{+}W^{-}$ coupling which is characteristic of the non abelian nature of the gauge theory. As already discussed, contribution of the $Z$ exchange diagram is crucial in curing  the bad high energy behavior of  the $e^{+} + e^{-} \rightarrow W^{+} W^{-}$ cross-section. $W^{+} W^{-}$ pair production in $e^{+} e^{-}$ collisions was studied at LEP-II where the centre of mass energy was increased  from the $Z$-pole  value of $91 \UGeV$ to the two $W$ threshold of $161 \UGeV$ and then finally to $209 \UGeV$. Fig.~\ref{fig:lepeeww} shows the LEP-II data along with the theory prediction. 
The data is well described by the solid line which represents the sum of the contribution of the $\nu_{e}$ exchange diagram and $Z/\gamma$ exchange diagrams shown in the left and the central panel of Fig.~\ref{fig:eeWW}. One sees that the contribution to the cross-section of just the $\nu_{e}$ exchange diagram of the left most panel, shown by the blue dashed curve, rises very fast with energy. The cross-section after including contribution of the s-channel $\gamma$ exchange alone, where the $ZWW$ coupling is put to zero in the diagram in the central panel of Fig.~\ref{fig:eeWW}, is shown by the red dashed curve. This addition tames the bad high energy behavior to some extend but not completely. Only after adding the $s$-channel $Z$-exchange diagram does the cross-section have a good high energy behavior, shown by the blue-green solid curve which also describes the data well. Thus we see that the temperate energy dependence of the $e^{+} +  e^{-} \rightarrow W^{+} W^{-}$  cross-section shown by the data, is 'direct' proof of the $ZW^{+}W^{-}$ triple gauge boson coupling.

The threshold rise of this cross-section also offers an accurate determination of $W$ mass and the width~\cite{bib:lep2data}:
\[
M_{W} = 80.376 \pm 0.033 \UGeV, \Gamma_{W} = 2.195 \pm 0.083 GeV.
\]
The same experiment offered a precision measurement of the hadronic decay width of the $W$ as well. These measurements served later as an input to the precision analysis of the EW observables which we will discuss in the next section.  

Note further also that since the energy dependence of the total cross-section is crucially decided by the $ZW^{+}W^{-}$ coupling, it is possible to use the energy dependence and the angular dependence of the process to probe any possible deviations of the $ZWW$ vertex from the SM structure and value.  This process can therefore be successfully used to look for deviations of this coupling from the SM prediction. In view of the important role played by the $ZWW$ coupling in curing the bad high energy behavior of the $W$-pair production cross-section, it is theoretically very important to probe its possible deviations from the SM predictions so as to get indications, if any, of the physics beyond the SM (BSM physics). Measurements of the cross-section and angular distributions of the produced $W$ at LEP-II, constrained strongly any anomalous $ZWW$ couplings; \ie couplings which differ from the SM in either structure or strength.

\subsection{Precision testing of the SM}
Thus we see that the various lepton-lepton and lepton-hadron scattering experiments along with the $p \bar p$  experiments helped establish the correctness of GSW model predictions at the tree level. These tested the tree level SM predictions  for the new NC couplings of the $Z$ boson with all the known fermions as in terms of the single 'free' parameter of the model. The prediction of $SU(2)_{L}$ symmetry for the structure and strength of the $ZWW$ vertex was also tested. Last but not the least the experiments also tested the correctness of the tree level predictions for the $W$ and $Z$ masses. This indeed established the $SU(2)_{L} \times U(1)_{Y}$ structure of the EW gauge theory. However, even with the somewhat imprecise determined values of $W,Z$ masses, the need for including the effects of loop corrections, an essential feature of QFT's, on all these tree level predictions was already clear.  Since the effect of radiative corrections on the extraction of $\sin^{2} \theta_{W}$ is different for different processes, it is necessary to correct the experimentally extracted value for these effects, before the $\sin^{2} \theta_{W}$ extracted from various observables can be compared at high precision. 
 
\subsubsection{Radiative corrections and $\rho$/$\sin^{2}\theta_{W}$ determination.}
\label{sec:radcorr}
In case of the SM, a QFT with SSB, renormalisability of the theory guarantees that the loop corrections to the tree level relations such as  given by Eqs.~\ref{eq:egrel},\ref{eq:rhodef} and \ref{eq:WZmasses}, will be finite and can be computed order by order in perturbation theory. Precision measurements can then test these corrected relations and hence the correctness of these  calculations of loop effects. This can then help establish the renormalisability of the $SU(2)_{L} \times U(1)_{Y}$  gauge theory of EW interactions. Below follows an extremely sketchy discussions of the issues involved.

Some of the one loop diagrams contributing to the corrections to 
\begin{figure}[hbt]
\begin{center}
\includegraphics*[width=16cm,height=8cm]{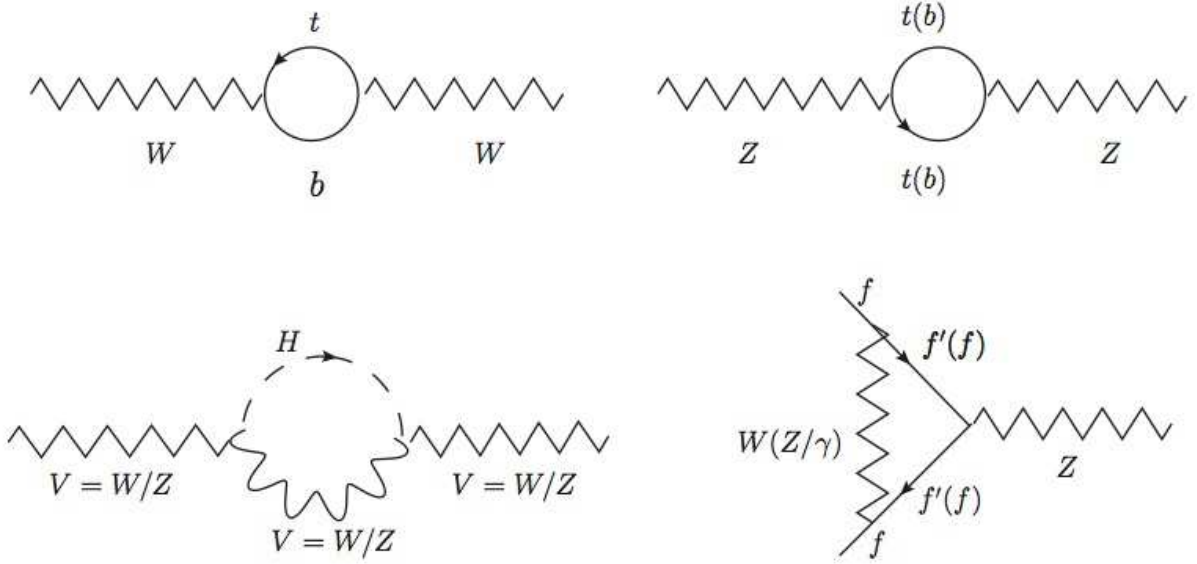}
\caption{Some of the one loop EW corrections to vertices and two point functions in the SM.}
\label{fig:radcorr}
\end{center}
\end{figure}
 the vertices and two point functions are shown in Fig.~\ref{fig:radcorr}. The two diagrams in the top row and the diagram on the left in the lower panel are the ones that need to be considered while calculating the loop corrections to the masses $M_{W}, M_{Z}$. The diagram on the right in the lower panel is an example of diagrams that give rise to  corrections to the $Z f \bar f$ vertex. 
The dominant corrections come from loops containing quarks of the third generation viz. $t,b$. We already notice that corrections to the $W$ and the $Z$ mass will be different, since the former involves a $tb$ loop where as the latter involves the $t\bar t, b \bar b$ loops. As a result the corrections to $\sin^{2} \theta_{W}$ from these diagrams, for example, will be different for the CC and NC processes. 
Let us recall Eq.~\ref{eq:mwsinthw}. We have used Eqs.~\ref{eq:egrel}  and \ref{eq:WZmasses} in deriving this. One needs to take into account radiative corrections to the weak processes used to extract  $\sin^{2} \theta_{W}$ as well as the energy dependence of the couplings and hence of $\sin^{2} \theta_{W}$ obtained via Eq.~\ref{eq:egrel}. The latter too is an integral part of QFT. The extraction of $\sin^{2} \theta_{W}$ from weak processes, taking into account all the weak corrections yielded~\cite{bib:radcorthw}
\[
\sin^{2} \theta_{W}(M_{W}) = 0.215 \pm 0.010 \pm 0.004.
\] 
In Eq.~\ref{eq:mwsinthw} one now needs to use ${\displaystyle\alpha_{\rm em} (M_{W}) = {1}/{127.49}}$ instead of the value ${\displaystyle \alpha_{\rm em} = {1}/{137.03}}$ used  therein. The expression for $M_{W} (M_{W})$ then becomes
\begin{equation}
M_{W} (M_{W}) = \sqrt{\frac{\pi}{\sqrt{2} G_{\mu}}  \frac{\alpha_{\rm em} (M_{W})}{\sin^{2} \theta_{W}(M_{W}) }} = \frac{38.6}{\sin \theta_{W} (M_{W})} \UGeV.
\label{eq:mwmw}
\end{equation}
This then gives, 
\begin{equation}
M_{W} = 83.5 \pm 2.2 \UGeV;~~M_{Z}  =  94.2 \pm 1.8 \UGeV.
\label{eq:mwmz}
\end{equation}
Thus loop effects change the predicted values from those in Eq.~\ref{eq:wzpred} by ${\cal O} (\sim 5 \%) $. This sets the scale for the precision with which one needs to measure the values of the masses of the $W,Z$ to be able to test theory at loop level.  The UA-1 and UA-2 measurements were clearly consistent with these predictions within the accuracy of the measurement as well as predictions. In Ref.~\cite{bib:ua2} these loop corrected predictions for $M_{W},M_{Z}$ were used to extract both $\sin \theta_{W} (M_{W})$ and $\rho$ just from the measured masses of the $W,Z$ in the UA-2 experiment, yielding 
\[
\sin^{2} \theta_{W} = 0.226 \pm 0.014, ~~~\rho = 1.004 \pm 0.052.
\]
This value of $\rho$ is consistent with the expectation of the SM
\ie the GSW model where $W/Z$ masses are generated via SSB. These values are also consistent with the corresponding determinations from the lower energy $\nu$ experiments (cf. Eq.~\ref{eq:rhothwnu}). Agreement of these two independent determinations of $\rho$ and $\sin^{2} \theta_{W} (M_{W})$ from two completely different sets of measurements, already showed consistency of the measurements with theory predictions at loop level. 

Diagrams shown in Fig.~\ref{fig:radcorr}  cause  $\rho$ to change from $1$, the prediction at tree level, since the corrections are different for $M^{2}_{W}$ and $M^{2}_{Z}$.  In fact, one can write
\[
\Delta \rho = \frac{\Sigma_{Z}(0)}{M_{Z}^{2}} - \frac{\Sigma_{W}(0)}{M_{W}^{2}}, 
\]
where $\Sigma_{V}$, $(V=W/Z)$ are the one loop corrections to the propagator. As emphasized above these are different for the $W$ and the $Z$ and hence $\Delta \rho$ is different from $0$. 
At one loop one gets, keeping  only the dominant corrections $\propto M_{t}^{2}$,
\begin{equation}
\rho_{\rm corr} = 1 + \Delta \rho \simeq 1 + \frac{3 G_{\mu} M_{t}^{2}}{8 \pi^{2} \sqrt{2}}
\label{eq:deltarho} 
\end{equation}
Thus one sees that the relation ${\displaystyle \rho = \frac{M_{W}^{2}}{M_{Z}^{2}\cos^{2}\theta_{W}} = 1}$ gets corrected by loop effects. The corrections are finite as advertised before: a result of the renormalisability of the EW theory.  Assuming the (at that time) unknown $M_{t}$ to be as large as the largest mass in the theory, $\sim {\cal O }(M_{W})$, one finds corrections to the tree level value of unity of $\rho$, to $\sim$ few parts in 1000. Thus one would need a high precision measurements of $M_{W}, M_{Z}$ to get a precision value of $\rho$ which can then be contrasted with above prediction given in Eq.~\ref{eq:deltarho}. This can then be used to estimate $M_{t}$ and comparing it with the experimentally observed value of the  $t$ quark mass would then constitute a precision test of the SM. 

In reality, indeed this is what happened. Recall the discussion around Fig.~\ref{fig:}. The precision measurements  at the $Z$ pole in $e^{+} e^{-} \rightarrow Z \rightarrow f \bar f$,  to be discussed momentarily, indicated a value for the top mass $M_{t} \simeq 2 M_{W}$ {\it before} the top quark was actually discovered. Agreement of the measured mass of the $t$ at the Tevatron with this value was then a big success story, testing the SM at loop level. For the much higher value of the mass that the $t$ quark has in real life compared to the $M_{W}$ taken in the numerical estimation above, corrections to $\rho$ in reality are about 1 part in 100 and hence measurable in precision experiments. For future reference, let me also add here that the corrections to $M_{V}^{2}$, from the third diagram in Fig.~\ref{fig:radcorr} involving the $VH$ loop, depend on the Higgs mass $M_{h}$ only logarithmically. 

A detailed discussion of the theoretical significance of the all important   quadratic dependence of these corrections on $M_{t}$, the logarithmic dependence on $M_{h}$ and  the non decoupling nature of the corrections to the $Z b \bar b$ vertex from the $t \bar t$ loop, are beyond the scope of the discussion in these lectures. The former comes from violation of the $SU(2)_{L}$ invariance, reflected in the mass difference between the two members of the doublet : the $t$ and the $b$. $\Delta \rho$ is in fact proportional to $M_{t}^{2} -M_{b}^{2}$. The loops involving $h$ and the $V$ give contributions to $\Delta \rho$ which depend on the Higgs mass, but the accidental Custodial Symmmetry (cf. section~\ref{sec:custodial}), guarantees that this dependence will be only logarithmic. This is consistent with the so called Veltman screening theorem~\cite{bib:veltman}. The corrections to the $Z b \bar b $ vertex, originating from the triangle diagram, one of which is shown in Fig.~\ref{fig:radcorr}, also depend on $M_{t}$ quadratically. This quadratic dependence, on the other hand has a different source. It arises from contributions of the longitudinal $W$ bosons in the loop. In a non-unitary gauge this can be seen as coming from the unphysical Goldstone bosons $\phi^{\pm}$, which are 'eaten up' to become the longitudinal degree of freedom of the $W$-boson. This then clearly explains the non decoupling nature of the correction, coming from the proportionality of $t \phi^{\pm}$ coupling $h_{t}$ or equivalently $M_{t}$.  Even when we do not discuss these issues in detail, suffice it be said that the $M_{t}^{2}$ dependence of the vertex correction is the tell tale sign of the SSB via the Higgs mechanism. Since the origins of the $M_{t}^{2}$ dependence, or equivalently the non-decoupling nature of the corrections, are quite different for the $\Delta \rho$ and $\delta g_{Z\mu \mu}$  and further only the $\Delta \rho$ receives contribution from the Higgs, it is quite important to confirm both of these independently. Let us now follow the story of precision measurements and comparison with the precision predictions further. 

Note here that these corrections can be calculated only if theory is renormalisable.  The renormalisability of a gauge field theory with SSB was proved by 't Hooft~\cite{bib:'thooft}. This theory necessarily has a physical scalar, the Higgs boson in the spectrum. As we will see shortly, the precision measurements at the LEP-I of the $Z$ properties along with weak neutral current couplings of all the fermions,  as well as precision measurements of the properties of the $W$ at LEP-200, tested these corrections. 
A test at the loop level of the various relations such as Eq.~\ref{eq:egrel} or Eq.~\ref{eq:WZmasses}, could then  indicate the need for a finite mass for the Higgs and thus  could be  an indirect  proof for the Higgs! However, we have seen that even with a quadratic dependence of $\Delta \rho$ on $M_{t}$ and the large mass $M_{t}$, the effects are only 1 part in 100, it is clear that with the logarithmic dependence of these corrections on $M_{h}$, this program would require indeed very high precision measurements.

\subsubsection{Precision measurements at LEP}
Let us first begin by a discussion of precision measurements of the mass and the coupling of the $Z$ boson at LEP 1 and the SLC in $e^{+} e^{-} \rightarrow Z \rightarrow f \bar f$. The four LEP experiments studied decays of about 17 Million $Z$, whereas the SLC studied about 600,000 $Z$ decays, but with polarized $e^{+}/e^{-}$ beams. These precision studies of the $Z$ have been summarised in Ref.~\cite{bib:lep1data}.  At the end of the day these experiments determined  the mass and the width of the $Z$ boson and also the values of $\rho$ and effective value of $\sin^{2} \theta_{W}$, to a great accuracy using only the leptonic sector. The  use of 'effective' implies that radiative corrections have been suitably included while extracting these values. 
\begin{eqnarray}
M_{Z}& =& 91.1875 ± 0.0021 GeV, \qquad\qquad \Gamma_{Z} = 2.4952 ± 0.0023 GeV, \nonumber\\
\rho_{l}& =& 1.0050 ± 0.0010,\qquad\qquad ~~~~~sin^{2} \theta^{eff}_{lept} = 0.23153 ± 0.00016.
\label{eq:lep1numb}
\end{eqnarray}
As already explained these  high precision measurements require also high precision calculations, to test the SM at high accuracy. Higher order QCD corrections play a highly important and nontrivial role while using results from the hadronic decays of the $Z$. One also requires an excellent understanding of QCD to calculate correctly the observables from quark final states in terms of what  the detectors actually observe {\it viz.} the jets. This ushered in an era of extremely close  and extensive collaboration  between experimentalists and theorists resulting in a number of  LEP Yellow Reports. These provide the best summary of both the theoretical and experimental issues involved in studies at LEP.

\begin{figure}[hbt]
\begin{center}
\includegraphics*[width=10cm,height=8cm]{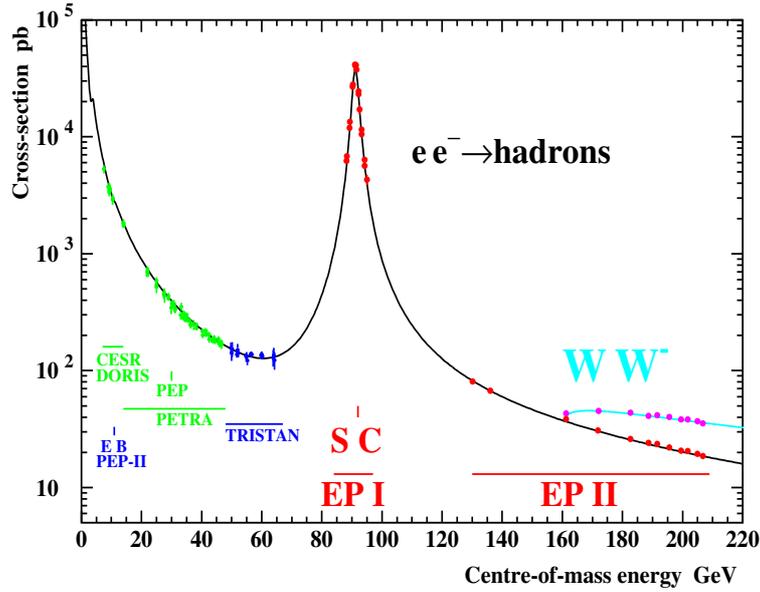}
\caption{The figure shows summary of the data on $e^{+} e^{-} \rightarrow {\rm hadrons}$ over a wide energy range taken 
from \protect\cite{bib:lep1data}.}
\label{fig:lep1csec}
\end{center}
\end{figure}

Fig.~\ref{fig:lep1csec} shows a compilation of the cross-section for the process $e^{+} + e^{-} \rightarrow {\rm hadrons}$, spanning the entire energy range from PEP/PETRA to LEP II. Solid line is theory prediction, including the electromagnetic and the QCD radiative corrections. 
Recall the expression for the cross-section for $e^{+} e^{-} \rightarrow \mu^{+} \mu^{-}$ given in Eq.~\ref{eq:eemumu2}. The initial fall off of the cross-section reflects the {$\displaystyle \frac{1}{s}$ } dependence of the first $\gamma$ exchange diagram in Fig.~\ref{fig:eemumu}. One can then see the onset of the rise in the cross-section due to interference between the $\gamma$ and $Z$ exchange contributions. Recall that it is these interference terms, at energies quite far away from the $Z$ resonance, that had allowed the first glimpse of effects of weak neutral current in the process $e^{+} e^{-} \rightarrow \mu^{+} \mu^{-}$. Thus we see that  the $Z$ resonance makes its presence felt  much before the resonant energy is reached, by just the shape of the cross-section curve. This line shape of the $Z$ resonance depends on $\Gamma_{Z}, M_{Z}$, partial decay width $\Gamma(Z \rightarrow f \bar f)$ and through them on $g_{V},g_{A}$ of the electron and the fermions in the final state being considered. The extremely accurate measurements of  $M_{Z}, \Gamma_{Z}$ mentioned above, were extracted by fitting the shape of this curve near resonance, taking into account effects such as the initial state radiation etc. This precision study of the line shape of $Z$ was made possible by the unprecedented energy resolution of the collider LEP-I.  The thin solid line is then the theoretical prediction for the cross-section including the QED and QCD radiative correction. The asymmetric shape of the curve near the resonance is the effect of the initial state radiation. The agreement between the data and theory needs no comment. 

Recall now the discussion in Sec.~\ref{sec:mwmzsnthw} and Eqs.~\ref{eq:eemumu2} -\ref{eq:FBasym}. One can extend constructions of these asymmetries of Eqs.~\ref{eq:eemumu2}-\ref{eq:FBasym}, for all the fermionic final states accessible in the $Z$ decay, {\it viz.} the leptons $e,\mu,\tau$ and the quarks $b,c$. Looking at the expressions in Eqs.~\ref{eq:eemumu2} -- \ref{eq:FBasym} one can see that a precision measurement of these asymmetries as well as partial widths, lead to an accurate determination of $g_{V}^{f},g_{A}^{f}$. The $Z$-decay data from SLC, which employed linearly polarised $e^{-}/ e^{+}$ beams, allowed for constructing polarisation asymmetries just like the forward-backward asymmetry of Eq.~\ref{eq:FBasym}. This too is a measure of parity violation, with the additional advantage that it involves $g_{V}$ linearly instead of the quadratic dependence in Eq.~\ref{eq:FBasym}. This linear dependence is similar to the case of polarization asymmetries in case of polarized electron-Deuterium scattering mentioned before. Recall also that for the value of $\sin^{2} \theta_{W}$ of Eq.~\ref{eq:rhothwnu} which  is rather close to $0.25$, 
the vector coupling of the electron involving  $(4 \sin^{2} \theta_{W} -1)$ is very small. Hence  this linear dependence of the asymmetries on $g_{V}$ allowed the experiments at the SLC  to reach a competitive  accuracy for the extraction of $g_{A},g_{V}$  with the much smaller luminosity and hence smaller number of the $Z$ decays (600000 versus 17 million at LEP) available there.

\begin{figure}[hbt]
\begin{center}
\includegraphics*[width=10cm,height=8cm]{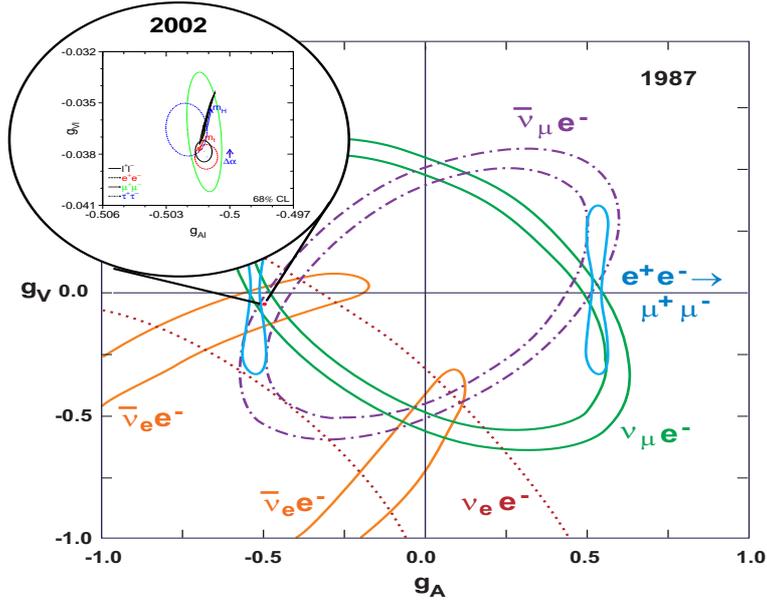}
\caption{
the plot shows determination of $g_{V},g_{A}$ of the electron using $Z$ decays taken from \protect\cite{bib:lep1data}.}
\label{fig:lep1gvga}
\end{center}
\end{figure}

Fig.~\ref{fig:lep1gvga} shows values of $g_{V}^{e},g_{A}^{e}$ obtained using the LEP-I data, juxtaposed  with the data from elastic $\nu$ scattering from 1987. 
The latter is a more refined version of the of the plot of Fig.~\ref{fig:earlync}. To truly appreciate the phenomenal improvement, compare the  size of the region in the $g_{V},g_{A}$ plane selected by all the measurements (shown in an inset at the left of the figure, blown up by roughly a factor of 1000)  with the size of the corresponding region in Fig.~\ref{fig:earlync}. Thus we see that at the $Z$ pole the weak NC couplings of the $Z$ with the fermions, were tested to about one part in 1000.

It goes without saying that with such precision in measurements,
if one were to repeat the earlier exercise of extracting the value $\sin^{2} \theta_{W}, \rho$ from them, such as given in Eq.~\ref{eq:lep1numb}, one HAS to use theoretical predictions which include all the relevant higher order corrections. This was already discussed in Sec.~\ref{sec:radcorr}. Since these corrections have a dependency on the masses of the particles like the $W,t$ and the Higgs, if the measurements are precise enough then they can be sensitive to these masses. We already saw this for the mass of $t$ quark and the radiative corrections to the $\rho$ parameter. The precision measurements of EW observables then  indicate  'indirectly', {\it in the framework of the SM}, the values of the masses of these particles preferred by the precision EW data. A comparison of these  masses determined 'indirectly',  with the ones measured directly, can then be a powerful precision test of the SM. 

\subsubsection
{Precision testing and indirect bounds}
Let us describe the logical steps in such a program to perform precision testing of the SM. In principle the EW part of the SM has following free parameters: $g_{1},g_{2},v$ and $\lambda$. In addition to this of course there is the QCD coupling $g_{3}$, the nine masses (or equivalently the Yukawa couplings) of the   massive charged leptons and quarks, the four parameters of the 
CKM matrix and the strong phase $\theta_{QCD}$.
At tree level all the couplings of the gauge bosons to fermions as well as to each other and their masses are completely given in terms of the first three  parameters in this list, {\it viz.} $g_{1}, g_{2}$ and $v$. In section ~\ref{sec:WZmasses} we already discussed  an analysis where we traded these three for the more accurately known $\alpha_{\rm em}$, $G_{\mu}$ and one free parameter $\sin \theta_{W}$ (cf. Eq.~\ref{eq:mwsinthw}). With the very precise knowledge of $M_{Z}$ provided by the LEP-I, it made sense to trade the $g_{1},g_{2}$ and $v$ for $M_{Z},\alpha_{\rm em}$ and $G_{\mu}$. As before, one can then use the relationships such as given by Eqs.~\ref{eq:egrel}, \ref{eq:WZmasses} etc., of course corrected for radiative effects, to express all the EW observables as functions of these three chosen quantities.  

A really large number of EW observables have been measured very accurately, beginning from the total width of $Z$ boson, $\Gamma_{Z}$, the various forward-backward and polarisation asymmetries on the $Z$-pole, masses $M_{W}, M_{t}$,  polarised $e$-Deuterium scattering, atomic parity violation etc. All these observables depend on $G_{\mu}, M_{Z}$ and $\alpha_{\rm em}$ through their dependencies on $g_{A}^{f}, g_{V}^{f}, M_{V}$ as well as on $\alpha_{s}$ and $M_{t}, M_{h}$  through the higher order QCD and EW corrections. 

Precision calculation for all these EW observables, including the 1 loop EW radiative corrections in the framework of the SM, are available. The idea is to make then a fit to the measured values of the EW observables  and test the SM predictions. In these fits, one keeps $M_{t}, M_{W}$ and $M_{h}$ as free parameters. As already noted the radiative corrections depend on $M_{t}$ quadratically and $M_{h}$ logarithmically.  Then compare the $M_{W},M_{t}$ values so obtained with experimentally determined values of the same, thus providing a test of the SM. Afterwords one can perform the exercise by varying the Higgs mass, find the value of $M_{h}$ that minimises the $\chi^{2}$ and then find the limits on the Higgs mass for which the data will be consistent  with the predictions of the SM.

\begin{figure}[hbt]
\begin{center}
\includegraphics*[width=12cm,height=12cm]{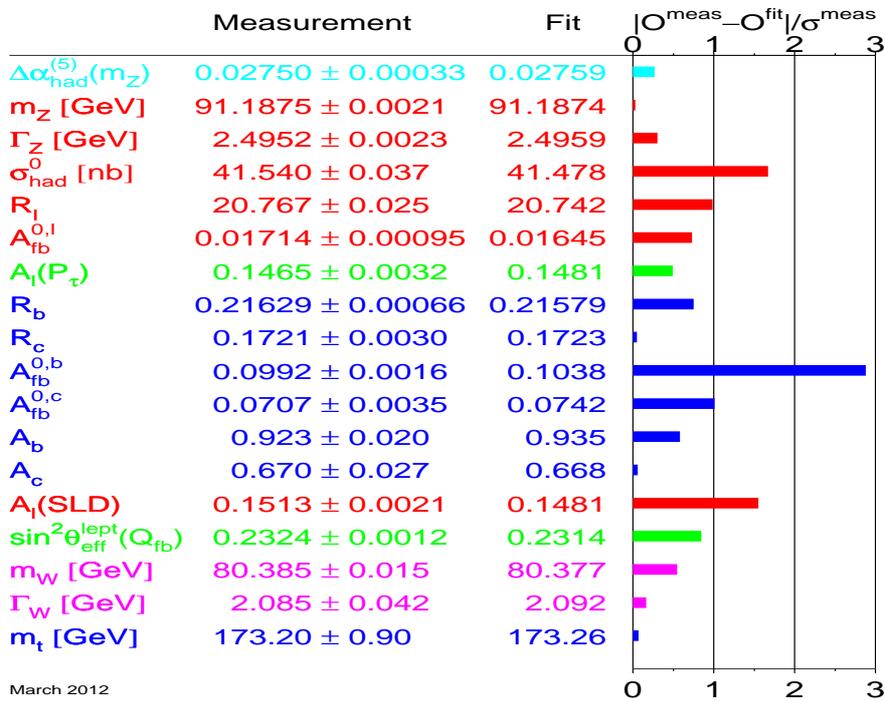}
\vspace{-2cm}
\caption{Pull for the SM fit for the totality of the EW precision observables. Taken from \protect\cite{bib:ewwg}.}
 \label{fig:pull}
\end{center}
\end{figure}

Fig.~\ref{fig:pull}, taken from the url of the LEP EW working group~\cite{bib:ewwg}, shows the result of such an exercise. The figure lists the measured values of a variety of EW obsevables, most of which we have discussed. The various $R$-ratios: $R_{b},R_{c}, R_{l}$ etc. are a measure of the relative production of the various final states and hence of the  partial decay width of the $Z$ into them. $A_{l} (P_{\tau})$ is the polarisation asymmetry for the $\tau$'s produced in $e^{+} e^{-} \rightarrow Z \rightarrow \tau^{+} \tau^{-}$ on the $Z$--pole. 
The second column shows the result of the SM fit for the observable and the third column the pull which is the difference between the measurement and the fit value normalized by the error of the measurement. The pull is less than  three for all the observables and above 2 for only one of the measurements {\it viz.} $A_{FB}^{b}$.  
This particular fit is the last one {\it \bf before} the discovery of the Higgs at the LHC, using the most accurate measurement of $M_{W}$ from the Tevatron, which has an error of $0.15 \UGeV$, again a 'one per mille' measurement. The $\chi^{2}$ of this fit is not very small, mainly due to the discrepancy between the best fit values and measured values for $A_{b}$ from LEP as well as at the SLC. Hence before the 'direct' discovery of the Higgs there were a few physicists who used to be a little uncomfortable about the goodness of the fit and accepting this as 'the proof'  for the correctness of the SM at loop level.

Note the values in the last two rows. The measured values and the best fit values of $M_{W},M_{t}$ agree with each other to a great precision and the pull is is rather small, providing thus a stringent test of the SM at loop level. This is the agreement between the $M_{t}$ predicted 'indirectly' from the LEP EW precision measurements and the 'direct' measurement from the Tevatron, that was alluded to before a few times. In fact this spectacular agreement was the QFD (Quantum Flavour Dynamics) equivalent of testing the $(g-2)_{\mu}$ prediction with the measurement in QED. The important role played by renormalisabilty and loop corrections in this context can be understood by doing a small numerical exercise of predicting $M_{W}$ from the very accurately measured values $\alpha_{em} =1/137.0359895(61),~~ G_{\mu} =1.16637(1)\times 10^{-5} \UGeV^{-2},~~ M_{Z}=91.1875 \pm 0.0021 \UGeV$ and the tree level relations given by the SM among these quantities and $M_{W}$. Notice that Eq.~\ref{eq:mwsinthw} can be written as, 
\[
\frac {G_{\mu}}{\sqrt{2}} = \frac{g_2^2}{8M_W^2} = \frac{\pi \alpha_{\rm em}} {2 M_W^2 (1 -M_{W}^2/M_{Z}^{2})}
\]
by using the tree level relation {$\displaystyle M_{Z} = \frac{M_{W}}{\cos \theta_{W}}$}. This gives,
$M_{W}^{tree} =  80.939 \UGeV$. Compare this now with the value of $M_{W}$ given in the second column of Fig.~\ref{fig:pull}, 
$M_{W}^{expt} = 80.385 \pm 0.015 \UGeV$. Of course, this points out the need for calculating loop corrections to the tree level relations. Renormalisability guarantees that all the corrections are finite and can be computed. Hence the value of $M_{W}$ obtained 'indirectly' from the fits using theoretical predictions which  {\it include} these loop corrections, then famously agrees with the 'direct' measurement as shown in Fig.~\ref{fig:pull}.
Agreement with the SM prediction would have been impossible unless 
the predicted values included higher order corrections calculated in perturbation theory.

The fit values and the pull for $M_{t}, M_{W}$ depends on the value of $M_{h}$, albeit very weakly, due to the logarithmic dependence on $M_{h}$ of the EW corrections to $M_{W},M_{Z}$ etc.
\begin{figure}[hbt]
\begin{center}
\includegraphics*[width=7cm,height=6cm]{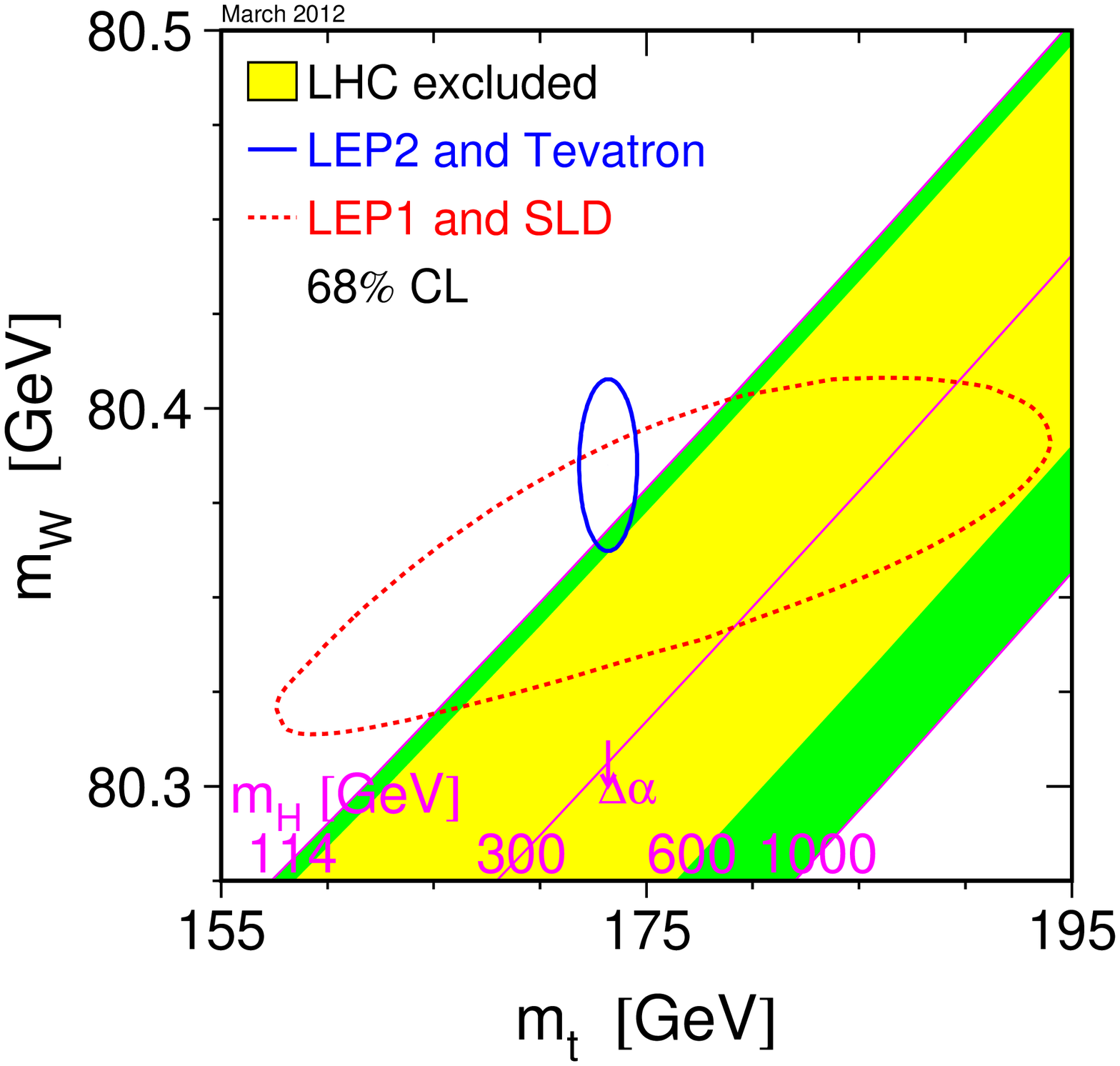}
\includegraphics*[width=7cm,height=6cm]{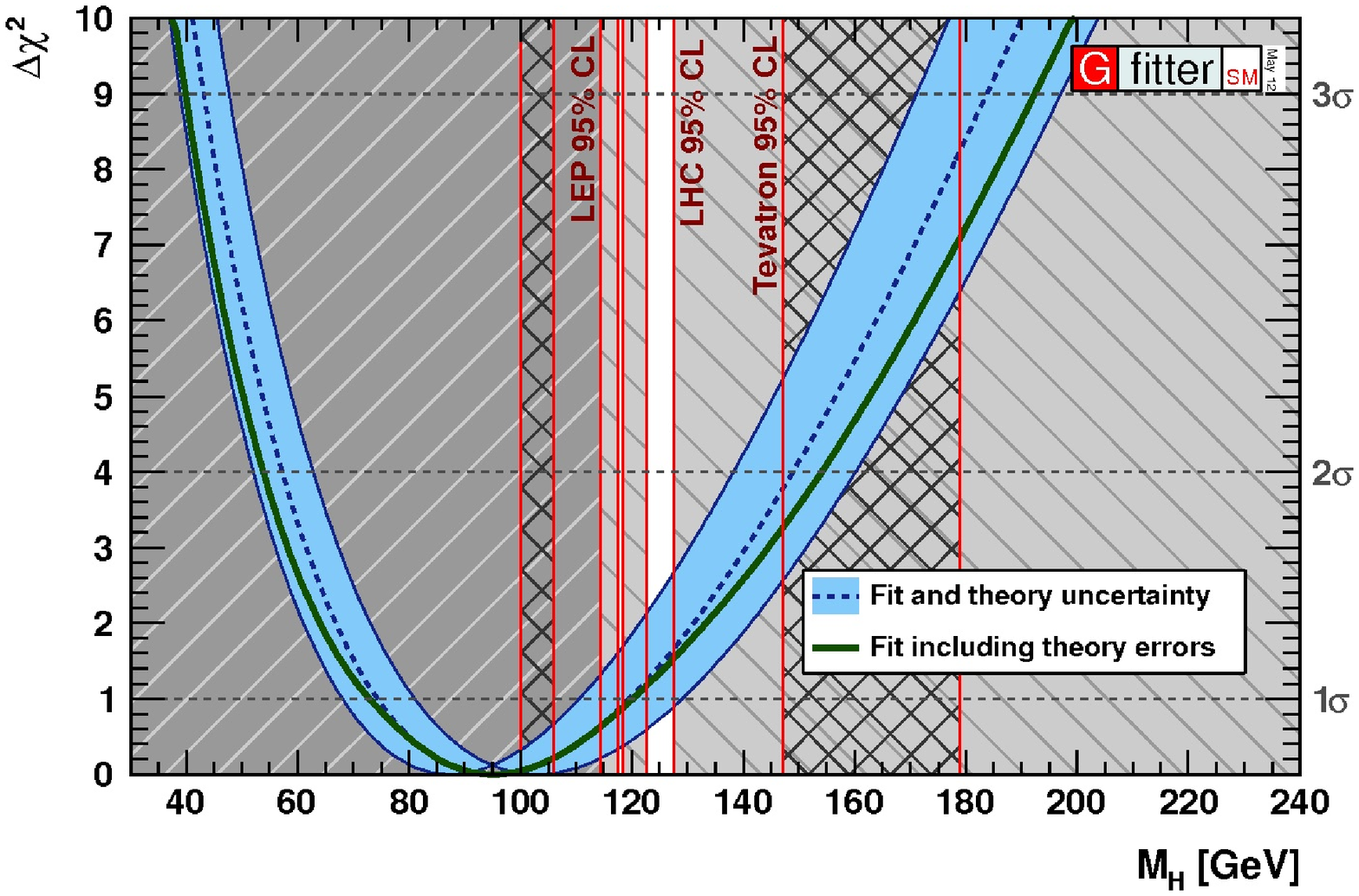}
\caption{Left panel shows the dependence on $M_{h}$ of the $M_{W}$--$M_{t}$ values obtained from the EW precision data. Taken from \protect\cite{bib:ewwg}. The right panel shows the status of the  'indirect' limits on $M_{h}$ obtained by fits to the EW precision data. This is taken from \protect\cite{bib:gfitter}. Both these are from the eve of the Higgs discovery,~~March 2012.}
 \label{fig:mwmt}
\end{center}
\end{figure}
Some of these effects can be seen from the two panels in Fig.~\ref{fig:mwmt}. The  plot in the left panel shows the dependence of the fit values for $M_{W},M_{t}$ for different values of $M_{h}$.  The long lopsided ellipse used the EW observables measured at LEP-I and the SLC, to determine allowed regions in the $M_{t}$--$M_{W}$ plane at $95\%$ c.l. Using the $M_{W}$ measurements at the LEP-II/Tevatron as input,  one now obtains the small blue ellipse which is consistent with the precision measurements. The dark green (grey) region and the large red ellipse show  that with results from LEP-I alone, the measurements were not sensitive to $M_{h}$ at all. On the other hand, the highly accurate LEP-II/Teavtron measurements of $M_{W}$ and the Tevatron measurement of $M_{t}$ is consistent with somewhat small values of the Higgs mass at the left most boundary of the green(grey) region. This was also consistent with the exclusion (from direct searches at the LHC) of a SM Higgs over a very large range as indicated by the $M_{h}$ values labeling the inclined lines in the region shaded in yellow (a shade of lighter gray). 

The right panel shows the same information in a  different format, where we show a plot of $\Delta \chi^{2}$ as a function of $M_{h}$. In fact the fact that this minimum of $\Delta \chi^{2}$ occurs at a nonzero, finite mass $M_{h}$ is already an indication of the 'existence' of the Higgs and hence a feather in the cap of the SM.  The dotted and solid black lines are the best fit with and without including the theory errors. The  region shaded in light blue (grey) indicates effect of the theoretical uncertainties as well as uncertainties  in the EW fit.
In the absence of any information from 'direct' searches for the Higgs, the indirect constraints will allow a region around the minimum of $\chi^{2}$ ($M_{h} \simeq 90-100 \UGeV$) upto $M_{h}$ values where $\Delta \chi^{2}$ is $9$: the $3 \sigma$ value.
Remaining values of $M_{h}$ will be disfavored by this 'indirect' search. The $\Delta \chi^{2} \le 9$ corresponds to an allowed mass range $40-45 \LTS\ M_{h} \LTS\ 180-200 \UGeV$ at $3 \sigma$.  However a lot of this 'allowed' region is ruled out from direct searches at the LEP, at the Tevatron and at the LHC. These bounds are indicated by the vertical red lines in this figure. The region ruled out by LEP is indicated by the dark grey region hatched with slanted lines. The region ruled out by the hadronic collider Tevatron is indicated by the cross-hatched region.  The above mentioned red lines mark the edges of these regions giving us the pre-LHC exclusion. The region excluded by the LHC in March 2012 is indicated by light grey region marked by lines slanted in a direction opposite to the LEP exclusion region.  

As one can see from this figure, before the LHC direct search constraints, the allowed mass range for the Higgs was $115 \le M_{h} \LTS 150-160$ and $180 \LTS M_{h} \LTS 200 \UGeV$. The LHC experiments ruled out existence of an SM Higgs in a major part of this range. As a result in March 2012, the mass value allowed for a SM Higgs by a combination of the EW precision measurements and 'direct' collider constraints was as indicated by the  small white slit around $125 \UGeV$. Failure to find a Higgs in this small 'allowed' mass range would then have meant the death for the SM.  Indeed a new boson  was found  with properties very similar to a SM Higgs in precisely this mass range. This discussion should make it very clear to us that the value of the mass of the observed Higgs boson itself tested the SM at loop level to a very great accuracy.

In fact it won't be out of place to recapitulate at this point how the SM was validated and tested at various levels by discovery of new particles whose masses were predicted : either in terms of a free parameter of the model which could be determined from experiments OR `indirectly' by comparing loop effects on physical observables with their precision measurement.

\begin{itemize}
\item Observation of suppression  of FCNC implied that the quarks must come in isospin doublets. Thus charm was predicted since the existence of the $s$ quark was known and top was {\it predicted} to be  present once the $b$ was found. Further, the very demand of cancellation of anomalies so as avoid these spoiling the   renormalisability,  implied  existence of third generation of quarks {\it AND} leptons  once the $\tau$ was found.

\item One could get indirect information on $M_c,M_t$ from flavour changing  neutral current processes induced by loops. Agreement of this 'indirect' information with 'direct' measurements 'proved' the correctness of description of EW interactions in terms of a gauge theory.

\item {\cal CP} violation in meson systems could  be explained in 
terms of the SM parameters and measured CKM mixing in quark sector
{\it only if} three generations of quarks exist.

\item $M_W,M_Z$ was predicted in terms of $\sin \theta_W$  and direct observation of the $W,Z$ at the predicted mass tested the particle content and tree level coupling of the matter fermions with the gauge bosons $W,Z$.

\item Study of energy dependence of the $e^{+} e^{-} \rightarrow W^{+} W^{-}$ process gave {\it direct} evidence for the tree level $ZWW$ coupling and also for the role played by this vertex  in taming the bad high energy behaviour of the cross-section. So in that sense, Fig.~\ref{fig:lepeeww} gives evidence for the gauge symmetry ($ZWW$ coupling as indicated by symmetry) and the symmetry  breaking (nonzero $W$ mass) as well.  

\item  Further, Teavtron found evidence for 'direct' production of the top quark at the mass $M_t$ which was in ageeement with the value obtained `indirectly' from precision measurement of  $M_W,M_Z$, considering effect of radiative corrections to these masses.

\item Last but not the least the existence of a minimum of $\Delta \chi^{2}$ at a finite nonzero mass for the SM fits to the EW precision measurements, gave an 'indirect' proof of the existence of the Higgs. Before the 'direct' discovery of the Higgs this was also an 'indirect' probe  of  the couplings of the Higgs with gauge bosons and the $t$ quarks.  Further, the same fits gave an 'indirect' determination of $M_{h}$ which now agrees completely with the measured mass of the observed Higgs.
\end{itemize} 

Now we can turn once again to the discussion of Fig.~\ref{fig:fig2}. As was already indicated by the right panel of  Fig.~\ref{fig:mwmt}, the 'directly' measured  value of the Higgs mass $M_{h} = 125.09 \pm 0.24 \UGeV$ is right in the 'allowed' white slit  and indeed confirms the SM at loop level most spectacularly. At this point, it is worth noting that if we improve upon the accuracy of measurements of $M_{t},M_{w}$ and $M_{h}$ we can indeed hope to look for effects by loops of heavy  particles which are not present in the SM but are expected to exist in various extensions of the SM, which are  in turn postulated to address various shortcomings of the SM! 

As already mentioned, the  Higgs mass range allowed by the EW precision measurements can  change when  one goes away from the SM. In fact before the 'direct' discovery of the Higgs,
a lot of effort had gone on, in constructing models which would allow one to avoid these constraints, should experiments reveal a Higgs boson not consistent with the bounds from the EW precision measurements. Of course, not only that many of these are not required, but some are now even ruled out, by the observation of  the light state. An example of one such model is the SM  with a fourth sequential generation of fermions, leptons and quarks. Since in the SM there is no guiding principle for total number of generations of fermions, except that they should be the same for quarks and leptons, this in principle is the simplest extension of the SM by addition of more matter particles to it. Observation of the low mass $\sim 125 \UGeV$  scalar ruled out this extension  very conclusively.

\section{Observed mass of  Higgs and the SM}
As we saw above the EW precision measurements did put 'indirect' bounds on the Higgs mass. However, theoretically there is no information on the mass of the Higgs in the SM, as it is determined by $\lambda$ an arbitrary parameter. Recall  $M_{h}$ and $\lambda$ are related by $M_{h}^{2} = 2 \lambda v^{2}$ . The observed mass of the Higgs determines the self coupling $\lambda$:
\[
\lambda = 0.5 M_{h}^{2}/v^{2}  \simeq 0.13
\]
This is the last free parameter of the SM that needed to be determined. Thus the only part of the scalar potential now that needs to be experimentally verified 'directly' is the triple Higgs and the quartic Higgs coupling in Eq.~\ref{eq:LhV}. Now that one 'knows' the value of $\lambda$ one can assess the possibilities of measuring it at current and future colliders.
One might ask the question whether this is the only nontrivial information about the SM that we can extract from the observed value of the mass of the Higgs. Asked differently, can one use this observed value of $M_{h}$ to infer something about the SM as well as the physics beyond the SM, {\it viz.} the BSM.  Since in these lectures we restrict ourselves to the SM, I will only talk about the possible implication of the observed Higgs mass for the SM itself.

While the SM has no 'prediction' for $M_{h}$, requirement of theoretical consistencies imply bounds on the same. These  theoretical limits on the mass of the Higgs boson come from  demanding good high energy behavior of scattering amplitudes in the $SU(2)_{L} \times U(1)_{Y}$ gauge theory and from the quantum corrections that the self coupling $\lambda$ of Eq.~\ref{eq:Lscalar} receives.  These limits are thus essentially an artifact of the quantum field theoretical description. Let us discuss this one by one. 

\subsection{Unitarity bound}
\begin{figure}[hbt]
\begin{center}
\includegraphics*[width=12cm,height=6cm]{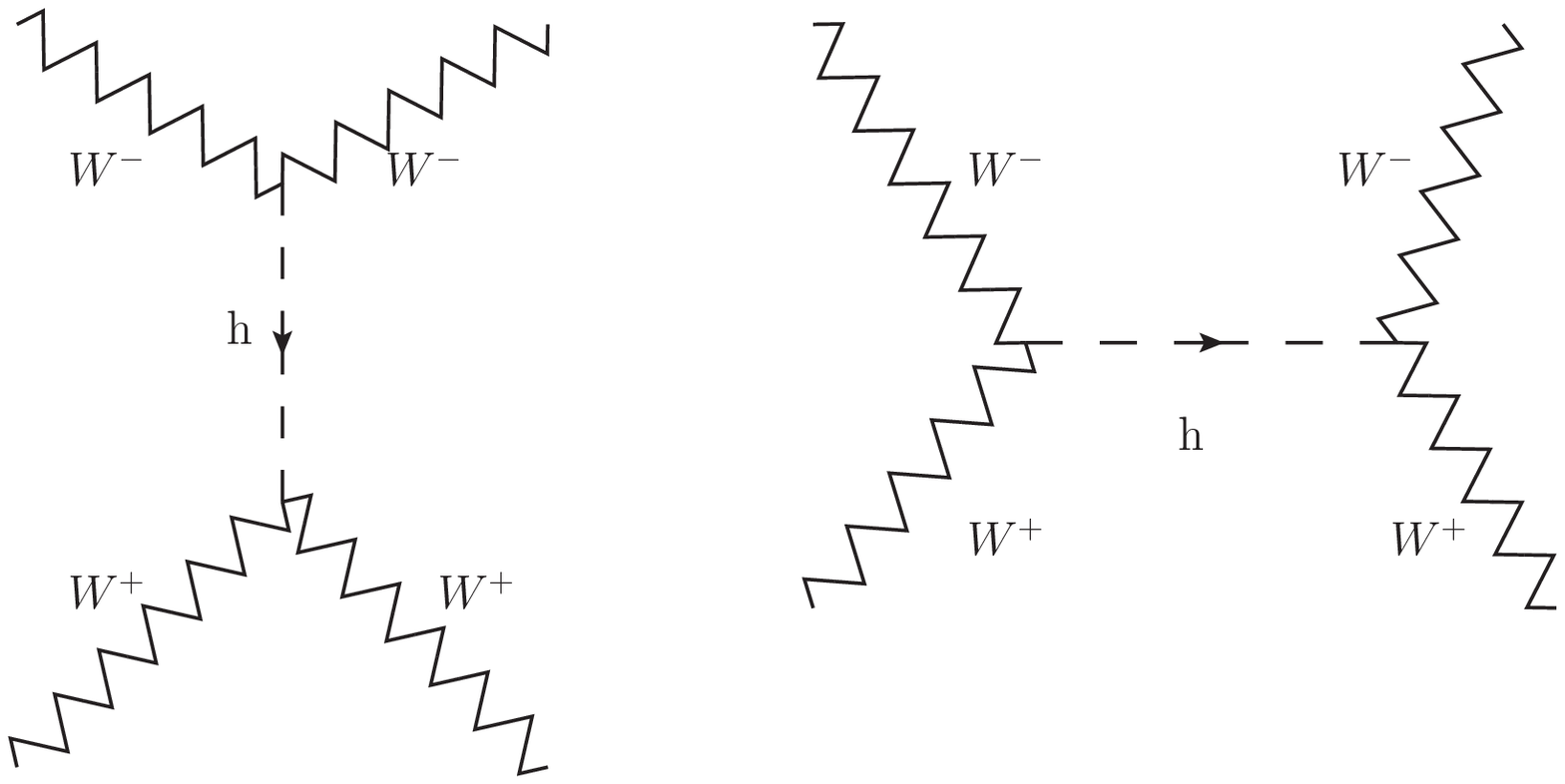}
\includegraphics*[width=12cm,height=6cm]{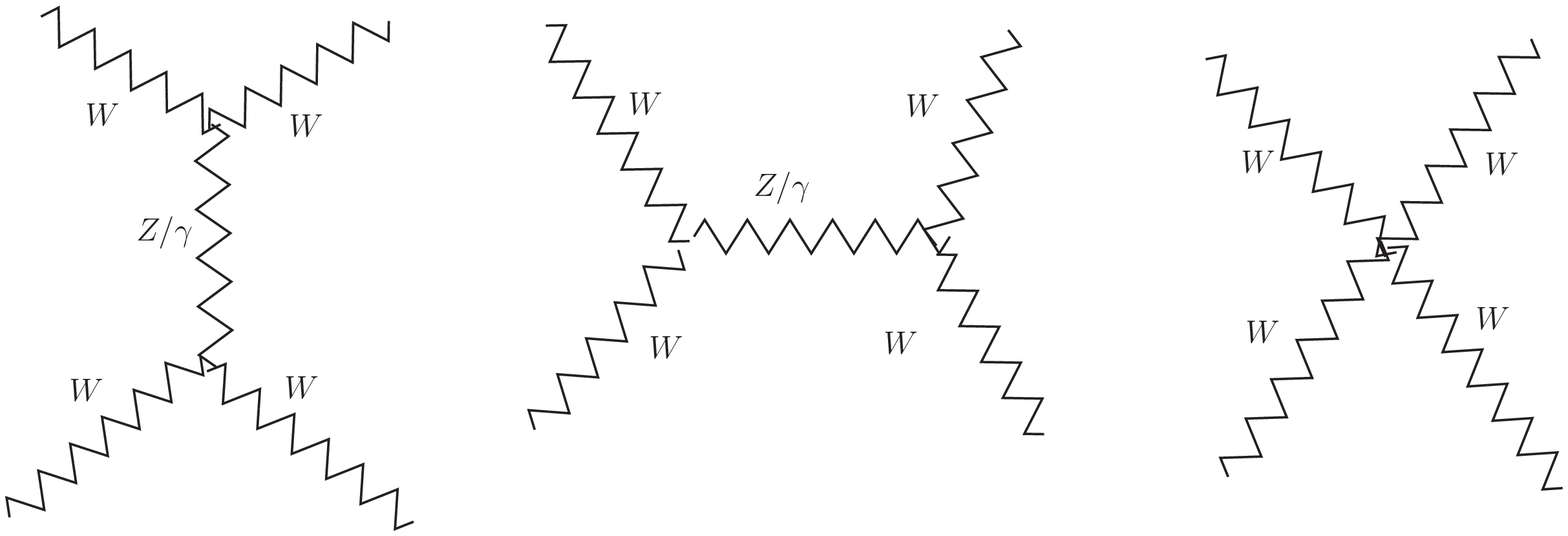}
\caption{The upper panel shows digrams involving $h$ bosons contributing to $W + W \rightarrow W W$ scattering. The $s$-channel diagram will of course contribute only for $W^{+} W^{-} \rightarrow W^{+} W^{-}$ scattering. The lower panel shows the all the diagrams which involve exchange of the gauge bosons $Z$ and $\gamma$ as well as the one involving pure gauge vertex.}
\label{fig:wwww}
\end{center}
\end{figure}

Recall our discussion in section~\ref{sec:HE} of the high energy behaviour of scattering amplitudes. We discussed therein the high energy behavior of the scattering amplitude $W^{+} W^{-} \rightarrow W^{+} W^{-}$. Various contributing diagrams are shown in Fig.~\ref{fig:wwww}. 
Each of these diagrams gives a contribution which grows as $s^{\alpha}$ with $\alpha = 1,2 $ where $s$ is the centre of mass energy of the $WW$. This divergence appears in the scattering of longitudinal $W$'s.  However in the SM  all the divergent terms in the $W W \rightarrow W W $ amplitude cancel  among each other after adding the contributions of all the diagrams shown in Fig.~\ref{fig:wwww}. The contribution of the $h$ exchange diagrams as well as the that from the diagrams with pure gauge vertices play an essential role in this cancellation as mentioned before. The cancellation  of the power divergences is independent of the Higgs mass and thus the requirement of non-divergent behavior does not single out any scale. Among the non divergent part of the amplitude ${\cal A}(WW \rightarrow WW)$, left over after all this cancellations, the  contributions of the Higgs exchange diagrams shown in the top panel of Fig.~\ref{fig:wwww} dominate and are dependent on the Higgs mass. These  were investigated in ~\cite{bib:LeeThacker} and they showed that though not divergent these can become non negligible for large values of $M_{h}$. The non-divergent part of this  invariant amplitude can be written as~\cite{bib:LeeThacker}
\[
{\cal A} (W_{L}^{+}W_{L}^{-} \rightarrow W_{L}^{+}W_{L}^{-}) = -\sqrt{2} G_{\mu} M_{h}^{2} \left ({\frac {s}{s-M_{h}^{2}}} + \frac{t}{t-M_{h}^{2}} \right).
\] 
From a partial wave analysis of this amplitude one can show that this amplitude will violate tree level unitarity if
\[
M_{h} > \left({\frac{8 \pi \sqrt{2}}{3 G_{\mu}}} \right)^{1/2} \sim  1000 \UGeV .
\]
Thus, the theory will be strongly interacting if $M_{h}$ were to exceed this value. As things stand, the observed value of $M_{h}$ implies $\lambda \simeq 0.13$, far from the strongly interacting region and also safe from any unitarity violation. Thus the observed mass of the Higgs boson satisfies the unitarity bound.

\subsection{Triviality and Stability bound}
\begin{figure}[hbt]
\begin{center}
\includegraphics*[width=14cm,height=4cm]{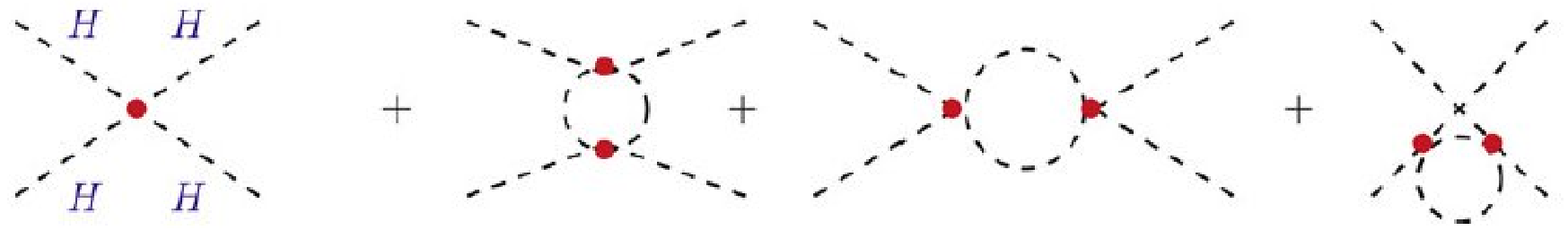}
\includegraphics*[width=14cm,height=4cm]{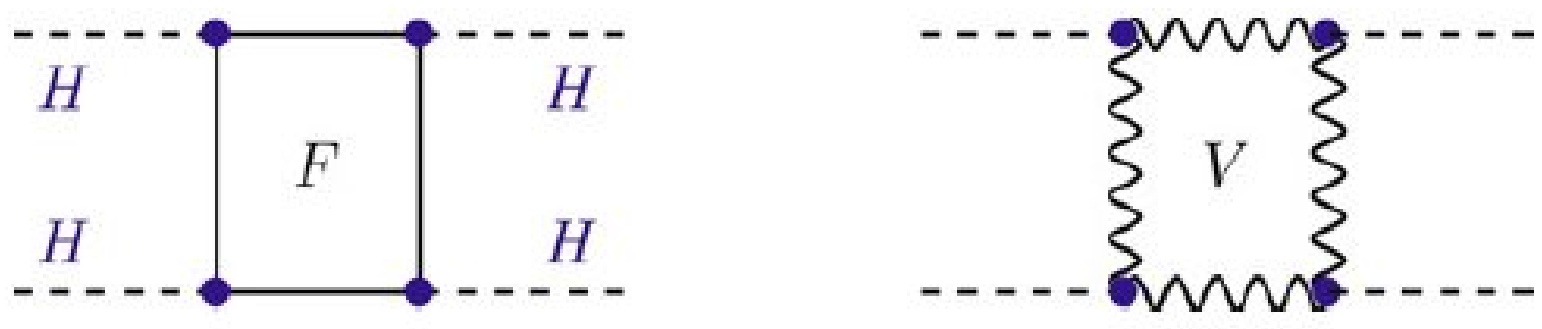}
\caption{The top panel shows loop corrections to the quartic coupling $\lambda$ from the Higgs sector itself. The diagrams in the lower panel show contributions to the running of $\lambda$ from fermion and gauge loops.}
\label{fig:trivial-stable}
\end{center}
\end{figure}

Effect of loop corrections to the self coupling $\lambda$ in a scalar field theory, in the presence of a high scale and additional interactions of the scalar with gauge bosons and matter, was first studied decades ago~\cite{bib:early-vacstab} with an aim to examine whether one could constrain the scalar mass and other high scale masses from pure theoretical considerations. 
Triviality bound results from considering loop corrections to the scalar potential in Eq.~\ref{eq:LhV}. One demands that the quartic coupling $\lambda$ in the  Higgs potential from Eq.~\ref{eq:LhV} reproduced below, 
\[
V_{h}= \lambda v h^{3} + \lambda/4 h^{4},
\]
remains perturbative as well as positive at all energy scales under loop corrections. The corrections come from two sets of diagrams shown somewhat schematically in Fig.~\ref{fig:trivial-stable}. 
The top panel shows loop corrections to the quartic coupling $\lambda$ from the Higgs sector itself whereas the diagrams in the lower panel show contributions to the running of $\lambda$ from fermion and gauge loops. So the diagrams shown in the top panel are applicable to any scalar with quartic self interaction. The ones in the lower panel are specific to a gauge theory. 

\subsubsection{Triviality Bound}
The triviality bound comes from demanding that $\lambda$ should always remain perturbative. To understand the origin of this bound let us consider the case of large $M_{h}$.
Since $M^2_h = \lambda v^2$, at large $m_h$ and hence large $\lambda$,  loop corrections are dominated by the $h$--loops shown in the top panel of Fig.~\ref{fig:trivial-stable}. A straightforward evaluation of this gives us
\begin{equation}
{\frac{d \lambda(Q^2)}{d\log Q^2}}  = {\frac{3}{4 \pi}} \lambda^2(Q^2) 
\end{equation}
Solving this,  one gets
\begin{equation}
\lambda(Q^2) = {\frac{\lambda(v^2)}{[1 - {3\over{4 \pi^2}} \lambda(v^2) \log ({Q^2\over v^2})]}}.
\label{eq:trivial}
\end{equation}
A look at Eq.~\ref{eq:trivial} shows us that at large  $Q^2 \gg v^2$,  $\lambda(Q^2)$ can develops a pole, the so called Landau pole, at some high scale $Q$ depending on the value $\lambda$ at the EW scale $v$.  If we demand that  $\lambda$ remains always in perturbative regime, then the ONLY solution would be $\lambda =0$. This would then mean that the theory will be trivial. That of course does not make for a sensible theory. Thus the starting  value of $\lambda(v)$ and hence $M_{h}$ is not allowed by these considerations.

One can understand this in yet another way.
If we demand that the scale at which $\lambda$ blows up is above a given scale $\Lambda$, then using Eq.~\ref{eq:trivial} we find that for a given value of $M_{h}$ and hence $\lambda(v)$, the scale at  which the Landau pole lies 
will be given by
\begin{equation}
\Lambda_C = v \; \exp \left({{2 \pi^2}\over {3 \lambda}}\right) = v \exp \left({{4 \pi^2 v^2}\over{3 M_h^2}}\right).
\end{equation}
Thus, for example, using $\Lambda_{C} = \Lambda = 10^{16}$ GeV, we will find $M_h~~ \LTS\/~~ 200$ GeV. 

This bound is called the triviality bound. In simple terms it means that the value of $\lambda$ at the EW scale (and hence the mass $M_{h}$) should be small enough so that $\lambda(Q^{2)}$ does not develop a pole up to a scale $Q = \Lambda_{C}$. Hence, if $M_{h}$ were found to have a mass larger than the triviality bound, it would have meant existence of new physics  below the scale $\Lambda_{C}$. This thus tells us that just the mass of the $h$ can give us an indication about the scale at which SM must be complemented by additional new physics. The  mass of the Higgs being only $125.09 \UGeV$ this is rather an academic discussion as this small value of the coupling $\lambda$  at the EW scale, implies that the loop effects will not be driving the self coupling $\lambda$ toward the Landau pole at an energy scale of interest. There are other issues that we need to address given that the observed mass is so small. But we will not discuss them here.

\subsubsection{Stability bound}

When $M_h$ is small and $\lambda$ is not large, the fermion/gauge boson loops  are important. Even more important is that the fermions loops come with a negative sign. This means that if the fermion mass is large enough the loop corrections may drive $\lambda$ negative at some scale, unless the starting value of $\lambda(v)$ is large enough. These considerations will imply a lower bound for $\lambda(v)$ and hence for $M_{h}$. This limit on $M_{h}$ is called the vacuum stability bound. Now one works in the limit of small $\lambda$, opposite to the one used when considering the triviality bound. Hence the contribution of the $h$-loops shown in the upper panel of Fig.~\ref{fig:trivial-stable}
can be neglected. Hence the equation for energy dependence of $\lambda$ now can be written as:
\begin{eqnarray}
{\frac{d \lambda(Q^2)}{d \log(Q^2)}} \simeq  {1 \over 16 \pi^2} 
[12 \lambda^2 + 6 \lambda f_t^2 {\bf - 3 f_t^4 - {3\over 2} \lambda  (3 g_2^2 + g_1^2)} \nonumber \\
 + {3 \over 16} (2 g_2^4 + (g_2^2 + g_1^2)^2) ]
 \label{eq:gaugetop}
 \end{eqnarray}
${\bf \displaystyle f_t = \frac{\sqrt{2} M_{t}}{v}}$ is the Yukawa coupling for the top. Since $M_{t} \sim 173 \UGeV$ and $v \simeq 246 \UGeV$, one can see that the Yukawa coupling is $\simeq 1$.  Thus it will dominate the scale dependence of $\lambda$. 
At small $M_h$ and hence small  $\lambda(v)$, $\lambda$ can turn negative at some value of $Q$.  Recall the Higgs potential. A  negative value of $\lambda$  will mean an unbounded potential and clearly the vacuum will be unstable. The condition for non negativity of $\lambda$ and hence vacuum stability, is
\begin{equation}
M_h^2 > {v^2 \over 8 \pi^2} \log(Q^2/v^2) \left[ 12 {m_t^2}/ {v^4} - 
{3 \over 16} (2 g_2^4 +(g_2^2 +g_1^2)^2)\right] .
\label{eq:stab-bound}
\end{equation}
Again, depending upon  the scale up to which we demand the potential to be positive definite, we find that the starting value $\lambda(v)$ (and hence $M_{h}$) has to be above a critical value dependent on the scale. If we demand that the $\lambda(Q)$ is positive up to $\Lambda_C$ we then get a  lower bound on $M_{h}$. For example choosing,
$\Lambda_C = 10^3 GeV$ we get  $M_h ~~\GTS~~ 70$ GeV. This bound is called the stability bound. 

\begin{figure}[htb]
\begin{center}
\includegraphics*[width=8cm,height=6cm]{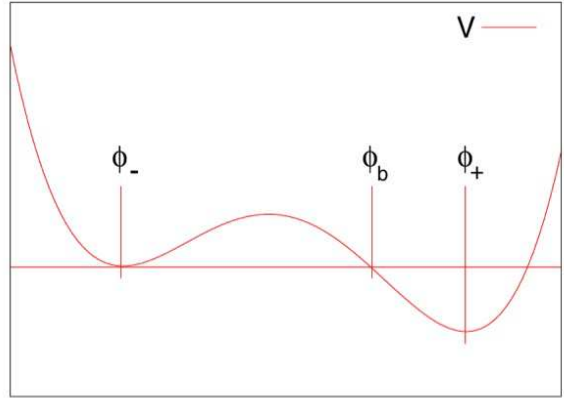}
\caption{Cartoon of a field configuration that would give rise to metastable vacuum.}
\label{fig:metastab}
\end{center}
\end{figure}

In the above analysis we have demanded that $\lambda(\Lambda)$ does not become negative so that the potential is stable. This is the condition for absolute stability of vacuum. However, 
Planck scale dynamics might stabilise the vacuum for $|\Phi| >> {v}$ and we might be living in a metastable vacuum which has a life time bigger than that of the Universe. The cartoon shown in 
Fig.~\ref{fig:metastab}
indicates such a situation. One can then obtain lower bounds on $M_{h}$ demanding that vacuum is metastable with a life time bigger than the life time of the Universe. Clearly evaluation of these bounds can not be presented in the simplistic analysis that we have given here.

\begin{figure}[htb]
\begin{center}
\includegraphics*[width=10cm,height=8cm]{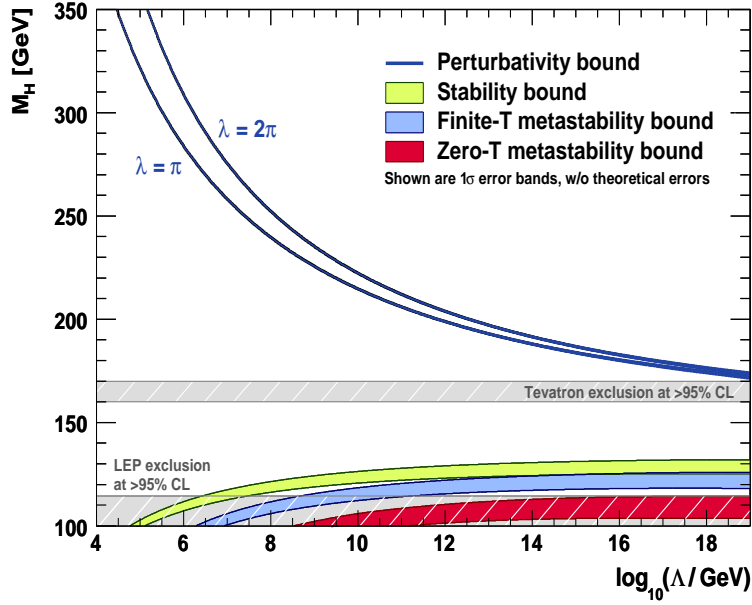}
\caption{The vacuum stability bound on $M_{h}$ as a function of the scale. Bounds are shown for absolute stability as well as
metastability. Taken from ~\protect\cite{bib:ellis}.}
\label{fig:vacstab} 
\end{center}
\end{figure}

A complete and sophisticated analysis of Ref.~\cite{bib:ellis} in fact gives the vacuum stability bounds on the Higgs mass taking into account the effect of renormalisation group evolution(RGE) as well as that of metastability of the vacuum. 
Fig.~\ref{fig:vacstab} taken from Ref.~\cite{bib:ellis} shows the stability bounds, indicated by the pale yellow green area, as  a function of scale at which the instability sets in. The spread is due to the theoretical uncertainties, major ones being the top mass uncertainty and the missing higher order contributions to the equations. RGE takes into account not just the one loop corrections shown in Fig.~\ref{fig:trivial-stable} but also includes the resummation of leading logarithmic corrections.  
As one can can see even from the simple minded analysis presented here, the bound depends critically on the value of $f_{t}$ and hence on $M_{t}$. If one overlays the bounds on the Higgs mass of Fig.~\ref{fig:mwmt} obtained 'indirectly' from the EW precision analysis as well as the LEP/Teavtron/LHC searches then we realise that the thin white silver which was still allowed by March 2013 corresponds to the boundary of the pale yellow-green region indicating the stability bound. Due to the finite width of these bands caused by various uncertainties mentioned above, the observed mass of the Higgs $M_{h}$ may or may not be consistent with the hypothesis that the SM remains consistent  all the way  to Planck scale. Given that everything depends logarithmically on different scales and with the high accuracy of the experimental measurement of $M_{h}$, the need to do the evolution of $\lambda$ taking into account higher order effects is thus clear.

In fact the need for more accurate calculation was already apparent, even before the Higgs discovery,  with the rather low values of $M_{h}$ indicated by the 'indirect' limits. To appreciate this, look at Fig.~\ref{fig:mwmt} again disregarding the vertical red lines corresponding to the LHC $95\%$ bound, which delineate the pale grey region hatched with inclined lines. The $3 \sigma$ region around the minimum of $\Delta \chi^{2}$ and hence preferred by the EW precision data, allowed by Tevatron data, $115 \leq M_{h} \leq 150 \UGeV$, covers the range of masses where the stability bound is operative and the upper limits on the possible scale of new physics indicated by the vacuum (in)stability interesting. The need for accuracy in the theoretical prediction of stability bound is thus very apparent.  In May 2012, with the discovery of the Higgs imminent, an NNLO analysis of the problem became available~\cite{bib:giudice}, which reduced the theoretical error on the bounds coming from the unknown higher order corrections  to $\sim 1 \UGeV$.
\begin{figure}[hbt]
\begin{center}
\includegraphics*[width=8cm,height=6cm]{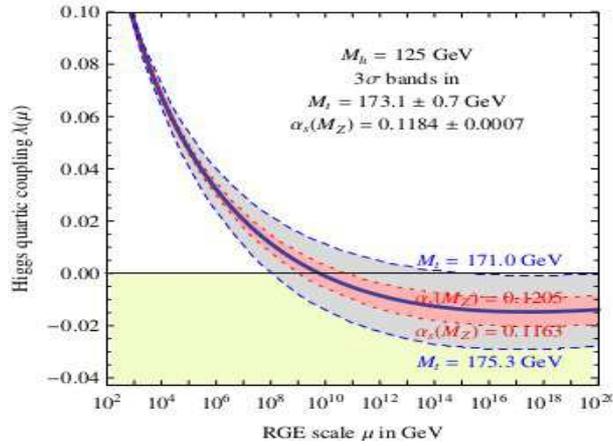}
\caption{$\lambda(\mu)$ as a function of scale for different values of $\alpha_{s}, M_{t}$ varied within the experimental errors. The plot is taken from~\protect\cite{bib:giudice}.}
\label{fig:h125}
\end{center}
\end{figure}  

However, there still remains a sizable error due to the errors in experimentally determined parameters $M_{t}, \alpha_{s}$. Fig.~\ref{fig:h125}, taken from \cite{bib:giudice}, shows behavior of $\lambda(\mu)$ as a function of the energy scale $\mu$. One now sees clearly that the scale at which $\lambda$ becomes zero and hence the vacuum unstable, depends critically on $M_{t}$ and 
the strong coupling $\alpha_{s}$. For example, for the central value of $M_{t}$ used, $\mu$ value at which $\lambda$ becomes zero changes by at least an order of magnitude as $\alpha_{s}$ is varied within errors. The dependence on $M_{t}$ is even stronger. We will comment later on the range of $M_{t}$ used in this analysis.
According to this analysis the absolute stability of the vacuum
up to Planck scale $M_{pl}$ is guaranteed for,
\begin{equation}
M_h~[{\UGeV}]
> 129.4 + 1.4 \left( \frac{M_t~[{\rm GeV}] -173.1}{0.7} \right)
-0.5\left( \frac{\alpha_s(M_Z)-0.1184}{0.0007}\right) \pm 1.0_{\rm ~th} .
\label{eq:mhbound}
\end{equation}
In this analysis the error on  pole mass of the top was taken to be $\Delta m_t = \pm 0.7 \UGeV$. Taking into account the errors, Eq.~\ref{eq:mhbound} then means that  for $m_h < 126 \UGeV$,  vacuum stability of the SM all the way to Planck Scale is excluded at $98 \%$ c.l. Clearly, this value is far too close to the observed value of $125.09 \pm 0.24 \UGeV$ to require careful considerations of various issues before we draw conclusions about the validity of the SM at high scale. For the measured value of the Higgs mass, the exact scale where $\lambda$ crosses zero, though not $M_{pl}$ seems close to it and depends entirely on the exact value of $M_{t}$ and $M_h$. Indeed these considerations may be relevant for consideration of BSM or models of inflation etc.

\begin{figure}[hbt]
\includegraphics*[width=17cm,height=8cm]{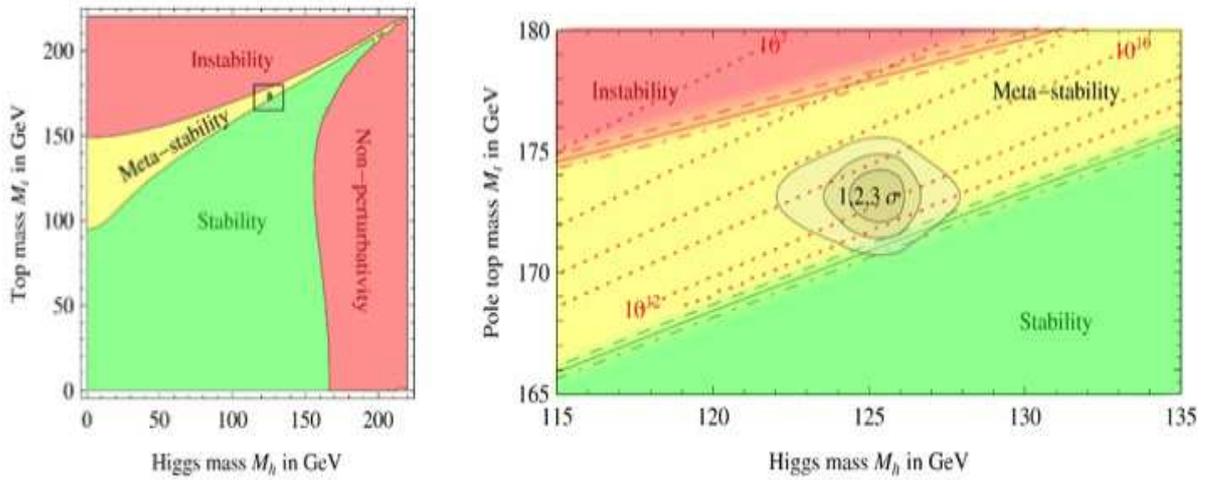}
\caption{The left  panel shows the regions in the $M_{t}$--$M_{h}$ plane where the vacuum is absolutely stable, metastable and unstable. Right panel shows the zoom-in the region of values preferred experimentally. The grey areas show allowed regions at 1,2 and 3 $\sigma$. The three curves on the boundary of two regions correspond to three values of $\alpha_{s}$. Superimposed on it are the contours of constant value of the high scale where the instability occurs. The plot is taken from \protect\cite{bib:giudice}}
\label{fig:mtmh}
\end{figure}

The same can be seen clearly from Fig.~\ref{fig:mtmh} taken from~\cite{bib:giudice}. This shows the results of this NNLO analysis
of the region in $M_{h}$--$M_{t}$ plane from the vacuum stability considerations.  The left  panel shows the regions in the $M_{t}$--$M_{h}$ plane where the vacuum is absolutely stable, metastable and unstable. To understand the role and size of various 'experimental' uncertainties the right panel shows a zoom in of the region around the experimentally determined $M_{h}$--$M_{t}$ values. The grey areas show allowed regions at 1,2 and 3 $\sigma$. The three curves on the boundary of two regions correspond to three values of $\alpha_{s}$. Superimposed on it are the contours of constant value of the high scale where the instability occurs. We see that the experimentally determined values lie right on the boundary of the stable/metastable region. The answer to the question as to whether or not, the experimentally determined value of $M_{h}$ (known now  to a high accuracy $M_{h} = 125.09 \pm 0.24 \UGeV$) is consistent with SM vacuum being (meta)stable all the way to Planck scale, very much depends on $M_{t}$ values.

Let us discuss this issue in a little more detail. The stability bounds given in \cite{bib:giudice} used errors on $m_{t}$ as measured at the hadronic colliders the Tevatron and the LHC. This is the so called Monte Carlo or kinematic mass, which is a parameter in the Monte Carlos used while analysing the data and studying the top quark production at the colliders. Conversion of this parameter into the pole mass, which is the parameter required in these theoretical considerations and for the RGE, has uncertainties coming from hadronisation and fragmentation models, underlying event etc. These are typically non perturbative in  character. Another way to extract the pole mass in a well defined manner is to extract $M_{t}^{\overline {MS}}$, the mass of the top quark in the $\overline {MS}$ scheme from the measurement of the top quark cross-sections at the Tevatron  and the NNLO calculation of the same. The procedure to convert this mass to the pole mass $M_{t}(M_{t})$, leads to uncertainties in $M_{t}$ larger than the $0.7 \UGeV$ taken in Eq.~\ref{eq:mhbound}. This exercise, using the available information in 2012 led to an
estimate of the pole mass for the top~\cite{bib:abdel-sven}:
\[
M_t^{pole} = 173.3 \pm 2.8 \UGeV. 
\]
Compare this with the error of $0.7 \UGeV$  that was used in the estimate obtained in~\cite{bib:giudice}. The vacuum stability constraint now becomes $M_h > 129.4 \pm  5.6 \UGeV$ instead of the one in Eq.~\ref{eq:mhbound}. This observation then can weaken the conclusion about the high scale upto which the SM remains valid without getting into conflict with stability. The future International Linear Collider(ILC) can measure the top mass $M_{t}$ to a high accuracy of $100 \UMeV$. What is more important is the fact 
\begin{figure}[hbt]
\begin{center}
\includegraphics[height=8.0cm,width=12.0cm]{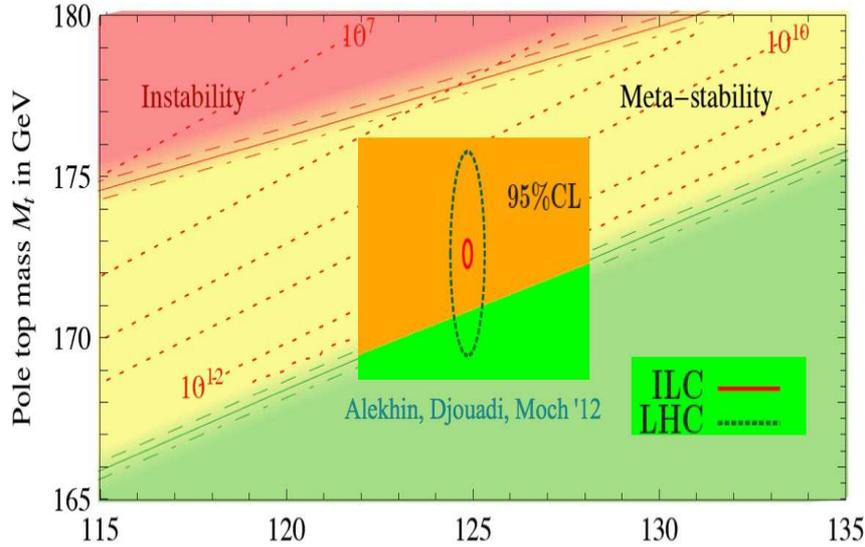}
\caption{This is the same figure as in the right panel of Fig.~\ref{fig:mtmh},  where the  zoomed  region around experimentally determined values from \protect\cite{bib:giudice} has ben overlaid with the uncertainties  of $M_{t}$ determination as extracted in~\protect\cite{bib:abdel-sven}. This was done by G. Isidori in his talk at SUSY 2014.}
\label{fig:vac-stab-ilc}
\end{center}
\end{figure}
that the determination of the $t$ mass at the ILC comes directly from measurement of the $t \bar t$ production  cross-section in $e^{+}e^{-}$ collisions,  near the $t \bar t$ threshold. This can be measured very accurately and has been computed theoretically to a high precision as well. This measurement can be converted into the pole mass in an unambiguous way. Fig.~\ref{fig:vac-stab-ilc}
shows how such a  precision measurement of the mass at the ILC can really shed  light on  whether the currently measured higgs mass points to the NEED of BSM physics at any {\bf particular high scale}. In the above figure, the bigger blue circle has been drawn assuming an LHC accuracy of $t$ mass measurement of $1$ GeV. However, a reduction of this error to about $500 \UMeV$ looks possible and is an active area of research. These kind of investigations are just the next logical step in our efforts to test the SM through a combination of the 'direct' and 'indirect' observations. 

\section{Concluding remarks}
In any case the days of Standard Model are coming to an end in
some sense!  Hopefully it will be the case of  'The King is Dead' and 'Long live the King'! We have, however, not much idea what particular BSM option, if any, would be  the new king. As we have discussed above, already the mass of the observed state can be used to answer the question about the scale upto which the SM is valid. In fact, this has been one of the most impressive facts about the SM. It has held the ability to ask and answer questions about its own consistency within its structure.  Just like the {\bf gauge principle} and the {\bf unitarity} were the 
guiding principle so far, now the {\bf small} mass of the discovered Higgs ($\sim {\cal O}$ weak scale) might be the guiding principle for future  theoretical developments! This will be discussed in other lectures at the school. We should  get a peek  at the BSM land through  the 'window' of measurement of the properties of the Higgs and the top quark! Exciting days are ahead for sure!
\begin{figure}[hbt]
\begin{center}
\includegraphics*[width=12cm,height=8cm]{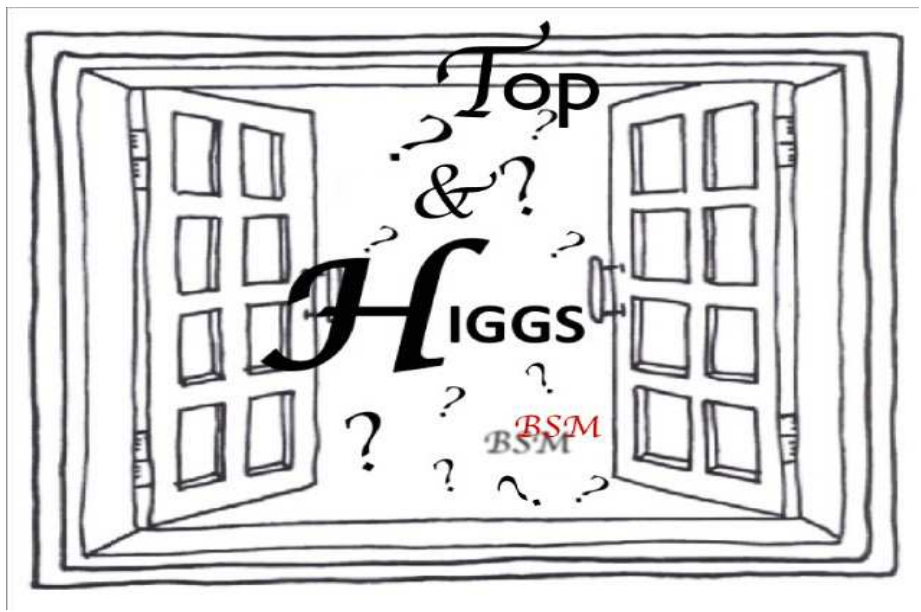}
\caption{The Higgs and Top portal for BSM physics.}
\end{center}
\end{figure}
If 14 TeV LHC should also fail to find 'direct' evidence for the
BSM physics we would really have to understand what is so special about the Standard Model. Precision measurements of the observed Higgs mass and Higgs couplings will be then our window to this world of physics beyond the SM.

\section{Acknowledgements}
I would like to acknowledge an ongoing collaboration with Sunil Mukhi to write a book on the Standard Model which is in preparation. I will also like to express my appreciation of the patience shown by Martijn Mulders in waiting for my lecture notes. I acknowledge hospitality of the Abdus Salam International Centre for Theoretical Physics where a substantial part of these notes was completed. Financial support from the Department of Science and Technology, India under Grant No. SR/S2/JCB-64/2007 under the J.C. Bose Fellowship scheme is gratefully acknowledged.
Last but not the least, I would like to acknowledge the AEPSHEP school organisers for asking me to give these lectures and the students for listening with patience! I would also like to acknowledge help from Ms. Anuja Thakar for providing some of the artwork and Dr. Gaurav Mendiratta for providing the drawing for Fig.~\ref{fig:mexicanhat}.

\section{About References}
In the bibliography I have listed the original theory papers and the early experimental papers which are referred to in the text from time to time. After that I list a large number of very good text books where one can find detailed discussions of many of the issues involved.

\end{document}